\documentclass[review]{elsarticle}
\usepackage[letterpaper,margin=1in]{geometry} \usepackage{graphicx} \usepackage[dvipsnames]{xcolor} \usepackage{tikz} \usetikzlibrary{backgrounds} \usetikzlibrary{arrows,shapes} \usepackage{amsmath} \usepackage{amsthm} \usepackage{amssymb} \usepackage{algpseudocode} \usepackage{mathtools, nccmath} \usepackage{wrapfig} \usepackage{comment} \usepackage{diagbox} \usepackage{algpseudocode} \usepackage{makecell} \usepackage{arydshln} \usepackage{algpseudocode} \usepackage{bm} \usepackage{threeparttable} \usepackage{braket} \usepackage[table]{xcolor}
\usepackage{xurl}
\usepackage{blindtext} \definecolor{navy}{RGB}{0,0,128}
\usepackage{placeins}
\usepackage{graphicx}
\usepackage{xspace}

\usepackage{array} \usepackage{ragged2e} \newcolumntype{P}[1]{>{\RaggedRight\hspace{0pt}}p{#1}} \newcolumntype{X}[1]{>{\RaggedRight\hspace*{0pt}}p{#1}}

\usepackage{tcolorbox} \usepackage{pgfplots} \pgfplotsset{compat=1.18}

\usepackage{tikz}
\usetikzlibrary{arrows,shapes,positioning,shadows,trees,mindmap}
\usepackage[edges]{forest} \usetikzlibrary{arrows.meta} \colorlet{linecol}{black!75} \usepackage{xkcdcolors} % xkcd colors

\usepackage{tikz} \usetikzlibrary{backgrounds} \usetikzlibrary{arrows,shapes} \usetikzlibrary{tikzmark} \usetikzlibrary{calc}
 \newcommand{\thicktoprule}{\Xhline{4\arrayrulewidth}} \newenvironment{reviewcomment}{\begin{quote}\itshape\sloppy}{\end{quote}} \newcommand{\thickbottomrule}{\Xhline{4\arrayrulewidth}} 
\colorlet{mhpurple}{Plum!80}

\usepackage{lineno,amsmath,amsfonts,amssymb,amsthm,mathrsfs} \usepackage[hyperfootnotes=false]{hyperref} \hypersetup{colorlinks=true,linkcolor=blue} \usepackage[linesnumbered,ruled,vlined,onelanguage]{algorithm2e} \usepackage{booktabs} \usepackage[skip=0pt]{caption,subcaption} \usepackage{float} \usepackage{longtable} \usepackage{setspace} \usepackage{tabularx} \usepackage{booktabs} \usepackage{tablefootnote}  
 \usepackage{todonotes} \journal{Transportation Science} \biboptions{authoryear}

\begin{document}
\begin{frontmatter}

\title{\textcolor{black}{Improving Feasibility in the Quantum Approximate Optimization Algorithm for Vehicle Routing via Constraint-Aware Initialization and Hybrid \(XY\)--\(X\) Mixing}}

\author[1]{Yuan-Zheng Lei} \author[1]{Yaobang Gong} \author[1]{Xianfeng Terry Yang\corref{cor1}} \ead{xtyang@umd.edu} \author[1]{Nii Attoh-Okine}

\cortext[cor1]{Corresponding author} \address[1]{Department of Civil \& Environmental Engineering, University of Maryland, 1173 Glenn Martin Hall, College Park, MD 20742, United States}
\begin{abstract}
{\color{black}The Quantum Approximate Optimization Algorithm (QAOA) is a leading framework for quantum combinatorial optimization. The vehicle routing problem (VRP), a core problem in logistics and transportation, is a natural application target but poses a major feasibility challenge for standard QAOA: feasible solutions typically occupy only a tiny fraction of the full binary search space, and the conventional Pauli-$X$ mixer can disrupt partial solution structures that already satisfy key local constraints. To address this issue, we propose a constraint-aware QAOA framework with two complementary components. First, we design an initialization strategy that encodes selected constraints into the initial state. This reduces the size of the initial superposition without explicitly constructing the complete feasible set and concentrates probability mass on states that satisfy the selected constraints. Second, we introduce a hybrid \(XY\)--\(X\) mixer whose preservation property is intentionally partial. The \(XY\)-exchange terms preserve only the selected structures chosen for protection, while the \(X\)-mixer terms retain exploration on qubits outside those structures, and the remaining VRP constraints are addressed through QUBO penalty terms and variational optimization. We evaluate the proposed framework on one three-node benchmark instance and two four-node benchmark instances under three regimes: exact-statevector objective optimization with noiseless finite-shot final sampling, ideal finite-shot optimization and sampling, and noisy finite-shot optimization and sampling. The experiments compare the proposed method with standard QAOA across all cases and with a block-wise XY-QAOA baseline for the four-node cases. Across the tested instances and regimes, the proposed method increases optimal-solution probability and reduces expected energy gap relative to standard QAOA. The four-node comparisons further show that block-wise XY-QAOA is a stronger feasibility-aware baseline than standard QAOA, but the proposed overlap-aware initialization and hybrid mixer generally provide a better balance between selected-constraint preservation and search flexibility. The advantage of the proposed ansatz is attenuated under hardware-inspired noise, in part because the structured mixers introduce additional two-qubit operations. These results highlight that the practical benefit of the proposed method depends on gate fidelity, readout quality, state-preparation cost, and circuit execution reliability.
}
\end{abstract}
\end{frontmatter}

\section{Introduction}
\par In both classical transportation and intelligent transportation systems (ITS), efficient routing has played and will continue to play a critical role in enabling better decision-making for transportation networks and supply chains. \textcolor{black}{One fundamental problem that has attracted sustained attention is the vehicle routing problem (VRP; \citep{toth2002vehicle}), which extends the traveling salesman problem (TSP; \citep{flood1956traveling,clarke1964scheduling}) and includes variants such as the capacitated vehicle routing problem (CVRP; \citep{ralphs2003capacitated}). These problems seek routes that minimize costs such as distance, fuel, or time and play an important role in urban logistics and mobility services.}
\par \textcolor{black}{VRP variants are NP-hard, and exact algorithms become computationally impractical as the problem size grows. Consequently, classical solvers often rely on heuristics or approximation algorithms to satisfy runtime requirements. This motivates the exploration of alternative computational algorithms and frameworks for VRPs and related combinatorial problems with substantial real-world impact.} \textcolor{black}{Quantum computing has motivated research into whether quantum superposition, entanglement, and interference can help tackle NP-hard problems by searching large solution spaces more effectively.} Although current quantum devices operate in the noisy intermediate-scale quantum (NISQ) regime with only tens to hundreds of qubits and significant noise, the field is advancing rapidly (\citep{preskill2018quantum}), and in recent years, we have witnessed significant improvements in gate and readout fidelity (\citep{li2023error, chen2023transmon, marxer2025above,wang202499}).
\par \textcolor{black}{A prominent quantum approach to combinatorial optimization is the Quantum Approximate Optimization Algorithm (QAOA; \citep{farhi2014quantum}). QAOA is a hybrid quantum--classical algorithm that seeks high-quality approximate solutions to hard combinatorial optimization problems by encoding an objective in a parameterized quantum circuit and using a classical optimizer to tune its parameters. By alternately applying a problem-specific cost unitary and a mixing unitary on \(n\) qubits, standard QAOA produces a parameterized quantum state. With appropriate angles and a sufficiently large depth \(p\), measuring this state can yield high-quality solutions among the \(2^n\) computational-basis bitstrings.}\footnote{\textcolor{black}{With the usual mixer \(H_M=\sum_i X_i\) and the initial state \(\lvert+\rangle^{\bigotimes n}\), the circuit typically assigns nonzero probability to all \(2^n\) bitstrings. In contrast, constraint-preserving mixers, such as the \(XY\) mixer, and related variants can restrict the evolution to a smaller subspace, such as a fixed-Hamming-weight sector or a feasible set, so some bitstrings cannot be reached.}} \textcolor{black}{QAOA has been applied to several optimization problems, including Maximum Cut (MaxCut; \citep{farhi2014quantum}), Maximum Independent Set (MIS; \citep{choi2019tutorial,zhou2020quantum}), the Binary Paint Shop Problem (BPSP; \citep{streif2021beating}), Binary Linear Least Squares (BLLS; \citep{borle2021quantum}), multi-knapsack problems \citep{awasthi2023quantum}, vehicle-routing problems \citep{azfar2025quantum,carmo2025warm}, and more generally, Quadratic Unconstrained Binary Optimization (QUBO) problems \citep{moussa2022unsupervised}.}\footnote{\textcolor{black}{For a detailed introduction to QAOA, see the survey by \citep{blekos2024review}.}}
\par \textcolor{black}{A common step in implementing QAOA for the VRP is to formulate the problem as a QUBO, or equivalently as an Ising model \citep{ising1925beitrag}.}\footnote{\textcolor{black}{A QUBO can be mapped directly to an Ising model and vice versa \citep{barahona1989experiments}.}} \textcolor{black}{This formulation typically converts the constrained optimization problem into an unconstrained one by incorporating constraints as penalty terms. Ideally, the encoding captures all constraints without introducing additional binary variables, and hence qubits, or substantially increasing circuit complexity. Given the limited qubit counts on near-term devices, formulations that use as few binary decision variables as possible are especially valuable. A substantial body of literature explores alternative QUBO/Ising encodings for the VRP \citep{palackal2023quantum}, as well as practical guidelines for choosing penalty weights \citep{montanez2024unbalanced} and parameter-search strategies. These choices directly affect both resource requirements and the quality of solutions produced by QAOA. Our focus is a different but equally important issue: how to promote solution feasibility more effectively and robustly in QAOA-based approaches. We therefore do not provide an in-depth discussion of alternative formulations or penalty-weight design and instead refer interested readers to the existing literature on QUBO encodings and penalty selection.}
\par When formulating the VRP for QAOA in an Ising-model framework, the binary decision variables $x_i\in\{0,1\}$ are mapped to spin variables $z_i\in\{-1,+1\}$ via $z_i=1-2x_i$ (equivalently, $x_i=(1-z_i)/2$). A standard QAOA then starts from the uniform superposition $\lvert \psi(0)\rangle=\lvert +\rangle^{\bigotimes n}$, which assigns nonzero amplitude to every computational-basis bit string in $\{0,1\}^n$ and therefore includes all feasible solutions, as well as the optimal one, whenever the instance is feasible. The drawback is that, for VRP encodings, feasibility typically occupies only a tiny fraction of the $2^n$ bit strings. For example, consider a toy VRP with two vehicles and three nodes (including the depot) under a directed link-based encoding with six binary arc variables, so that each candidate solution corresponds to a $6$-qubit string $\lvert x_1x_2x_3x_4x_5x_6\rangle$. Among the $2^6=64$ possible strings, only a handful satisfy the routing constraints, so the feasible fraction can be as small as $\frac{1}{64}$.
\par \textcolor{black}{This issue is further exacerbated by the choice of mixer. The most common choice is the transverse-field (Pauli-\(X\)) mixer, which applies identical single-qubit \(X\) rotations and thus enables independent bit flips across all qubits. Since such independent flips do not generally respect problem constraints, feasibility is not preserved during QAOA evolution. For instance, if the problem structure imposes a constraint \(x_5+x_6=1\), then the feasible configurations on qubits \((5,6)\) form the subspace \(\mathrm{span}\{\lvert01\rangle_{5,6},\lvert10\rangle_{5,6}\}\). An equal-weight state in this subspace is \((\lvert01\rangle_{5,6}+\lvert10\rangle_{5,6})/\sqrt{2}\). However, flipping either qubit can map a feasible configuration to \(\lvert00\rangle_{5,6}\) or \(\lvert11\rangle_{5,6}\), thereby destroying feasibility.}
\par Constraint-preserving alternatives, such as the XY mixer, operate by swapping $\lvert 01\rangle\leftrightarrow \lvert 10\rangle$ on selected qubit pairs while conserving the total number of ones (the Hamming weight). This conservation law can help maintain certain classes of constraints, but it also implies that each Hamming-weight sector evolves independently. Consequently, if feasible VRP solutions reside in a particular weight sector $k$ (e.g., $k=4$ in the above six-variable toy encoding), then any fixed-weight initialization with a mismatched weight (such as a $W$-state with weight one) cannot reach the feasible sector under an XY mixer. Moreover, because the XY mixer cannot change the Hamming weight of any basis component, any amplitude initially assigned to bit strings outside the feasible weight sector can never be steered into feasibility, even if those strings already satisfy a subset of the constraints. In this respect, the uniform Pauli-$X$ mixer can be advantageous: by allowing independent bit flips and thus changing Hamming weight, it at least preserves the \emph{possibility} of moving from an arbitrary bit string to a feasible one. For example, in the same six-variable setting, a nontrivial portion of the initial superposition may already satisfy some constraints (e.g., those involving $x_3,x_4,x_5,x_6$), and a strictly weight-preserving evolution would fail to leverage such partial feasibility if the remaining violations require changing the total number of ones. These observations motivate our emphasis on developing mixers and initialization strategies that more robustly promote solution feasibility, rather than providing an exhaustive discussion of alternative Ising/QUBO encodings or penalty-weight design.
\par \textcolor{black}{In this work, we address the feasibility bottleneck from two complementary angles.\label{sec:21}\label{line:21} First, instead of initializing from a uniform superposition over all \(2^n\) bitstrings, we construct a constraint-aware initial state by restricting the superposition to a carefully chosen subset of states that is consistent with the problem structure. This restriction substantially reduces the number of basis states that receive nonzero amplitude while retaining all feasible solutions for the tested instances, as verified by exhaustive enumeration, thereby increasing the feasible-to-infeasible ratio at initialization. Unlike the Grover-mixer variant of QAOA in \citep{bartschi2020grover}, whose initialization is supported only on fully feasible states, the proposed support is specified through selected local constraints and does not require explicit construction of the complete feasible set. This distinction concerns the logical support construction; the gate-level cost of preparing the prescribed superposition depends on the chosen state-preparation decomposition and is not included in the present noise model. Second, we propose a hybrid \(XY\)--\(X\) mixer with a selected-constraint-preserving rather than fully feasible-subspace-preserving guarantee. The initialization encodes only selected local constraints, the \(XY\)-exchange terms preserve the selected one-hot or overlap structures, the \(X\)-mixer terms retain exploration on qubits outside those protected structures, and the remaining VRP constraints are handled by QUBO penalties and subsequent variational optimization. Across the tested small VRP instances and simulation regimes, the proposed initialization and mixer improve feasibility-oriented performance, although the relative advantage is attenuated under noisy finite-shot simulation.}

\section{Literature review}
\par \textcolor{black}{Over the past several decades, quantum computing has attracted attention as a potential means of addressing computational tasks that are difficult for classical methods \citep{shor1994algorithms,steane1998quantum,preskill2023quantum}. The Hilbert-space dimension of an \(n\)-qubit system grows exponentially with \(n\), a property often described as exponential parallelism \citep{rieffel2000introduction}. An \(n\)-qubit register can therefore be prepared in a coherent superposition over up to \(2^n\) computational-basis states, and a unitary operation acts on all components of that state. This does not permit direct readout of \(2^n\) classical results in one run; rather, quantum algorithms exploit interference and entanglement to amplify the probability of desirable outcomes upon measurement \citep{james2001measurement,chatterjee2021semiconductor}. Quantum-computing applications have also been explored in transportation research \citep{cooper2021exploring,zhuang2024quantum,somvanshi2026quantum,udekwe2025q,ke6431107quantum,ke2026impact,azfar2026hardware}. Variational quantum algorithms (VQAs), including QAOA and the Variational Quantum Eigensolver (VQE) \citep{peruzzo2014variational,wang2019accelerated}, are designed for current hardware by combining a parameterized quantum circuit with a classical optimizer that minimizes an objective estimated from circuit measurements. Such algorithms typically use relatively shallow circuits, which can reduce their exposure to noise on NISQ devices \citep{blekos2024review}. This paper focuses on QAOA because it is directly aligned with the present optimization objective. For broader discussions of VQAs, including VQE, see \citep{tilly2022variational}.}
\par \citep{harwood2021formulating} focuses on how different mathematical formulations of routing problems affect quantum solvability, rather than modifying the QAOA ansatz itself: using the vehicle routing problem with time windows (VRPTW) as the main testbed, it compares multiple modeling routes that map the problem into binary optimization form, including QUBO-based encodings and an ADMM-based decomposition that yields QUBO subproblems, and discusses how these choices change the number of binary variables, coupling structure, and ultimately circuit resources, while employing standard QAOA as a representative heuristic solver for the resulting QUBO instances. Along a similar modeling-to-quantum-solver pipeline, \citep{ak2025quantum} studies a multi-vessel LNG ship-routing problem and integrates a digital-twin data layer with quantum optimization: the routing task is formulated as an MILP and then converted into a QUBO by adding quadratic penalty terms that enforce flow conservation, origin--destination constraints, and shared-edge constraints; the resulting QUBO is solved using a standard QAOA-style variational circuit implemented in Qiskit, and the paper primarily emphasizes the modeling-to-QUBO translation and the role of penalty tuning and data-driven cost updates, rather than proposing a new initialization or mixer design. In contrast, \citep{azad2022solving} directly formulates the VRP as an Ising Hamiltonian and solves it via QAOA with the canonical uniform-superposition initialization and the conventional transverse-field Pauli-$X$ mixer, emphasizing that performance depends strongly on factors such as circuit depth, the classical optimizer, parameter initialization, and problem characteristics; \citep{azfar2025quantum} follows the same Ising/QUBO encoding and standard-QAOA circuit structure, but evaluates it on real quantum hardware and highlights how penalty scaling, coefficient normalization, and circuit depth jointly affect feasibility under hardware noise. Related traffic applications also adopt this standard QAOA recipe: \citep{harikrishnakumar2021smart} applies QAOA to a bike-sharing rebalancing problem, where bikes are transported between surplus and deficit stations to reduce system imbalance while minimizing the travel cost of the rebalancing vehicle; the problem is encoded as a QUBO with quadratic penalty terms for operational constraints and is solved using the uniform-superposition initialization and the transverse-field Pauli-$X$ mixer, with Qiskit-based simulations illustrating the sensitivity of feasibility and solution quality to circuit depth and penalty scaling. Complementary to these application-driven studies, \citep{egger2021warm} proposes warm-start QAOA, where a classical relaxation guides both initialization and mixing: instead of $|+\rangle^{\bigotimes n}$ with an $X$ mixer, it uses a biased product-state initialization whose single-qubit marginals match the relaxed solution together with a matching warm-start mixer, and introduces an $\epsilon$-regularization to avoid frozen dynamics, ensure nonzero overlap with every computational-basis state, and provide a smooth interpolation back to standard QAOA, with particular benefits in the low-depth regime and approximation guarantees when the warm start arises from randomized rounding. Returning to routing problems under realistic conditions, \citep{mohanty2023analysis} adopts the same penalty-based QUBO/Ising encoding philosophy as \citep{azad2022solving,azfar2025quantum} and still uses the standard QAOA-style alternating structure with the uniform-superposition initialization and the transverse-field Pauli-$X$ mixer, but shifts the emphasis to robustness by systematically quantifying how noisy channels and circuit depth affect solution quality and feasibility when parameters are tuned by classical optimizers in the presence of noise. In a related shared-mobility setting, \citep{onah2025quest} studies a Windbreaking-as-a-Service matching problem and formulates the surfer--breaker assignment as a binary optimization model that can be mapped to a QUBO/Ising Hamiltonian; using this encoding, the authors again apply standard QAOA with the uniform-superposition initialization and the transverse-field Pauli-$X$ mixer, and evaluate the approach against classical baselines on small instances. Beyond standard QAOA, several works modify the ansatz to promote feasibility more directly: building on warm-start QAOA while explicitly addressing combinatorial feasibility, \citep{carmo2025warm} makes the warm-start idea constraint-aware for routing-style subproblems by initializing directly within the one-hot subspace and using an $XY$ mixer that preserves Hamming weight within each register, so evolution remains within the intended one-hot structure; the warm-start signal is derived from a Goemans--Williamson MaxCut relaxation and biases the superposition over valid one-hot states, increasing the fraction of valid tours without relying solely on penalty scaling. \citep{bartschi2020grover} takes feasibility-by-design further by replacing both the full-space initialization and local mixers with a Grover-mixer variant of QAOA: it assumes a state-preparation routine that produces an equal-weight superposition over feasible computational-basis states only and employs a Grover-style selective phase-shift mixer defined with respect to this feasible superposition, which performs global mixing entirely within the feasible set, so the entire QAOA evolution remains strictly supported on feasible states at every layer rather than depending on penalty terms or post-selection. Building on the broader idea of Grover-style mixing for constrained optimization \citep{bartschi2020grover}, \citep{picariello2025quantum} applies QAOA to TSP instances augmented with logistics-motivated constraints for urban delivery settings: the paper introduces a Grover-inspired mixer that enforces the canonical one-city-per-step one-hot constraint by construction, so the quantum evolution remains within the corresponding structured subspace, while the remaining constraint that each city is visited once is still handled via a penalty in the cost Hamiltonian; to improve scalability beyond small instances, the authors further propose a clustering variant (Cl-QAOA) that decomposes large instances into smaller subproblems, solves them with QAOA, and then recombines them, enabling experiments on much larger, data-driven instances. More broadly, QAOA has been applied to a wide range of combinatorial optimization problems beyond routing, including tail assignment in airline scheduling \citep{vikstaal2020applying} and portfolio optimization \citep{baker2022wasserstein}. Given the diversity of application domains and the rapidly growing number of QAOA variants, a comprehensive survey is beyond the scope of this work; interested readers are referred to \citep{blekos2024review} for a detailed review of QAOA and its major variants.
\par \textcolor{black}{The remainder of this paper is organized as follows: in section \ref{section:3}, we review key concepts and definitions needed to understand quantum computing and QAOA. Section \ref{section:4} presents the proposed initialization and mixer design, and }\textcolor{blue}{\textbf{Section}}~\ref{section:5}\textcolor{black}{ compares the proposed method with the baselines under three experimental regimes: exact-statevector objective optimization followed by noiseless finite-shot final sampling, finite-shot objective optimization and final sampling, and noisy finite-shot optimization and sampling with gate and readout errors. Finally, section \ref{section:6} discusses the observed advantages and their limitations under hardware imperfections.}

\section{\textcolor{black}{Conceptual review}} \label{section:3}
\par \textcolor{black}{Quantum computing differs fundamentally from classical computing. This section introduces the concepts and definitions needed to understand the proposed method.}
\par \textcolor{black}{The first concept is the tensor product, which combines vector spaces to form larger vector spaces and is central to the quantum mechanics of multipartite systems \citep{nielsen2010quantum}. To illustrate the tensor product concretely, let \(A\) be an \(m\times n\) matrix and \(B\) be a \(p\times q\) matrix. Their tensor product has the following matrix representation:}
\begin{equation}
A\bigotimes B \equiv
\overbrace{
\left[
\left.
\begin{array}{cccc}
A_{11}B & A_{12}B & \cdots & A_{1n}B\\
A_{21}B & A_{22}B & \cdots & A_{2n}B\\
\vdots  & \vdots  & \ddots & \vdots\\
A_{m1}B & A_{m2}B & \cdots & A_{mn}B
\end{array}
\right]
\right\}^{mp}
}^{nq} \label{eq:1}
\end{equation}
where each block $A_{ij}B$ is the matrix $B$ scaled by the scalar $A_{ij}$. \textcolor{black}{For example, the tensor product of the matrices} $A = \left[ \begin{array}{cc}
    0 & 1 \\
    1 & 0
\end{array} \right]$ and $B = \left[ \begin{array}{cc}
    0 \ & -i \\
    i & 0
\end{array} \right]$ is:
\begin{equation}
A\bigotimes B =  \left[ \begin{array}{cc}
    0 \cdot B & 1 \cdot B \\
    1 \cdot B & 0 \cdot B 
\end{array} \right]  = \left[ \begin{array}{cccc}
    0  & 0 & 0 & -i \\
    0  & 0 & i & 0  \\
    0  & -i & 0 & 0 \\
    i & 0 & 0 & 0
\end{array} \right] \label{eq:2}
\end{equation}
\par \textcolor{black}{A bit is the fundamental unit of classical computation and information \citep{nielsen2010quantum}. Quantum computation and quantum information are built on the analogous quantum bit, or qubit. A classical bit is in one of the two states \(0\) and \(1\). A qubit has the computational-basis states \(\lvert0\rangle\) and \(\lvert1\rangle\), but it can also occupy a linear combination of these states, called a superposition:}
\begin{equation}
    |\psi\rangle = \alpha |0\rangle + \beta|1\rangle \label{eq:3}
\end{equation}
\textcolor{black}{where \(\alpha\) and \(\beta\) are complex numbers. A computational-basis measurement returns \(0\) with probability \(\lvert\alpha\rvert^2\) and \(1\) with probability \(\lvert\beta\rvert^2\). Therefore, \(\alpha\) and \(\beta\) satisfy the normalization relation}
\begin{equation}
    |\alpha|^{2} +  |\beta|^{2} = 1 \label{eq:4}
\end{equation}
\textcolor{black}{because the probabilities must sum to one. The states \(\lvert0\rangle\) and \(\lvert1\rangle\) are the computational-basis states and can be represented as column vectors, where \(\lvert0\rangle = \left[ \begin{array}{c}
     1  \\
    0
\end{array}\right]\) and \(\lvert1\rangle = \left[ \begin{array}{c}
     0  \\
    1
\end{array}\right]\). An equal-amplitude superposition of the two computational-basis states is}
\begin{equation}
    \frac{1}{\sqrt{2}} |0\rangle + \frac{1}{\sqrt{2}}|1\rangle \label{eq:5}
\end{equation}
which is sometimes denoted as $|+\rangle$ (plus state).
\par
\begin{figure}
    \centering
    \includegraphics[width=0.6\linewidth]{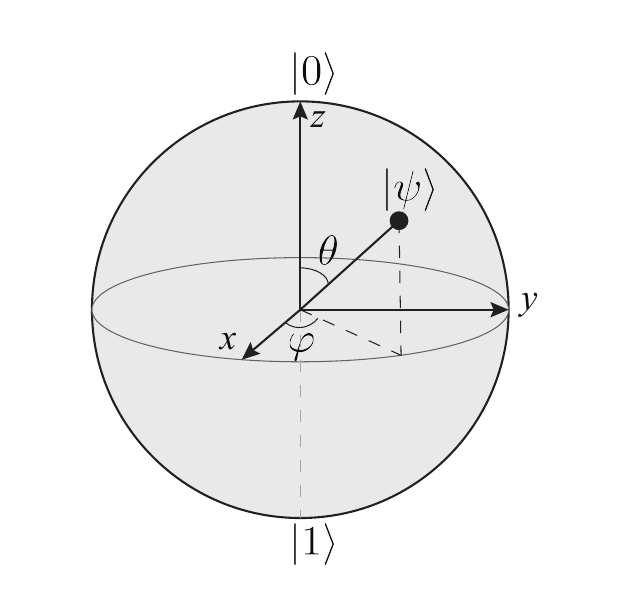}
    \caption{Bloch sphere representation of a qubit}
    \label{fig:1}
\end{figure}

 \textcolor{black}{A pure single-qubit state also admits a geometric representation. By Euler's formula,}
\begin{equation}
\label{eq:6}
    e^{ix} = \cos{x} + i\sin{x}
\end{equation}
\textcolor{black}{After removing an unobservable global phase, the state in \eqref{eq:3} can be written as}
\begin{equation}
    |\psi\rangle = \cos{\frac{\theta}{2}}|0\rangle + e^{i\varphi}\sin{\frac{\theta}{2}}|1\rangle \label{eq:7}
\end{equation}
\textcolor{black}{where \(0\leq\theta\leq\pi\) and \(0\leq\varphi<2\pi\) specify a point on the unit two-sphere embedded in three-dimensional space. This sphere is the Bloch sphere shown in }\textcolor{blue}{\textbf{Figure}}~\ref{fig:1}\textcolor{black}{, which provides a geometric representation of a pure single-qubit state up to global phase.}
\par Having introduced the tensor product and the basic notions of a single qubit, we next describe quantum gates and circuits, which specify how quantum states are manipulated in practice. A quantum gate is a physical operation that transforms quantum states via a unitary matrix, and a quantum circuit is a sequence of such gates applied to one or more qubits. Unlike classical logic gates, quantum gates must be reversible and therefore unitary, which makes them suitable for coherent evolution and quantum interference.
\par Common single-qubit gates include the Pauli gates $X$, $Y$ and $Z$ and the Hadamard gate $H$. Their matrix representations are given by
\begin{equation}
X=
\begin{bmatrix}
0 & 1\\
1 & 0
\end{bmatrix}
\quad
Y=
\begin{bmatrix}
0 & -i\\
i & 0
\end{bmatrix}
\quad
Z=
\begin{bmatrix}
1 & 0\\
0 & -1
\end{bmatrix}
\quad
H=\frac{1}{\sqrt{2}}
\begin{bmatrix}
1 & 1\\
1 & -1
\end{bmatrix} \label{eq:8}
\end{equation}
The Pauli-$X$ gate flips the computational-basis states, i.e., $X|0\rangle=|1\rangle$ and $X|1\rangle=|0\rangle$ (as illustrated in \eqref{eq:9}), analogous to a classical NOT operation. The Pauli-$Z$ gate keeps $|0\rangle$ unchanged and adds a phase of $-1$ to $|1\rangle$, i.e., $Z|0\rangle=|0\rangle$ and $Z|1\rangle=-|1\rangle$ (as illustrated in \eqref{eq:10}), which modifies relative phase without changing measurement probabilities in the computational basis. The Pauli-$Y$ gate combines a bit-flip with a phase shift and corresponds to a rotation about the $y$ axis on the Bloch sphere. The Hadamard gate is widely used to create superposition, for example $H|0\rangle=|+\rangle$ and $H|1\rangle=|-\rangle$ (as illustrated in \eqref{eq:11}).

\begin{equation}
X|0\rangle=
\begin{bmatrix}
0 & 1\\
1 & 0
\end{bmatrix}
\begin{bmatrix}
1\\
0
\end{bmatrix}
=
\begin{bmatrix}
0\\
1
\end{bmatrix}
=
|1\rangle
\quad
X|1\rangle=
\begin{bmatrix}
0 & 1\\
1 & 0
\end{bmatrix}
\begin{bmatrix}
0\\
1
\end{bmatrix}
=
\begin{bmatrix}
1\\
0
\end{bmatrix}
=
|0\rangle  \label{eq:9}
\end{equation}

\begin{equation}
Z|0\rangle=
\begin{bmatrix}
1 & 0\\
0 & -1
\end{bmatrix}
\begin{bmatrix}
1\\
0
\end{bmatrix}
=
\begin{bmatrix}
1\\
0
\end{bmatrix}
=
|0\rangle
\quad
Z|1\rangle=
\begin{bmatrix}
1 & 0\\
0 & -1
\end{bmatrix}
\begin{bmatrix}
0\\
1
\end{bmatrix}
=
\begin{bmatrix}
0\\
-1
\end{bmatrix}
=
-|1\rangle \label{eq:10}
\end{equation}

\begin{equation}
\color{black}
\begin{aligned}
H|0\rangle
&=
\frac{1}{\sqrt{2}}
\begin{bmatrix}
1 & 1\\
1 & -1
\end{bmatrix}
\begin{bmatrix}
1\\
0
\end{bmatrix}
=
\frac{1}{\sqrt{2}}
\begin{bmatrix}
1\\
1
\end{bmatrix}
=
|+\rangle
\\
H|1\rangle
&=
\frac{1}{\sqrt{2}}
\begin{bmatrix}
1 & 1\\
1 & -1
\end{bmatrix}
\begin{bmatrix}
0\\
1
\end{bmatrix}
=
\frac{1}{\sqrt{2}}
\begin{bmatrix}
1\\
-1
\end{bmatrix}
=
|-\rangle
\end{aligned}
\label{eq:11}
\end{equation}
\par A convenient continuous family of single-qubit gates is given by rotations about the Bloch-sphere axes
\begin{equation}
\textcolor{black}{
R_X(\theta)=e^{-i\theta X/2}
\quad
R_Y(\theta)=e^{-i\theta Y/2}
\quad
R_Z(\theta)=e^{-i\theta Z/2}
}
\label{eq:12}
\end{equation}
\textcolor{black}{Geometrically, $R_X(\theta)$, $R_Y(\theta)$, and $R_Z(\theta)$ rotate the Bloch vector by angle $\theta$ around the $x$, $y$, and $z$ axes, respectively. In variational quantum algorithms, these parameterized rotations provide tunable controls and are commonly used to encode adjustable angles in the circuit.}
\par To couple qubits and create entanglement, two-qubit gates are required. The controlled-NOT (CNOT) gate flips a target qubit if and only if the control qubit is in state $|1\rangle$, while the controlled-$Z$ (CZ) gate applies a $Z$ phase conditioned on the control. A widely used parameterized two-qubit interaction is the $ZZ$-rotation
\begin{equation}
R_{ZZ}(\theta)=e^{-i\theta (Z\bigotimes Z)/2}  \label{eq:13}
\end{equation}
which correlates the phases of two qubits and serves as a convenient primitive for implementing Ising-type couplings.
\par \textcolor{black}{These gates provide the circuit-level implementation of QAOA. In particular, the mixer unitary generated by $\sum_i X_i$ can be realized by applying single-qubit $R_X$ rotations to each qubit, whereas the cost unitary for an Ising/QUBO objective can be implemented using $R_Z$ gates for single-qubit $Z_i$ terms and two-qubit constructions such as $R_{ZZ}$ (or equivalent decompositions using CNOT and $R_Z$ gates) for pairwise $Z_iZ_j$ couplings.}

% \begin{equation}
% \begin{aligned}
% |\psi(0)\rangle
% &=
% \frac{1}{\sqrt{\binom{m-1}{K}}}
% \sum_{\substack{S\subseteq \{1,2,\ldots,m-1\}\\ |S|=K}}
% \left(\ \bigotimes_{i\in S} |10\rangle_i\right)
% \bigotimes
% \left(\ \bigotimes_{i\in \{1,2,\ldots,m-1\}\setminus S} |01\rangle_i\right) \\
% &=
% \frac{1}{\sqrt{\binom{m-1}{K}}}
% \sum_{\substack{S\subseteq \{1,2,\ldots,m-1\}\\ |S|=K}}
% \bigotimes_{i=1}^{m-1}
% \begin{cases}
% |10\rangle_i, & i\in S,\\
% |01\rangle_i, & i\notin S,
% \end{cases}
% \end{aligned}
% \end{equation}

\section{Methodology} \label{section:4}
\par \textcolor{black}{This section first introduces the standard vehicle-routing formulation and QAOA, and then explains the proposed constraint-aware initialization and hybrid \(XY\)--\(X\) mixer in depth.}

\subsection{Vehicle routing problem}
\par The VRP admits multiple modeling paradigms, including link-based and route-based formulations. Since our numerical experiments adopt a link-based formulation that maps naturally to a binary (QUBO/Ising) encoding, we briefly introduce the link-based model and omit route-based approaches for concision.
\par \textcolor{black}{In this study, we assume that each customer's demand is fulfilled upon the vehicle's arrival at the corresponding node. Under this assumption, the objective is to minimize the total travel cost, i.e., the sum of the distances (or costs) over all traversed links. Let \(x_{i,j}\in\{0,1\}\) denote whether the directed link \((i,j)\) between nodes \(i\) and \(j\) in the node set \(\mathcal{N}\) is used, let \(w_{i,j}\) denote the corresponding link distance (or cost), and let \(N_{\mathrm{node}}\) denote the number of nodes. The resulting link-based VRP can be formulated as follows:}
\par \noindent\textbf{Input:}
\begin{itemize}
    \item \textcolor{black}{\(\mathcal{N}\): Set of nodes, indexed by \(0,1,\ldots,N_{\mathrm{node}}-1\)}
    \item \textcolor{black}{\(w_{i,j}\in\mathbb{R}^{+}\): Distance or cost of the directed link \((i,j)\)}
    \item $k$: The total number of vehicles.
    \item \textcolor{black}{\(\mathcal{S}\): Any nonempty subset of the node set \(\mathcal{N}\)}
\end{itemize}
\textbf{Decision Variables:}
\begin{itemize}
    \item \textcolor{black}{\(x_{i,j}\in\{0,1\}\): \(1\) if the directed link from node \(i\) to node \(j\) is selected}
\end{itemize}
\textbf{Mathematical Formulation:}
\begin{subequations}
\label{eq:14}
\begin{align}
  & \textcolor{black}{\min_{\mathbf{x}}\sum_{i\in\mathcal{N}}\sum_{\substack{j\in\mathcal{N}\\ j\neq i}}w_{i,j}x_{i,j}} \label{eq:14a}
    \\
  &\text{subject to}
    \textcolor{black}{\sum_{\substack{i\in\mathcal{N}\\ i\neq j}}x_{i,j}=1\quad \forall j\in\mathcal{N}\setminus\{0\}} \label{eq:14b}
    \\
  &\qquad
    \textcolor{black}{\sum_{\substack{j\in\mathcal{N}\\ j\neq i}}x_{i,j}=\sum_{\substack{j\in\mathcal{N}\\ j\neq i}}x_{j,i}\quad \forall i\in\mathcal{N}\setminus\{0\}} \label{eq:14c}
    \\
  &\qquad
    \textcolor{black}{\sum_{j\in\mathcal{N}\setminus\{0\}}x_{0,j} = k\quad \sum_{i\in\mathcal{N}\setminus\{0\}}x_{i,0} = k} \label{eq:14d}
    \\
  &\qquad
    \textcolor{black}{\sum_{i \in \mathcal{S}}\sum_{j \in \mathcal{N}\setminus\mathcal{S}}x_{i,j} \geq 1\quad \forall \mathcal{S} \subseteq \mathcal{N}\setminus\{0\}\quad 2 \leq |\mathcal{S}| \leq N_{\mathrm{node}} - 1} \label{eq:14e}
\end{align}
\end{subequations}
\textcolor{black}{Here, \eqref{eq:14a} defines the total travel-cost objective, \eqref{eq:14b} enforces the customer-visit requirement, and \eqref{eq:14c} imposes flow balance so that any vehicle entering a customer node must also depart from it. Constraint \eqref{eq:14d} ensures that exactly \(k\) vehicles leave the depot and return to it. Finally, \eqref{eq:14e} provides subtour-elimination constraints to prevent disjoint loops that do not include the depot.}
\subsection{QAOA formulation}
\par
\begin{figure}
    \centering
    \includegraphics[width=1.0\linewidth]{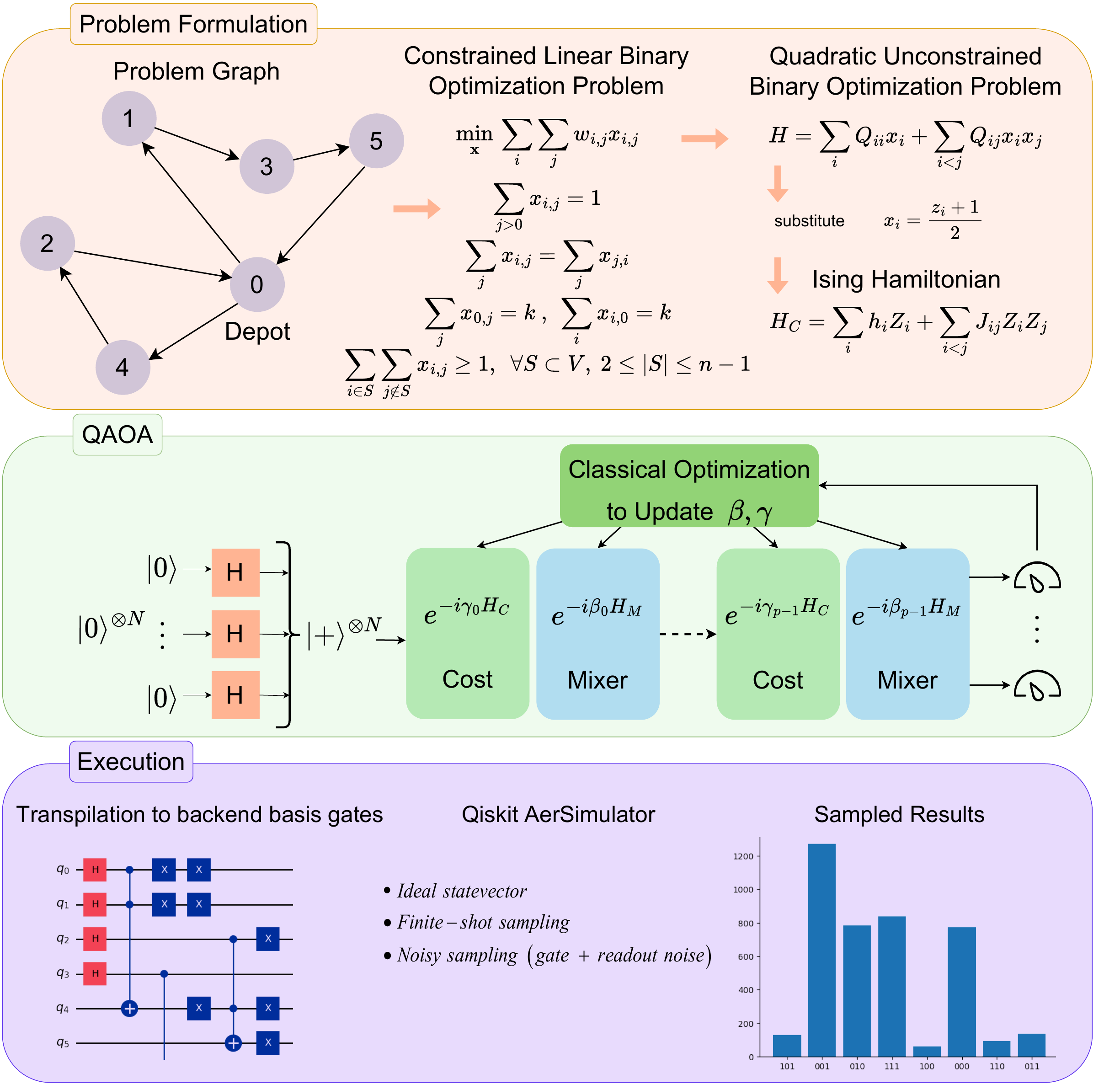}
    \caption{QAOA Pipeline (Adapted from \citep{azfar2025quantum})}
    \label{fig:2}
\end{figure}

 \textcolor{black}{As a hybrid quantum--classical algorithm, QAOA combines quantum state preparation and measurement with a classical outer-loop optimizer, as illustrated in }\textcolor{blue}{\textbf{Figure}}~\ref{fig:2}\textcolor{black}{. In the standard QAOA workflow, a first and often essential step is to rewrite the original binary optimization problem in an Ising form. The reason is practical: in a quantum circuit, the most natural diagonal observable is the Pauli-$Z$ operator, whose computational-basis eigenvalues are $\pm 1$. Therefore, instead of working directly with binary variables $x_i\in\{0,1\}$, it is convenient to introduce classical spin variables $z_i\in\{-1,+1\}$ corresponding to the eigenvalues of the Pauli-$Z$ operator. The two representations are connected by the one-to-one affine mapping}
\begin{equation}
\label{eq:15}
\color{black}
z_i=1-2x_i
\qquad
x_i=\frac{1}{2}\left(1-z_i\right)
\end{equation}
Accordingly, $x_i=0$ corresponds to $z_i=+1$, whereas $x_i=1$ corresponds to $z_i=-1$. This convention is consistent with the computational basis because $Z_i|0\rangle=+|0\rangle$ and $Z_i|1\rangle=-|1\rangle$. At the operator level, the binary variable is therefore represented by $x_i\mapsto (I-Z_i)/2$, where $I$ denotes the identity operator and $Z_i$ denotes the Pauli-$Z$ operator acting on qubit $i$. Substituting this mapping into the original objective $C(\mathbf{x})$ yields an equivalent classical Ising energy function $E(\mathbf{z})$ defined over $\{-1,+1\}^n$. To use this energy function in QAOA, each classical spin variable is promoted to the corresponding Pauli operator through $z_i\mapsto Z_i$. This produces the full quantum cost operator $\widehat{C}$. Since the constant term contributes only a global phase to the cost unitary and does not affect measurement probabilities or the optimization of the variational parameters, it can be omitted when defining the QAOA cost Hamiltonian:
\begin{equation}
\label{eq:16}
\color{black}
\begin{aligned}
E(\mathbf{z})
&=
C_{\mathrm{I}}+\sum_i h_i z_i+\sum_{i<j}J_{ij}z_i z_j
\\
\widehat{C}
&=
C_{\mathrm{I}}I+\sum_i h_i Z_i+\sum_{i<j}J_{ij}Z_iZ_j
\\
H_C
&=
\widehat{C}-C_{\mathrm{I}}I
=
\sum_i h_i Z_i+\sum_{i<j}J_{ij}Z_iZ_j
\end{aligned}
\end{equation}
\textcolor{black}{Here, \(C_{\mathrm{I}}\), \(h_i\), and \(J_{ij}\) are real coefficients determined by the original QUBO formulation. The classical energy function \(E(\mathbf{z})\) is defined in terms of the scalar spin variables \(z_i\in\{-1,+1\}\), whereas \(\widehat{C}\) and \(H_C\) are quantum operators expressed in terms of the Pauli-\(Z\) operators \(Z_i\). Both \(\widehat{C}\) and \(H_C\) are diagonal in the computational basis. For any bitstring \(\mathbf{x}\) and its corresponding spin assignment \(\mathbf{z}\), the computational-basis state \(|\mathbf{x}\rangle\) is an eigenstate of \(\widehat{C}\) whose eigenvalue equals the original objective value \(C(\mathbf{x})=E(\mathbf{z})\). Because \(H_C\) differs from \(\widehat{C}\) only by the constant term \(C_{\mathrm{I}}I\), the two operators have the same minimizing computational-basis states. Consequently, minimizing the original objective can be recast as seeking low-energy eigenstates of \(H_C\), and QAOA targets this goal by applying the corresponding cost unitary \(e^{-i\gamma H_C}\) within its variational circuit.}
\par Similar to classical optimization methods that require one or more initial guesses, QAOA starts from an initial quantum state at $t=0$, which serves as the starting point of the variational evolution. The most common choice is the uniform superposition over all computational-basis states
\begin{equation}
|\psi(0)\rangle
= H^{\bigotimes n}|0\rangle^{\bigotimes n}
= \frac{1}{\sqrt{2^{n}}}\sum_{x=0}^{2^{n}-1}|x\rangle
= |+\rangle^{\bigotimes n} \label{eq:17}
\end{equation}
which assigns equal amplitude to every bitstring. For example, when $n=2$, this initialization yields
\begin{equation}
|\psi(0)\rangle=\frac{1}{2}\Big(|00\rangle+|01\rangle+|10\rangle+|11\rangle\Big) \label{eq:18}
\end{equation}
and thus each bitstring is observed with probability $(1/2)^2=1/4$ upon measurement. The advantage of this initialization is that it covers the entire search space and introduces maximal diversity. However, for constrained problems, feasible solutions may occupy only a small fraction of the full space, so the initial probability mass on feasible states can be low. In later sections, we will also discuss alternative initialization strategies; \textcolor{blue}{\textbf{Figure}}~\ref{fig:2} illustrates this standard uniform-superposition initialization.
{\color{black}
\par The role of the cost unitary in QAOA can be understood by comparing it with classical objective evaluation. In a classical optimizer, one selects a candidate solution \(\mathbf{x}\), substitutes it into the objective, and obtains a scalar value \(C(\mathbf{x})\) for comparison. In QAOA, the objective is encoded in the cost Hamiltonian \(H_C\), which is diagonal in the computational basis. As a result, applying the cost unitary does not compute and output \(C(\mathbf{x})\) as a number; instead, it encodes the shifted objective value into the quantum phase of each basis state. Specifically, for any computational-basis state \(|\mathbf{x}\rangle\),
\begin{equation}
H_C|\mathbf{x}\rangle = \widetilde{C}(\mathbf{x})\,|\mathbf{x}\rangle
\qquad
\widetilde{C}(\mathbf{x})=C(\mathbf{x})-C_{\mathrm{I}}
\label{eq:19}
\end{equation}
and thus
\begin{equation}
e^{-i\gamma H_C}|\mathbf{x}\rangle = e^{-i\gamma \widetilde{C}(\mathbf{x})}\,|\mathbf{x}\rangle
\label{eq:20}
\end{equation}
where \(\gamma\) is a variational parameter. The omitted constant \(C_{\mathrm{I}}\) would contribute only a global phase and therefore does not affect measurement probabilities. Thus, lower-cost and higher-cost bit strings acquire different relative phases. Importantly, this phase rotation alone does not change the measurement probability of \(|\mathbf{x}\rangle\) because the magnitude of its amplitude is preserved. The subsequent mixer unitary then couples different basis states so that these cost-dependent phase differences can interfere and translate into changes in amplitudes, thereby increasing the probability of sampling low-cost solutions after measurement.
}
\par \textcolor{black}{Putting these components together, a $p$-layer QAOA circuit applies the cost unitary followed by the mixer unitary in each layer, starting from the initial state $|\psi(0)\rangle$}
\begin{equation}
\textcolor{black}{
|\psi(\bm{\gamma},\bm{\beta})\rangle
=
e^{-i\beta_pH_M}e^{-i\gamma_pH_C}
\cdots
e^{-i\beta_1H_M}e^{-i\gamma_1H_C}
|\psi(0)\rangle
}
\label{eq:21}
\end{equation}
\textcolor{black}{where $\bm{\gamma}=(\gamma_1,\dots,\gamma_p)$ and $\bm{\beta}=(\beta_1,\dots,\beta_p)$ are variational parameters. The rightmost pair with index (1) acts first, matching the layer order used in the implementation. In standard QAOA, the mixer Hamiltonian is chosen as}
\begin{equation}
\textcolor{black}{H_M=\sum_{i=0}^{n-1}X_i}
\label{eq:22}
\end{equation}
where $X_i$ denotes the Pauli-$X$ operator acting on the $i$th qubit. The corresponding mixer unitary can therefore be written as
\begin{equation}
\textcolor{black}{
e^{-i\beta H_M}
=
e^{-i\beta\sum_{i=0}^{n-1}X_i}
=
\prod_{i=0}^{n-1} e^{-i\beta X_i}
=
\prod_{i=0}^{n-1} R_X^{(i)}(2\beta)
}
\label{eq:23}
\end{equation}
\textcolor{black}{where \(R_X^{(i)}\) denotes an \(x\)-axis rotation acting on qubit \(q_i\).} This shows that the standard mixer applies an $x$-axis rotation to every qubit. Its effect is to continuously mix the amplitudes of computational-basis states, allowing the algorithm to move between bitstrings that differ in one or more binary entries. This choice provides strong exploratory power because it enables broad movement over the full search space. However, precisely because the standard $X$-mixer acts on all qubits without explicitly respecting problem constraints, it may also disrupt structural patterns associated with feasible solutions. For constrained optimization problems, this means that probability mass can be transferred from feasible states to infeasible ones during the evolution. This limitation motivates the use of alternative initialization strategies or constraint-preserving mixers in later sections. Finally, the parameters $\bm{\gamma}$ and $\bm{\beta}$ are updated by a classical optimization routine using measurement outcomes from the quantum circuit, thereby forming the hybrid quantum-classical loop shown in \textcolor{blue}{\textbf{Figure}}~\ref{fig:2}.

\subsection{\textcolor{black}{Scalability and quantum-resource limitations}}
\label{section:4.3}
\label{sec:13}\label{line:13}

\begin{figure}[htbp]
\centering
\includegraphics[width=0.98\linewidth]{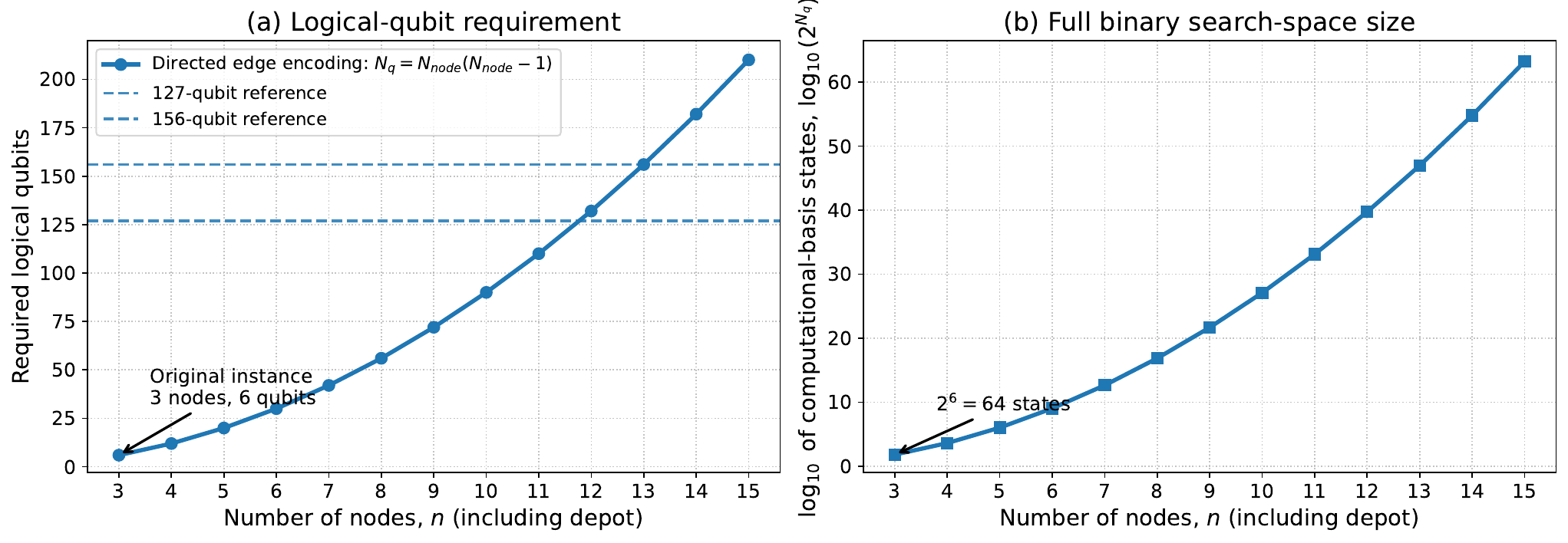}
\caption{\textcolor{black}{Scalability of the complete directed edge-based VRP encoding adopted in this study. Panel (a) shows the circuit-qubit requirement $N_q=N_{\mathrm{node}}(N_{\mathrm{node}}-1)$ as a function of the number of nodes \(N_{\mathrm{node}}\), including the depot. The horizontal lines at 127 and 156 qubits represent the nominal physical widths of representative IBM Eagle and IBM Heron r2/r3 processors, respectively, according to the IBM Quantum hardware specifications \protect\citep{ibmhardware2026}. These lines are numerical hardware-width references only and do not imply that QAOA circuits occupying all corresponding physical qubits are practically executable. The usable width may be smaller because of unavailable or poorly calibrated qubits, restricted connectivity, transpilation overhead, accumulated gate errors, and the distinction between noisy physical qubits and error-corrected logical qubits. Panel (b) shows the size of the associated computational-basis space, $2^{N_q}=2^{N_{\mathrm{node}}(N_{\mathrm{node}}-1)}$. The proposed constraint-aware initialization restricts the states receiving nonzero initial amplitude but does not reduce the circuit-qubit requirement of the underlying encoding.}}
\label{fig:3}
\end{figure}

\textcolor{black}{The scalability of QAOA-based approaches is fundamentally constrained by the number of binary variables introduced by the underlying optimization formulation. In the complete directed edge-based VRP formulation adopted in this study, one binary variable $x_{i,j}$ is assigned to each ordered pair of distinct nodes $(i,j)$. Therefore, for a graph containing \(N_{\mathrm{node}}\) nodes, including the depot, the required number of circuit qubits is}
\begin{equation}
\color{black}
N_q=N_{\mathrm{node}}(N_{\mathrm{node}}-1)
\label{eq:24}
\end{equation}
\textcolor{black}{Accordingly, complete directed instances with three, four, and five nodes require 6, 12, and 20 qubits, respectively. Although this requirement grows quadratically with the number of nodes, the associated computational-basis space grows much more rapidly:}
\begin{equation}
\color{black}
N_{\mathrm{basis}}
=
2^{N_q}
=
2^{N_{\mathrm{node}}(N_{\mathrm{node}}-1)}
\label{eq:25}
\end{equation}
\textcolor{black}{For example, the basis size increases from $2^6=64$ states for three nodes to $2^{12}=4096$ states for four nodes and $2^{20}=1{,}048{,}576$ states for five nodes. This does not imply that a quantum processor explicitly enumerates all basis states, but it illustrates the rapid growth of the encoded Hilbert space and the exponential memory requirement of exact statevector simulation (as shown in \textcolor{blue}{\textbf{Figure}}~\ref{fig:3}).}

\textcolor{black}{The qubit count alone does not fully characterize the practical scalability of QAOA. As the number of variables and quadratic QUBO interactions increases, the cost Hamiltonian generally requires more two-qubit operations. Restricted hardware connectivity can introduce additional routing and SWAP operations during transpilation, increasing the effective circuit depth and accumulated noise. A larger QAOA depth $p$ further repeats the cost and mixer layers and increases the number of variational parameters and circuit evaluations. Consequently, an instance may remain impractical even when its nominal circuit width is below the physical qubit count of the processor \citep{azfar2026hardware,du2026overcoming}.}

\textcolor{black}{The proposed method mitigates only part of this scalability challenge. Constraint-aware initialization concentrates the initial probability amplitude on states satisfying selected local constraints, thereby improving the utilization of the encoded state space. It does not remove any binary variables and therefore does not reduce the width \(N_q=N_{\mathrm{node}}(N_{\mathrm{node}}-1)\). Before hardware-specific decomposition and routing, the standard transverse-field mixer uses \(n\) single-qubit \(R_X\) gates and no two-qubit mixer gates per QAOA layer. The proposed three-node mixer contains two \(XY\)-exchange pairs, corresponding to four two-qubit gates \(R_{XX}\) and \(R_{YY}\), together with two \(R_X\) gates per layer. The proposed four-node mixer contains four exchange pairs, corresponding to eight \(R_{XX}\) and \(R_{YY}\) gates, together with two \(R_X\) gates per layer. The four-node block-wise XY-QAOA baseline contains twelve exchange pairs and therefore twenty-four \(R_{XX}\) and \(R_{YY}\) gates per layer. These counts exclude the \(R_{ZZ}\) gates in the cost layer, which are common to the compared ansatzes for a fixed instance, as well as any routing operations introduced by restricted hardware connectivity.}

\textcolor{black}{The prescribed constraint-aware state is injected directly as a logical state in the simulations. This isolates the variational dynamics of the initialization--mixer combination but does not provide a gate-level state-preparation circuit. On physical hardware, decomposing this state preparation into native gates may introduce additional depth and two-qubit operations. State-preparation gates and their noise are not included in the reported noise model. Consequently, the noisy finite-shot results characterize the post-initialization QAOA evolution and measurement rather than the complete end-to-end hardware cost. The feasibility improvements must therefore be interpreted together with both the mixer overhead quantified above and the unresolved cost of hardware-level state preparation.}

\begin{figure}[htbp]
\centering
\includegraphics[width=0.98\linewidth]{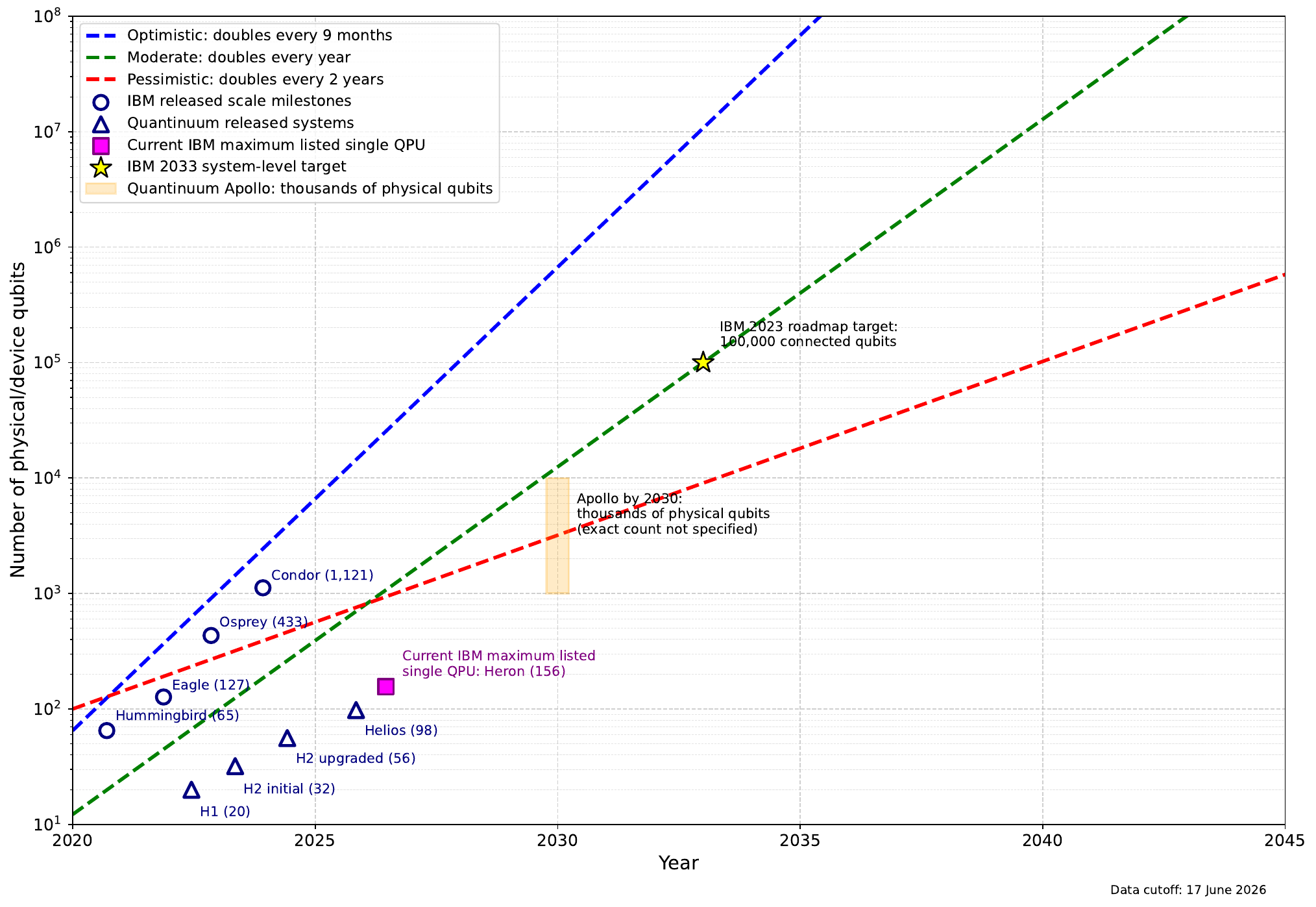}
\caption{\textcolor{black}{Illustrative qubit-growth scenarios and selected hardware milestones for gate-based quantum computing (Data collected from \citep{ibm2022,ibm100k2023,quantinuum2024roadmap,quantinuumH12022, quantinuumH22023,quantinuumH22024,quantinuum2024roadmap,quantinuumHelios2025}). The dashed curves represent reference scenarios in which the physical or device qubit count doubles every nine months, one year, and two years, respectively. The markers show reported physical-qubit counts for selected IBM and Quantinuum systems. The IBM roadmap point represents the announced objective of a system containing $100{,}000$ connected qubits by 2033, while the Quantinuum Apollo milestone is shown as a range because its roadmap specifies thousands of physical qubits rather than an exact count \protect\citep{ibm100k2023,quantinuum2024roadmap}. These physical-qubit milestones should not be interpreted as direct estimates of executable QAOA problem size because practical execution also depends on connectivity, fidelity, circuit depth, compilation overhead, and the physical resources required to realize error-corrected logical qubits.}}
\label{fig:4}
\end{figure}

\textcolor{black}{These problem-side limitations should be considered in conjunction with the continuing development of gate-based quantum hardware. Following the scenario-based perspectives in \citep{groenland2025introduction,du2026overcoming}, \textcolor{blue}{\textbf{Figure}}~\ref{fig:4} presents illustrative optimistic, moderate, and pessimistic trajectories corresponding to qubit-count doubling every nine months, one year, and two years, respectively. These curves are reference scenarios rather than statistically validated forecasts, since future progress also depends on gate fidelity, connectivity, control systems, interconnects, compilation, and quantum error correction.}

\textcolor{black}{The hardware roadmaps in \textcolor{blue}{\textbf{Figure}}~\ref{fig:4} provide a positive but uncertain long-term outlook. Larger processors, improved gate fidelities, modular architectures, and more effective error-correction techniques may gradually enable wider and deeper quantum circuits \citep{ibm2022,ibm100k2023,quantinuum2024roadmap}. Nevertheless, increasing the physical qubit count alone will not eliminate the connectivity, depth, sampling, and optimization bottlenecks of QAOA. Progress toward practically sized routing problems will also require more compact encodings, preprocessing to remove unnecessary links, hardware-aware circuit design, and decomposition strategies that divide a large VRP into smaller quantum-compatible subproblems.}
\par \textcolor{purple}{Alternative routing-native encodings illustrate this resource trade-off. The colored-permutation formulation of \citep{onah2026optimal} removes separate logical load variables by evaluating capacity diagonally on the routing register and proposes a binary-compressed representation scaling as \(N_{\mathrm{cust}}\lceil\log_2(KN_{\mathrm{cust}})\rceil\) for \(N_{\mathrm{cust}}\) customers and \(K\) vehicles, although retaining favorable symbol-preserving mixer dynamics after compression remains an implementation challenge.}

\subsection{Constraint-aware initialization and hybrid XY-X mixer}
\par
\begin{figure}
    \centering
    \includegraphics[width=0.7\linewidth]{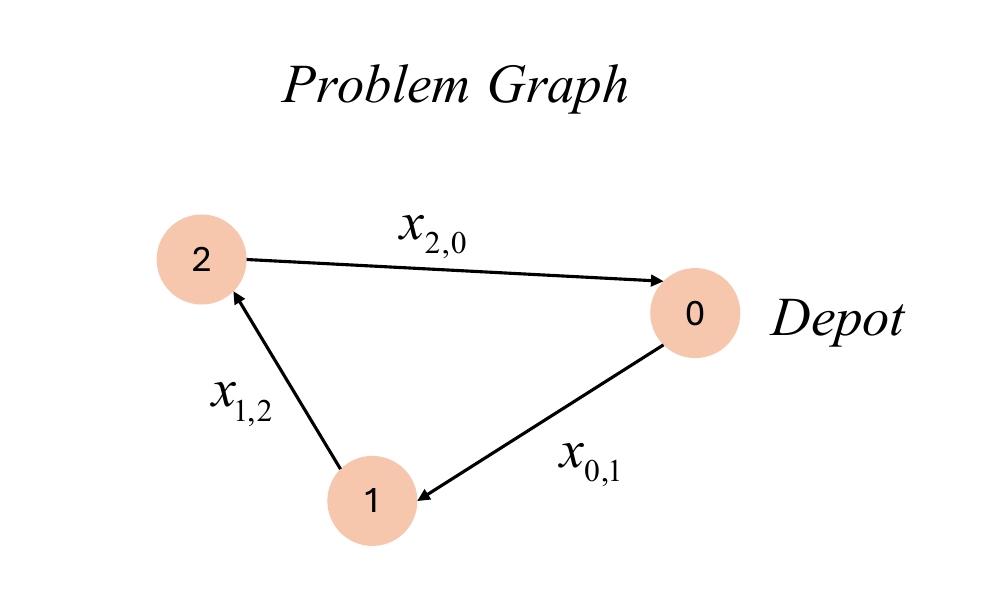}
    \caption{Problem Graph}
    \label{fig:5}
\end{figure}

 To better explain the motivation behind the proposed method, let us consider a simple illustrative example. As shown in \textcolor{blue}{\textbf{Figure}}~\ref{fig:5}, we study a small VRP instance with three nodes, including one depot, and two vehicles. Following a link-based formulation, let $x_{i,j}\in\{0,1\}$ denote whether the directed link $(i,j)$ between nodes $i$ and $j$ is selected. In this toy example, the decision vector contains six binary variables,
\begin{equation}
[x_{0,1},x_{0,2},x_{1,0},x_{1,2},x_{2,0},x_{2,1}] \label{eq:26}
\end{equation}
and therefore the corresponding QAOA encoding requires six qubits. If we use the standard uniform-superposition initialization in \eqref{eq:17}, the initial state is
\begin{equation}
|\psi(0)\rangle
=
\frac{1}{\sqrt{2^6}}
\Big(\underbrace{|000000\rangle+|000001\rangle+|000010\rangle+...+|111111\rangle}_{\times 2^{6} = 64}\Big) \label{eq:27}
\end{equation}
that is, an equal-weight superposition over all $2^6=64$ computational-basis states, ranging from $|000000\rangle$ to $|111111\rangle$. Under the constraints of this specific toy instance, there is only one feasible solution, which can be written as
\begin{equation}
|111010\rangle \label{eq:28}
\end{equation}
Hence, the feasible state accounts for only a $1/64$ fraction of the initial probability mass. If one further adopts the standard mixer in \eqref{eq:22}, whose action is to apply independent $x$-axis rotations to all qubits and thereby mix $|0\rangle$ and $|1\rangle$ on each position, the evolution explores the full Hilbert space without explicitly preserving the feasible structure of the solution. As a result, amplitude can easily flow from feasible states to infeasible ones, which is particularly unfavorable when feasible solutions are already extremely sparse. This helps explain why, in hardware implementations of standard QAOA such as those reported in \citep{azfar2025quantum}, the optimal solution may appear with relatively low rank and low sampling frequency in the final output distribution.
\par The key observation behind our method is that, for VRP-type problems, a substantial subset of the constraints, especially local equality constraints such as degree-balance constraints, can be exploited directly to shrink the search space before the QAOA evolution begins.\label{sec:22}\label{line:22} \textcolor{black}{This initialization is intentionally selective: it encodes only part of the local constraint structure and should not be interpreted as an encoding of all VRP constraints. The remaining customer-degree constraints, depot departure/return constraints, and other routing constraints remain in the QUBO penalty Hamiltonian.} In the toy example above, the six decision variables in \eqref{eq:26} correspond to qubits \(q_0\) through \(q_5\) of the encoded state. Consider first the constraint
\begin{equation}
x_{1,0}+x_{1,2}=1 \label{eq:29}
\end{equation}
which means that exactly one outgoing link must be selected from node $1$. Under the ordering in \eqref{eq:26}, this constraint involves qubits \(q_2\) and \(q_3\) only. Without imposing the constraint, these two positions admit all $2^2=4$ computational-basis configurations. After enforcing the constraint, however, only two assignments remain admissible, namely $|01\rangle_{(q_2,q_3)}$ and $|10\rangle_{(q_2,q_3)}$. A natural constraint-aware initialization over this reduced subspace is therefore
\begin{equation}
\label{eq:30}
\frac{1}{\sqrt{2}}\Bigl(|01\rangle_{(q_2,q_3)}+|10\rangle_{(q_2,q_3)}\Bigr)
\end{equation}
which already excludes the infeasible patterns $|00\rangle_{(q_2,q_3)}$ and $|11\rangle_{(q_2,q_3)}$. If we further impose a second constraint
\begin{equation}
\label{eq:31}
x_{0,2}+x_{1,2}=1
\end{equation}
then qubits \(q_1\), \(q_2\), and \(q_3\) are jointly restricted. Instead of all $2^3=8$ basis states, only two assignments satisfy both constraints, namely $|001\rangle_{(q_1,q_2,q_3)}$ and $|110\rangle_{(q_1,q_2,q_3)}$. Accordingly, the corresponding constraint-aware initialization over these three qubits becomes
\begin{equation}
\label{eq:32}
\frac{1}{\sqrt{2}}\Bigl(|001\rangle_{(q_1,q_2,q_3)}+|110\rangle_{(q_1,q_2,q_3)}\Bigr)
\end{equation}
\par Moreover, in this case, the only nontrivial subset that satisfies the subtour-elimination cardinality condition is
\begin{equation}
\label{eq:33}
2 \leq |\mathcal{S}| \leq 2
\quad
\mathcal{S}=\{1,2\}
\end{equation}
and the full set of constraints can therefore be written as
\begin{subequations}
\label{eq:34}
\begin{align}
  &\quad
    x_{1,0} + x_{2,0} = 2 \label{eq:34a}
    \\
  &\quad
    x_{0,1} + x_{0,2} = 2  \label{eq:34b}
    \\
  &\quad
    x_{1,0} + x_{1,2} = 1  \label{eq:34c}
    \\
  &\quad
    x_{0,1} + x_{2,1} = 1  \label{eq:34d}
    \\
  &\quad
    x_{2,0} + x_{2,1} = 1  \label{eq:34e}
    \\
  &\quad
    x_{0,2} + x_{1,2} = 1  \label{eq:34f}
    \\
  &\quad
    x_{1,0} + x_{2,0} \geq 1  \label{eq:34g}
\end{align}
\end{subequations}
\begin{figure}
    \centering
    \includegraphics[width=0.8\linewidth]{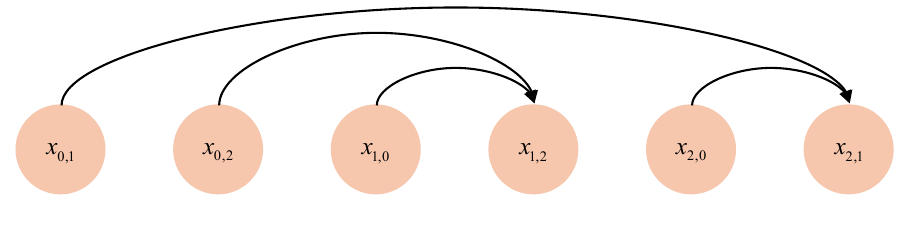}
    \caption{Relations of constraints. Each link refers to a one-hot constraint among two variables.}
    \label{fig:6}
\end{figure}
\par Among these constraints, also as shown in \textcolor{blue}{\textbf{Figure}} \ref{fig:6}, \eqref{eq:34c}, \eqref{eq:34d}, \eqref{eq:34e}, and \eqref{eq:34f} are particularly suitable for a constraint-aware initialization because they act locally on small groups of variables and immediately eliminate a large number of infeasible basis states. Under the ordering in \eqref{eq:26}, constraints \eqref{eq:34c} and \eqref{eq:34f} jointly restrict \(q_1,q_2,q_3\), leaving only two admissible patterns, namely $|001\rangle_{(q_1,q_2,q_3)}$ and $|110\rangle_{(q_1,q_2,q_3)}$. Similarly, constraints \eqref{eq:34d} and \eqref{eq:34e} jointly restrict \(q_0,q_4,q_5\), again leaving only $|001\rangle_{(q_0,q_4,q_5)}$ and $|110\rangle_{(q_0,q_4,q_5)}$. Following the same reasoning as in the previous paragraph, we can therefore construct the following constraint-aware initial state
\begin{equation}
|\psi(0)\rangle
=
\frac{1}{\sqrt{2}}\bigl(|001\rangle+|110\rangle\bigr)_{(q_1,q_2,q_3)}
\;\bigotimes\;
\frac{1}{\sqrt{2}}\bigl(|001\rangle+|110\rangle\bigr)_{(q_0,q_4,q_5)} \label{eq:35}
\end{equation}
which is supported on only four computational-basis states instead of all $2^6=64$ states in the standard uniform-superposition initialization. In other words, by explicitly incorporating these structurally informative constraints into the initialization stage, we can substantially reduce the size of the superposed state space and significantly increase the initial probability mass on states that already satisfy key VRP constraints. The remaining global constraints, such as \eqref{eq:34a}, \eqref{eq:34b}, and \eqref{eq:34g}, can then be handled during the subsequent optimization through the cost Hamiltonian.
\par \textcolor{black}{This example illustrates a selected-local-constraint initialization rather than a fully feasible-state initialization. The selected one-hot constraints remove many infeasible basis states and concentrate amplitude on more structured configurations, but they do not include all VRP constraints. This is different from the Grover-mixer variant of QAOA in \citep{bartschi2020grover}, whose initialization is designed to lie entirely in the feasible subspace and therefore requires a superposition over feasible states only. Our construction keeps the initialization simple by encoding only the selected local constraints that are straightforward to prepare, while the remaining VRP constraints are enforced through the QUBO penalty Hamiltonian and the subsequent variational search. Therefore, whenever we state that the proposed ansatz preserves constraint structure, the statement should be read in a partial sense. It preserves only those selected structures protected by the \(XY\)-exchange terms and does not guarantee invariance of the complete VRP feasible subspace.}
\par After applying the proposed constraint-aware initialization, part of the encoded solution structure already satisfies a subset of the VRP constraints. Since any feasible solution must also satisfy these constraints, it is natural to avoid destroying such favorable structure during the subsequent evolution. This is one of the main motivations for introducing an $XY$-type mixer. In essence, the $XY$ interaction couples basis states of the form $|01\rangle$ and $|10\rangle$, thereby mixing amplitudes within a fixed Hamming-weight subspace. As a result, if the initial state is prepared in a subspace corresponding to an exactly-one or one-hot type constraint, the $XY$ mixer can preserve that structural property during the evolution. \textcolor{black}{For a selected undirected exchange graph with edge set \(\mathcal{E}_{XY}\), the mixer is written as}
\begin{equation}
\label{eq:36}
\textcolor{black}{
H_M^{XY}
=
\sum_{\{i,j\}\in\mathcal{E}_{XY}}
\left(X_iX_j+Y_iY_j\right)
}
\end{equation}
However, the same property that makes the $XY$ mixer attractive also imposes an important limitation: it preserves the Hamming weight of the subspace on which it acts. For example, consider the computational-basis state $|010101\rangle$, whose Hamming weight is $3$. Under a pure $XY$ mixer, this state can only evolve within the weight-$3$ subspace. Therefore, if all feasible solutions have Hamming weight at least $4$, then $|010101\rangle$ can never evolve into a feasible solution. In the present toy example, the unique feasible solution is $|111010\rangle$, whose Hamming weight is $4$, so a standard $XY$ mixer alone is insufficient to connect such weight-mismatched initial states to the feasible set.
\par To address this issue, we introduce a hybrid mixer of the form
\begin{equation}
\label{eq:37}
\textcolor{black}{
H_M^{\mathrm{hyb}}
=
\sum_{\{i,j\}\in\mathcal{E}_{XY}}
\left(X_iX_j+Y_iY_j\right)
+
\lambda\sum_{k\in\mathcal{K}}X_k
}
\end{equation}
\textcolor{black}{where $\lambda>0$ is a weighting parameter and $\mathcal{K}$ denotes the set of qubits left outside the \(XY\)-protected structures. The \(XY\)-exchange terms preserve the selected one-hot structures chosen for mixer preservation. The \(X\)-mixer terms act only on the remaining positions so that the circuit keeps some ability to change Hamming weight and explore outside the fixed-weight subspaces. This is a partial preservation guarantee, not a full feasibility guarantee. Constraints not protected by the selected \(XY\) terms, including remaining customer-degree constraints and depot departure/return constraints, are still handled through the QUBO cost Hamiltonian and the variational search.} In this way, states such as $|010101\rangle$, whose Hamming weight does not match that of the feasible solution, are no longer trapped in an unreachable subspace and may evolve toward feasible solutions. In our example, the hybrid mixer takes the specific form
\begin{equation}
H_M
=
\left(X_2X_3 + Y_2Y_3\right)
+
\left(X_4X_5 + Y_4Y_5\right)
+
\lambda\left(X_0 + X_1\right) \label{eq:38}
\end{equation}
\textcolor{black}{Here, the $XY$ terms preserve only the two selected local one-hot structures associated with the pairs \((q_2,q_3)\) and \((q_4,q_5)\). The $X$ terms on \(q_0\) and \(q_1\) are not part of these \(XY\)-protected pairs and provide additional flexibility to change the Hamming weight. The mixer therefore protects the selected one-hot structures used for preservation in this example, but it does not by itself preserve the full VRP feasible set.}

\section{Experiments}
\label{section:5}
\par \textcolor{black}{In this section, we evaluate the proposed QAOA framework, which combines a constraint-aware initialization with a hybrid $XY$--$X$ mixer, on one three-node toy VRP instance and two four-node VRP benchmark instances. The three-node instance provides a transparent setting for explaining the proposed initialization and mixer, while the two four-node instances test whether the observed behavior remains stable when the number of qubits increases from six to twelve.}

{\color{black}
The experimental design has four main components. First, the two four-node VRP cases increase the full computational-basis space from \(2^6=64\) to \(2^{12}=4096\). Second, the depth study over \(p\in\{1,2,3\}\) examines the trade-off among solution quality, circuit depth, and optimization complexity. Third, the block-wise XY-QAOA baseline for the four-node instances provides a feasibility-aware comparison that is more directly related to the proposed ansatz than standard QAOA alone. Fourth, the ablation studies separate the contribution of the constraint-aware initialization from that of the hybrid mixer.
}

\subsection{Problem Setup}
\label{subsec:problem_setup}

\textcolor{black}{Throughout the manuscript, Case~I denotes the three-node VRP instance, Case~II denotes the balanced symmetric four-node VRP instance, and Case~III denotes the customer-cluster four-node VRP instance.}
\label{sec:1}
\label{line:1}

\textcolor{black}{For the numerical experiments, we use one small illustrative instance and two four-node benchmark instances. The three-node case follows \citep{azfar2025quantum} with a minor modification to the distance matrix and serves as a toy example for explaining the proposed constraint-aware initialization and hybrid XY--X mixer. The two four-node cases preserve the same two-vehicle routing structure but increase the circuit width to twelve qubits. All instances are converted to QUBO form via quadratic penalties and then to Ising Hamiltonians using the convention in \eqref{eq:41}. Detailed QUBO and Ising derivations for the three-node toy instance are given in Appendix~\ref{app:qubo_ising_derivation}; the corresponding details for the four-node benchmark instances are given in \textcolor{blue}{\textbf{Appendix}}~\ref{app:fournode_qubo_ising_derivation}.}

\subsubsection{Toy instance: three-node VRP}
\label{subsubsec:toy_three_node}

For the three-node instance, node \(0\) is the depot and nodes \(1\) and \(2\) are customer nodes. The distance matrix is shown in Table~\ref{table:1}.

\begin{table}[htbp]
\centering
\caption{\textcolor{black}{VRP distance matrix for the three-node instance}}
\label{table:1}
\begin{tabular}{cccc}
\toprule
 & 0 & 1 & 2 \\
\midrule
0 & 0    & 61.3 & 4.7  \\
1 & 61.3 & 0    & 42.9 \\
2 & 4.7  & 42.9 & 0    \\
\bottomrule
\end{tabular}
\end{table}

Let \(x_{i,j}\in\{0,1\}\) denote whether the directed link from node \(i\) to node \(j\) is selected. For this three-node, two-vehicle instance, the decision vector is ordered as \(\mathbf{x}=[x_{0,1},x_{0,2},x_{1,0},x_{1,2},x_{2,0},x_{2,1}]\). Under this ordering, the objective is

\begin{equation}
\min_{\mathbf{x}}
61.3x_{0,1}
+4.7x_{0,2}
+61.3x_{1,0}
+42.9x_{1,2}
+4.7x_{2,0}
+42.9x_{2,1}
\label{eq:39}
\end{equation}

The constraints are

\begin{subequations}
\label{eq:40}
\begin{align}
x_{1,0}+x_{2,0} &= 2 \label{eq:40a} \\
x_{0,1}+x_{0,2} &= 2 \label{eq:40b} \\
x_{1,0}+x_{1,2} &= 1 \label{eq:40c} \\
x_{0,1}+x_{2,1} &= 1 \label{eq:40d} \\
x_{2,0}+x_{2,1} &= 1 \label{eq:40e} \\
x_{0,2}+x_{1,2} &= 1 \label{eq:40f} \\
x_{1,0}+x_{2,0} &\geq 1 \label{eq:40g} \\
x_{i,j} &\in\{0,1\} \label{eq:40h}
\end{align}
\end{subequations}

The inequality in \eqref{eq:40g} corresponds to the only nontrivial subtour-elimination subset in this toy instance, namely \(\mathcal{S}=\{1,2\}\). Following \citep{harwood2021formulating,azfar2025quantum}, the penalty coefficient is set to twice the sum of all directed link costs, giving \(P=435.6\).

\textcolor{black}{Throughout the paper, we use the binary-to-Ising convention}\label{sec:17}\label{line:17}

\begin{equation}
\textcolor{black}{
x_i=\frac{I-Z_i}{2}
}
\label{eq:41}
\end{equation}

\noindent\textcolor{black}{where \(Z_i\) is the Pauli-\(Z\) operator acting on qubit \(q_i\). In the selected qubit order, \((q_0,q_1,q_2)\) corresponds to \((x_{0,1},x_{0,2},x_{1,0})\), while \((q_3,q_4,q_5)\) corresponds to \((x_{1,2},x_{2,0},x_{2,1})\).} After omitting the constant energy shift, the unscaled cost Hamiltonian is

\begin{equation}
\label{eq:42}
\begin{aligned}
H_C
&=
326.7Z_2Z_4
+217.8Z_0Z_1
+217.8Z_2Z_3
+217.8Z_0Z_5
+217.8Z_4Z_5
+217.8Z_1Z_3
\\
&\quad
+404.95Z_0
+433.25Z_1
+513.85Z_2
-21.45Z_3
+542.15Z_4
-21.45Z_5
\end{aligned}
\end{equation}

For numerical stability, the implementation uses the positively scaled Hamiltonian \(\widetilde{H}_C=H_C/435.6\), which preserves the ordering of all computational-basis energies.

Using the above qubit order, the proposed constraint-aware initialization prepares

\begin{equation}
\label{eq:43}
|\psi(0)\rangle
=
\frac{1}{2}
\left(
|000101\rangle
+
|100110\rangle
+
|011001\rangle
+
|111010\rangle
\right)
\end{equation}

Thus, instead of starting from the uniform superposition over all \(2^6=64\) basis states, the proposed initialization restricts the initial probability mass to a four-state structured subspace that satisfies selected local one-hot constraints and still contains the globally optimal feasible bitstring.

The corresponding hybrid XY--X mixer is

\begin{equation}
\label{eq:44}
H_M
=
\left(X_2X_3+Y_2Y_3\right)
+
\left(X_4X_5+Y_4Y_5\right)
+
\lambda\left(X_0+X_1\right)
\end{equation}

The two \(XY\) terms preserve the selected local one-hot structures on \((q_2,q_3)\) and \((q_4,q_5)\), while the single-qubit \(X\) terms on \(q_0\) and \(q_1\) provide additional exploration on qubits left outside these \(XY\)-protected pairs. This preservation statement applies only to the selected structures protected by the \(XY\) terms, not to every constraint used in the initialization and not to the full VRP feasible subspace.

{\color{black}
\subsubsection{Benchmark instances: four-node VRP}
\label{subsubsec:four_node_benchmark}
\label{sec:3}
\label{line:3}

The two four-node, two-vehicle VRP benchmark instances test whether the observed behavior is stable beyond the toy setting. In both cases, node \(0\) is the depot and nodes \(1\), \(2\), and \(3\) are customer nodes. The two cases use the same directed-arc encoding and the same degree-balance constraints, but they have different distance matrices and therefore different cost landscapes.

The first four-node instance, denoted Case~II, is a balanced symmetric instance, and its distance matrix is shown in Table~\ref{table:2}.

\begin{table}[htbp]
\centering
\caption{\textcolor{black}{VRP distance matrix for the four-node Case~II balanced symmetric instance}}
\label{table:2}
\begin{tabular}{ccccc}
\toprule
 & 0 & 1 & 2 & 3 \\
\midrule
0 & 0    & 21.7 & 34.2 & 28.6 \\
1 & 21.7 & 0    & 17.4 & 24.9 \\
2 & 34.2 & 17.4 & 0    & 19.8 \\
3 & 28.6 & 24.9 & 19.8 & 0    \\
\bottomrule
\end{tabular}
\end{table}

The second four-node instance, denoted Case~III, is a customer-cluster instance, and its distance matrix is shown in Table~\ref{table:3}.

\begin{table}[htbp]
\centering
\caption{\textcolor{black}{VRP distance matrix for the four-node Case~III customer-cluster instance}}
\label{table:3}
\begin{tabular}{ccccc}
\toprule
 & 0 & 1 & 2 & 3 \\
\midrule
0 & 0    & 42.3 & 26.8 & 38.5 \\
1 & 42.3 & 0    & 13.7 & 29.4 \\
2 & 26.8 & 13.7 & 0    & 16.2 \\
3 & 38.5 & 29.4 & 16.2 & 0    \\
\bottomrule
\end{tabular}
\end{table}

For four-node instances, the decision vector is ordered by origin node and then by destination node: first \([x_{0,1},x_{0,2},x_{0,3}]\), followed by \([x_{1,0},x_{1,2},x_{1,3}]\), \([x_{2,0},x_{2,1},x_{2,3}]\), and \([x_{3,0},x_{3,1},x_{3,2}]\). Hence, each four-node instance requires twelve qubits. Compared with the three-node case, the circuit width increases from six to twelve qubits, and the full computational-basis space increases from \(64\) to \(4096\) states.

The four-node constrained VRP enforces one incoming and one outgoing arc for each customer, together with two-vehicle depot departure and return constraints. The degree-balance constraints are

\begin{subequations}
\label{eq:45}
\begin{align}
x_{0,1}+x_{2,1}+x_{3,1} &= 1 \label{eq:45a} \\
x_{0,2}+x_{1,2}+x_{3,2} &= 1 \label{eq:45b} \\
x_{0,3}+x_{1,3}+x_{2,3} &= 1 \label{eq:45c} \\
x_{1,0}+x_{1,2}+x_{1,3} &= 1 \label{eq:45d} \\
x_{2,0}+x_{2,1}+x_{2,3} &= 1 \label{eq:45e} \\
x_{3,0}+x_{3,1}+x_{3,2} &= 1 \label{eq:45f} \\
x_{0,1}+x_{0,2}+x_{0,3} &= 2 \label{eq:45g} \\
x_{1,0}+x_{2,0}+x_{3,0} &= 2 \label{eq:45h}
\end{align}
\end{subequations}

The complete VRP formulation also includes subtour-elimination inequalities. In the QUBO used in the experiments, the travel-cost objective and the eight degree-balance constraints are penalized. Explicit subtour penalties are omitted because exhaustive enumeration confirms that, for the two four-node benchmark instances considered here, the bitstrings satisfying the eight degree-balance constraints correspond to valid two-route depot-to-depot solutions. The detailed QUBO and Ising derivation is given in Appendix~\ref{app:fournode_qubo_ising_derivation}.

The penalty coefficient is again set as twice the sum of all directed link costs, giving \(P=586.4\) for Case~II and \(P=667.6\) for Case~III. The qubit-to-variable correspondence follows the same ordered list, so \(q_k\) encodes the \(k\)-th entry of the four-node decision vector.

For the proposed four-node ansatz, we use an overlap-four initialization. Rather than preparing the uniform superposition over all \(4096\) basis states, we impose four selected overlapping degree constraints:

\begin{subequations}
\label{eq:46}
\begin{align}
x_{0,1}+x_{2,1}+x_{3,1} &= 1 \label{eq:46a} \\
x_{0,2}+x_{1,2}+x_{3,2} &= 1 \label{eq:46b} \\
x_{1,0}+x_{1,2}+x_{1,3} &= 1 \label{eq:46c} \\
x_{2,0}+x_{2,1}+x_{2,3} &= 1 \label{eq:46d}
\end{align}
\end{subequations}

These four constraints define a \(100\)-state initialization support, denoted by \(\mathcal{C}_{\mathrm{init}}\). The proposed initial state is

\begin{equation}
|\psi(0)\rangle
=
\frac{1}{\sqrt{100}}
\sum_{\mathbf{x}\in\mathcal{C}_{\mathrm{init}}}
|\mathbf{x}\rangle
\label{eq:47}
\end{equation}

This initialization does not enforce full VRP feasibility. Its role is to encode a selected subset of structurally informative local constraints and to reduce the initial support from \(4096\) basis states to \(100\) basis states while retaining all globally optimal feasible solutions verified by exhaustive enumeration. This distinction is important: the selected constraints define the initialized support, but the remaining VRP constraints are still represented through the QUBO penalty Hamiltonian.

The corresponding four-node hybrid mixer is\label{sec:18}\label{line:18}

\begin{equation}
\label{eq:48}
\begin{aligned}
H_M
&=
\left(X_0X_{10}+Y_0Y_{10}\right)
+
\left(X_1X_{11}+Y_1Y_{11}\right)
\\
&\quad
+
\left(X_3X_5+Y_3Y_5\right)
+
\left(X_6X_8+Y_6Y_8\right)
+
\lambda\left(X_2+X_9\right)
\end{aligned}
\end{equation}

The four \(XY\) exchange terms preserve only the selected overlap constraints in \eqref{eq:46a}--\eqref{eq:46d}.\label{sec:23}\label{line:23} The \(X\) terms act on \(q_2\) and \(q_9\), which are not included in these selected overlap constraints, so the ansatz retains limited exploration outside the overlap-preserving subspace. The overlap qubits \(q_4\) and \(q_7\) appear in two selected constraints each and are therefore excluded from the mixer. This exclusion means only that the mixer does not act on them; it does not mean that they are fixed to zero in the initial state. Other VRP constraints, including the remaining customer-degree constraints and the depot departure/return constraints, are still enforced through the QUBO penalty Hamiltonian. Thus, the four-node mixer is partially constraint-preserving for the selected overlap-4 constraints, not fully feasible-subspace-preserving. Appendix~\ref{app:overlap4_init_mixer} gives the full explanation of the \(100\)-state support, the overlap-sector counts, and the mixer-preservation property (\textcolor{blue}{\textbf{Table}}~\ref{table:4}).
}

\subsection{Experimental Settings and Evaluation Metrics}
\label{subsec:experimental_settings_metrics}

{\color{black}
We compare three QAOA variants. The first is standard QAOA, which uses the uniform-superposition initialization \(\ket{+}^{\bigotimes n}\) and the conventional transverse-field Pauli-\(X\) mixer. This is the canonical baseline for QUBO/Ising optimization. The second is the proposed QAOA framework, which replaces the full-space initialization with the constraint-aware initialization and replaces the standard mixer with the hybrid \(XY\)--\(X\) mixer. The third is a block-wise XY-QAOA baseline, abbreviated as XY-QAOA when the context is clear, for the four-node instances. We use the term \textit{block-wise} because the decision variables are partitioned into small arc blocks, and the XY mixer acts only inside each block to preserve its fixed Hamming weight. This provides a stronger feasibility-aware baseline than standard QAOA while remaining distinct from our proposed overlap-based initialization and hybrid mixer.

The experiments are conducted under three evaluation regimes. Regime~I uses exact statevector expectation values during variational optimization and noiseless finite-shot sampling for the independent final evaluation. Thus, its optimization objective has neither sampling noise nor hardware noise, although its reported final-distribution metrics retain finite-shot uncertainty. Regime~II is an ideal finite-shot simulation, where both optimization and final evaluation rely on measurement samples. Regime~III is a noisy finite-shot simulation, where the same finite-shot procedure is combined with a hardware-inspired gate and readout noise model. This three-regime design separates exact-objective optimization, finite-shot objective estimation, and hardware-inspired noise effects.

The noisy finite-shot regime uses symmetric readout error with \(\epsilon_{01}^{\mathrm{ro}}=\epsilon_{10}^{\mathrm{ro}}=0.001\), where \(\epsilon_{01}^{\mathrm{ro}}\) and \(\epsilon_{10}^{\mathrm{ro}}\) denote the \(0\to1\) and \(1\to0\) readout-flip probabilities. Single-qubit gates use a Qiskit Aer depolarizing channel with parameter \(\epsilon_{1q}=0.00015\), corresponding to an average single-qubit gate infidelity of \(7.5\times 10^{-5}\). Two-qubit gates use a depolarizing parameter \(\epsilon_{2q}=0.00125\), corresponding to an average two-qubit gate infidelity of \(9.375\times 10^{-4}\). For standard QAOA, this two-qubit noise is applied to \(R_{ZZ}\) gates. For the proposed method and the block-wise XY-QAOA baseline, it is applied to the relevant two-qubit gates, including \(R_{ZZ}\), \(R_{XX}\), and \(R_{YY}\). These values should be interpreted as optimistic laboratory-level reference settings motivated by recent superconducting-qubit progress, rather than as typical public-cloud hardware performance.

To study the role of QAOA depth, the main experimental comparisons are reported at \(p=2\), and a separate \(p\)-depth sensitivity study with \(p\in\{1,2,3\}\) examines the trade-off among solution quality, circuit depth, and optimization complexity. Under the single-angle-per-layer QAOA parameterization used in this study, a depth-\(p\) circuit optimizes \(2p\) variational angles, namely \(\gamma_1,\ldots,\gamma_p\) and \(\beta_1,\ldots,\beta_p\). The hybrid-mixer weight \(\lambda\) is treated as an external hyperparameter selected through sensitivity analysis rather than as an additional COBYLA-trained variational angle. Therefore, increasing \(p\) can improve expressivity, but it also increases circuit depth, two-qubit gate count, and classical optimization complexity.\label{sec:11}\label{line:11}

All variational parameters are optimized using multi-restart COBYLA. Each independent run uses \(10\) random restarts, and the best parameter vector among the restarts is retained. The initial \(\gamma\) parameters are sampled uniformly from \([-\pi,\pi]\), and the initial \(\beta\) parameters are sampled uniformly from \([0,\pi/2]\). The COBYLA initial trust-region radius is set to \(0.5\). The stopping tolerance is \(10^{-4}\) for exact statevector optimization and \(2\times 10^{-3}\) for finite-shot and noisy finite-shot optimization, where the objective estimate is stochastic. The complete multi-restart optimization workflow, including random initialization, COBYLA updates, and final selection of the best restart, is summarized in \textcolor{blue}{\textbf{Algorithm}}~\ref{algorithm:generalized-qaoa-cobyla-inner}.\label{sec:15}\label{line:15}

The maximum COBYLA iteration budget is selected by method, instance, and regime after preliminary tuning and is reported explicitly so that the computational budget of each comparison is transparent. In the \(p\)-depth studies, standard QAOA uses \(700\) iterations in the statevector regime and \(200\) iterations in the two finite-shot regimes. The proposed method uses \(1300\) statevector iterations for Case~I, \(2000\) for Case~II, and \(600\) for Case~III, while using \(200\) iterations in both finite-shot regimes. The block-wise XY-QAOA baseline uses \(1700\) statevector iterations and \(200\) finite-shot iterations for the four-node cases. These budgets are per restart. The main comparisons use the same instance definitions, QAOA depth, simulation regimes, shot protocol, noise model, and number of independent runs; the optimizer budgets and tuned ansatz hyperparameters are stated separately because they differ across methods and instances. For finite-shot optimization, the stochastic objective is estimated using \(2048\) shots per batch and two batches per objective evaluation. The independent final histogram uses \(4096\) shots for all three-node experiments, the primary four-node statevector comparisons, and the four-node statevector ablation evaluations. It uses \(8192\) shots for all four-node finite-shot and noisy finite-shot evaluations and for all four-node depth-sensitivity evaluations. In every regime, the reported optimal-solution probability, expected energy gap, and sampling rank are computed from this independent finite-shot histogram. Regime~I differs in that the parameters are optimized with exact statevector expectation values before the final noiseless sampling stage. All reported configurations are repeated over \(30\) independent runs.
\textcolor{black}{Here \(B\) denotes the number of independent shot batches used for one objective-function evaluation. Each batch contains \(S_{\mathrm{obj}}\) shots and produces one sample-mean energy estimate; the \(B\) estimates are then averaged. Thus, \(S_{\mathrm{obj}}=2048\) and \(B=2\) use \(4096\) shots in total for each finite-shot objective evaluation.}

The hybrid mixer includes the weighting parameter \(\lambda\). For Case~I, the \(\lambda\)-sensitivity sweep uses \(\lambda\in\{0.5,0.6,\ldots,2.0\}\). For the four-node overlap-four experiments, the final \(p\)-depth and ablation studies use \(\lambda=2.0\), selected from the corresponding full-method hyperparameter sweep and then held fixed for those follow-up comparisons. The full \(\lambda\)-sensitivity curves are reported separately, so the four-node comparison should be interpreted as a comparison with the tuned proposed ansatz rather than as an independent post-hoc choice of \(\lambda\) for each individual metric.
}

{\color{black}
\subsubsection{Block-wise XY-QAOA baseline}
\label{subsubsec:block_xy_baseline}\label{sec:8}\label{line:8}

We include the block-wise XY-QAOA baseline to provide a stronger and more relevant comparison than standard QAOA. It also separates the role of the mixer more clearly. Standard QAOA uses a Pauli-\(X\) mixer only, the block-wise baseline uses \(XY\) exchange terms only, and the proposed method combines both mechanisms through a hybrid \(XY\)--\(X\) mixer. Therefore, this baseline tests whether the improvement comes simply from adding an \(XY\)-type mixer, or from combining \(XY\)-based constraint preservation with controlled \(X\)-type exploration.

The block-wise XY-QAOA construction must be initialized in a format compatible with the \(XY\) mixer. An \(XY\) exchange swaps \(|01\rangle\) and \(|10\rangle\), so it moves the active bit inside a block but preserves the number of ones in that block. As a result, a pure \(XY\)-mixer baseline cannot change block weights during the QAOA evolution. Its initial state must therefore be prepared in block-wise fixed-Hamming-weight sectors that contain the feasible routes the baseline is intended to search. If a feasible solution lay in a different block-weight sector, that solution would be unreachable under an \(XY\)-only mixer. In this baseline, the selected block-weight pattern is \((2,1,1,1)\). All fully feasible four-node routes satisfy this outgoing-degree pattern and are therefore included in the initial support. The support is not fully feasible, however, because incoming-degree and depot-return constraints are still enforced by the QUBO penalty Hamiltonian.

Intuitively, a block with Hamming weight one behaves like a one-choice block. Exactly one arc in that block is active, and the \(XY\) mixer moves the active position from one arc to another without changing the number of active arcs. A block with Hamming weight two works similarly, except that two active arcs are preserved. This makes block-wise XY-QAOA a natural feasibility-aware baseline because it hard-codes simple outgoing-degree structures into both the initialization and the mixer. It is still different from our proposed method. Our initialization uses selected overlapping incoming and outgoing constraints, and our hybrid \(XY\)--\(X\) mixer preserves those selected overlap constraints while allowing controlled exploration through qubits left outside the selected overlap structures.

In the four-node implementation, the twelve qubits are divided into four three-qubit blocks. The first block corresponds to depot outgoing arcs, and the remaining three blocks correspond to outgoing arcs from customers \(1\), \(2\), and \(3\):
\begin{subequations}
\label{eq:49}
\begin{align}
B_0 &= (q_0,q_1,q_2)=(x_{0,1},x_{0,2},x_{0,3}) \label{eq:49a} \\
B_1 &= (q_3,q_4,q_5)=(x_{1,0},x_{1,2},x_{1,3}) \label{eq:49b} \\
B_2 &= (q_6,q_7,q_8)=(x_{2,0},x_{2,1},x_{2,3}) \label{eq:49c} \\
B_3 &= (q_9,q_{10},q_{11})=(x_{3,0},x_{3,1},x_{3,2}) \label{eq:49d}
\end{align}
\end{subequations}

The initialization fixes the Hamming weight of \(B_0\) to two, because two vehicles must depart from the depot. It fixes the Hamming weight of each customer-outgoing block \(B_1\), \(B_2\), and \(B_3\) to one, because each customer has exactly one outgoing arc. These fixed block weights are not merely an implementation detail; they define the reachable sector of the \(XY\)-only baseline. Thus, the initialized support satisfies

\begin{subequations}
\label{eq:50}
\begin{align}
q_0+q_1+q_2 &= 2 \label{eq:50a} \\
q_3+q_4+q_5 &= 1 \label{eq:50b} \\
q_6+q_7+q_8 &= 1 \label{eq:50c} \\
q_9+q_{10}+q_{11} &= 1 \label{eq:50d}
\end{align}
\end{subequations}

The corresponding initial state is the tensor product of one fixed-weight superposition for each block. Since the depot-outgoing block has three weight-two states and each customer-outgoing block has three weight-one states, the total initialization support size is \(3\times 3\times 3\times 3=81\), as shown as follows:
\begin{equation}
\label{eq:51}
\begin{aligned}
|\psi(0)\rangle
={}&
\frac{1}{\sqrt{3}}
\bigl(|110\rangle+|101\rangle+|011\rangle\bigr)_{(q_0,q_1,q_2)}\bigotimes
\frac{1}{\sqrt{3}}
\bigl(|100\rangle+|010\rangle+|001\rangle\bigr)_{(q_3,q_4,q_5)}
\\
&\bigotimes
\frac{1}{\sqrt{3}}
\bigl(|100\rangle+|010\rangle+|001\rangle\bigr)_{(q_6,q_7,q_8)}\bigotimes
\frac{1}{\sqrt{3}}
\bigl(|100\rangle+|010\rangle+|001\rangle\bigr)_{(q_9,q_{10},q_{11})}
\end{aligned}
\end{equation}
This is smaller than the full \(4096\)-state computational basis and is comparable in spirit to feasibility-aware QAOA methods that initialize inside structured Hamming-weight sectors. Because the complete selected sector is used, every fully feasible four-node route is present in the initial support, while infeasible states in the same outgoing-degree sector remain possible and are penalized by the cost Hamiltonian.

The block-wise XY mixer acts independently inside each block. For a three-qubit block \(B=(a,b,c)\), it applies \(XY\) exchange interactions on all three pairs \((a,b)\), \((b,c)\), and \((a,c)\). The full mixer is therefore

\begin{equation}
\label{eq:52}
H_M^{XY}
=
\sum_{B\in\{B_0,B_1,B_2,B_3\}}
\sum_{\{i,j\}\in\binom{B}{2}}
\left(X_iX_j+Y_iY_j\right)
\end{equation}
which can be further expressed as:
\begin{equation}
\label{eq:53}
\begin{aligned}
H^{XY}_{M}
&=
\left(X_0X_1+Y_0Y_1\right)
+
\left(X_0X_2+Y_0Y_2\right)
+
\left(X_1X_2+Y_1Y_2\right)
\\
&\quad
+
\left(X_3X_4+Y_3Y_4\right)
+
\left(X_3X_5+Y_3Y_5\right)
+
\left(X_4X_5+Y_4Y_5\right)
\\
&\quad
+
\left(X_6X_7+Y_6Y_7\right)
+
\left(X_6X_8+Y_6Y_8\right)
+
\left(X_7X_8+Y_7Y_8\right)
\\
&\quad
+
\left(X_9X_{10}+Y_9Y_{10}\right)
+
\left(X_9X_{11}+Y_9Y_{11}\right)
+
\left(X_{10}X_{11}+Y_{10}Y_{11}\right)
\end{aligned}
\end{equation}

This mixer preserves the Hamming weight of every block. As a result, the block-wise XY-QAOA baseline always remains inside the same block-wise fixed-Hamming-weight subspace during mixer evolution. This is its main strength, because it preserves important outgoing-degree structures by construction. It is also its main limitation, because the circuit cannot use transitions that leave this selected outgoing-degree sector, even if such transitions would be useful for exploring lower-penalty or lower-cost regions of the QUBO landscape. By contrast, the proposed hybrid mixer preserves only the selected overlap constraints and adds weighted \(X\) rotations on qubits not included in those selected overlap constraints. The comparison between block-wise XY-QAOA and the proposed method therefore tests whether controlled flexibility beyond block-wise Hamming-weight preservation improves performance.
}

{\color{black}
To quantify performance, we use three primary metrics. First, the optimal-solution probability is the total probability assigned to all computational-basis states that encode globally optimal feasible solutions. Its population value and finite-shot estimator are

\begin{equation}
\label{eq:54}
\begin{aligned}
p^*
&=
\sum_{\mathbf{x}\in\mathcal{X}_{\mathrm{feas}}^*}p(\mathbf{x})
\\
\widehat{p}^{*}
&=
\frac{1}{S_{\mathrm{final}}}
\sum_{\mathbf{x}\in\mathcal{X}_{\mathrm{feas}}^*}n(\mathbf{x})
\end{aligned}
\end{equation}

where \(\mathcal{X}_{\mathrm{feas}}^*\) is the set of feasible global optima and \(n(\mathbf{x})\) is the count of bitstring \(\mathbf{x}\) in the independent final histogram. Second, the expected energy gap and its finite-shot estimator are

\begin{equation}
\label{eq:55}
\begin{aligned}
\Delta E
&=
\sum_{\mathbf{x}\in\{0,1\}^{n}}p(\mathbf{x})C(\mathbf{x})-C_{\mathrm{feas}}^*
\\
\widehat{\Delta E}
&=
\frac{1}{S_{\mathrm{final}}}
\sum_{\mathbf{x}\in\{0,1\}^{n}}n(\mathbf{x})C(\mathbf{x})-C_{\mathrm{feas}}^*
\end{aligned}
\end{equation}

where \(C(\mathbf{x})\) is the full unscaled QUBO objective, including penalty terms, and \(C_{\mathrm{feas}}^*\) is the globally optimal feasible cost. A smaller gap indicates greater concentration on lower-QUBO-energy states, but it is not by itself a separate feasibility probability. Third, the sampling rank records the rank position of the highest-ranked optimal solution after sorting sampled bitstrings by empirical frequency. Rank \(1\) means that an optimal solution is the most frequently sampled bitstring. In all three regimes, the reported values of \(p^*\) and \(\Delta E\) are the finite-shot estimates \(\widehat{p}^*\) and \(\widehat{\Delta E}\); the hats are omitted in tables and figures for readability. For each metric, we report the sample mean, sample standard deviation, and normal-approximation \(95\%\) confidence interval over the \(30\) independent runs:

\begin{equation}
\bar{m}\pm 1.96\frac{s_m}{\sqrt{30}}
\label{eq:56}
\end{equation}
}

\subsection{Results Analysis}
\label{subsec:results_analysis}

{\color{black}
We analyze the numerical results under the three experimental regimes. The full numerical tables are provided in the appendix (\textcolor{blue}{\textbf{Tables}}~\ref{table:5}--\ref{table:13}), while \textcolor{blue}{\textbf{Figures}}~\ref{fig:7}, \ref{fig:8}, \ref{fig:9}, \ref{fig:10}, \ref{fig:11}, and~\ref{fig:12} summarize the cross-regime and hyperparameter-sensitivity trends. The ablation summaries in \textcolor{blue}{\textbf{Figures}}~\ref{fig:13}--\ref{fig:15} further isolate the contribution of the constraint-aware initialization and the hybrid mixer. The discussion below focuses on the main performance implications rather than reproducing every table entry.

\paragraph{Overall comparison across regimes}
\textcolor{blue}{\textbf{Figure}}~\ref{fig:7} shows the three-node results across the statevector-objective, ideal finite-shot, and noisy finite-shot regimes. The proposed method consistently improves over standard QAOA in optimal-solution probability and expected energy gap. This confirms that the constraint-aware initialization and hybrid mixer are beneficial when the objective is evaluated exactly and remain effective when objective sampling noise and gate/readout noise are introduced.

\begin{figure}[htp]
    \centering
    \begin{subfigure}[b]{0.45\linewidth}
        \centering
        \includegraphics[width=\linewidth]{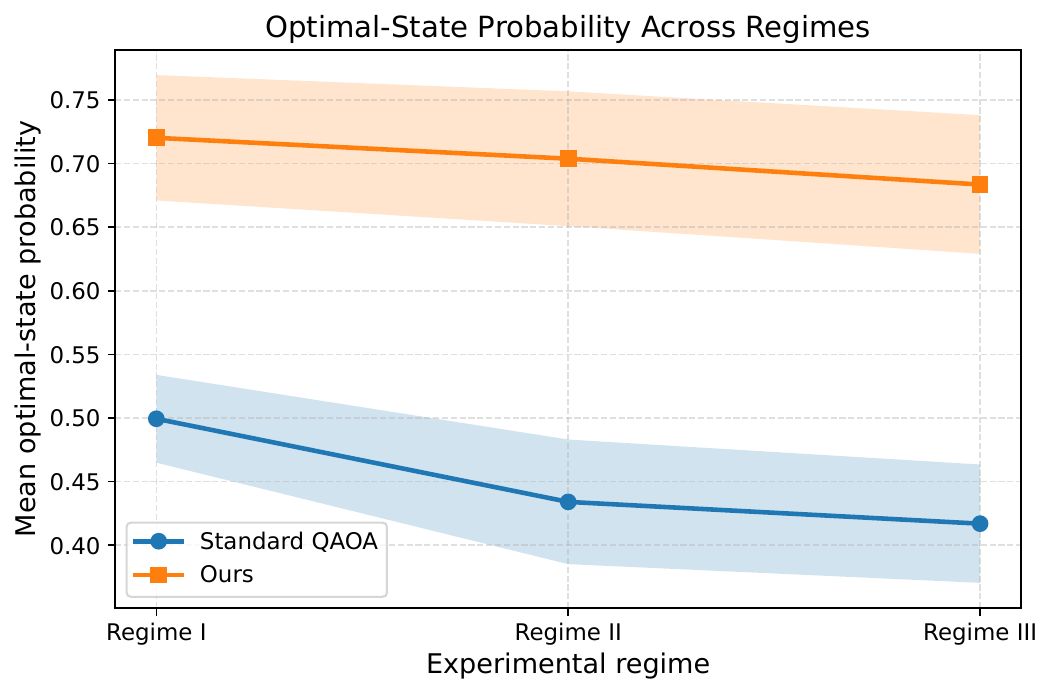}
        \caption{Optimal-solution probability across regimes}
        \label{fig:7a}
    \end{subfigure}
    \hfill
    \begin{subfigure}[b]{0.45\linewidth}
        \centering
        \includegraphics[width=\linewidth]{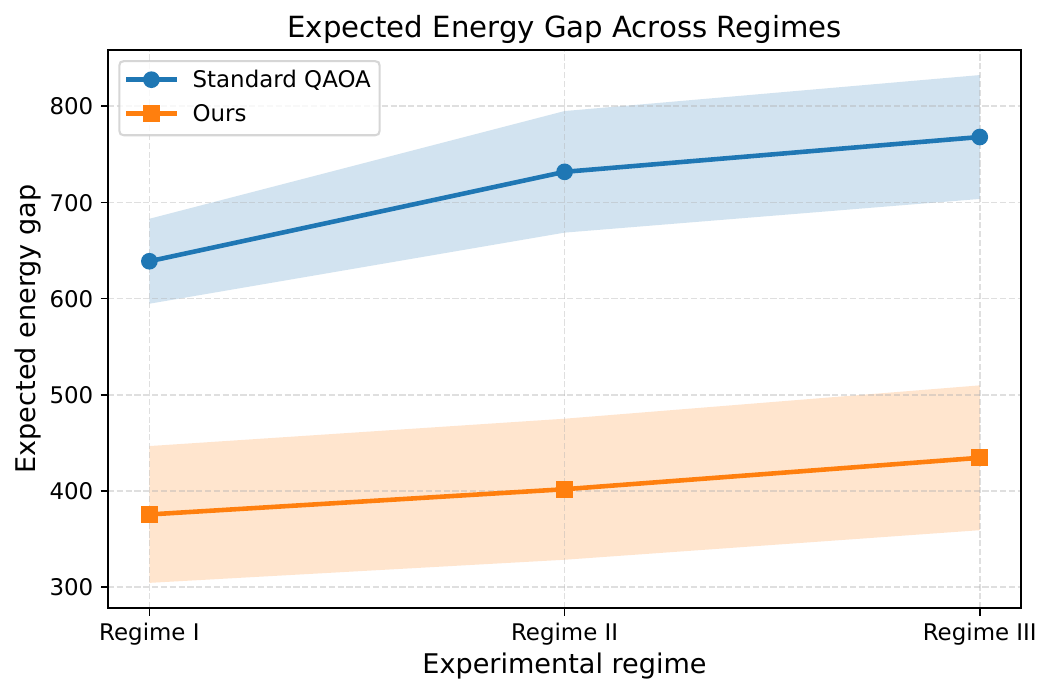}
        \caption{Expected energy gap across regimes}
        \label{fig:7b}
    \end{subfigure}

    \begin{subfigure}[b]{0.45\linewidth}
        \centering
        \includegraphics[width=\linewidth]{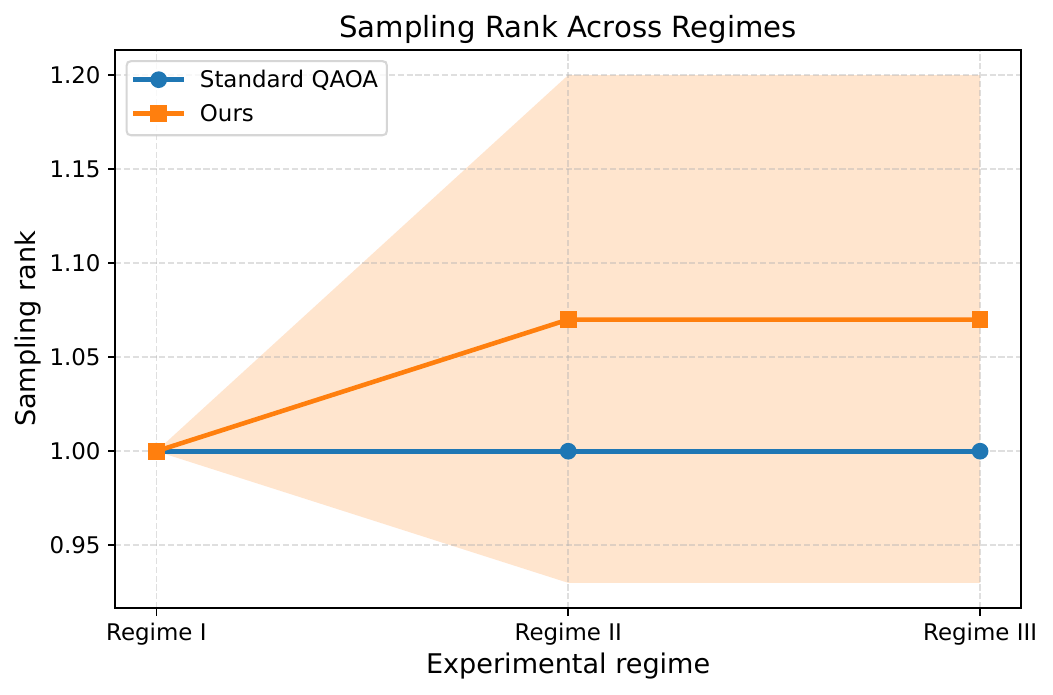}
        \caption{Sampling rank across regimes}
        \label{fig:7c}
    \end{subfigure}

    \caption{Performance comparison across the three experimental regimes. (Case I)}
    \label{fig:7}
\end{figure}
\FloatBarrier

For Case~I, the improvement is particularly clear in the optimal-solution probability. At \(p=2\), the proposed method increases the mean optimal-solution probability from roughly \(0.50\) to \(0.72\) in the statevector regime and from roughly \(0.42\) to \(0.68\) in the noisy finite-shot regime. The expected energy gap also decreases substantially, with reductions of about \(41.2\%\), \(45.1\%\), and \(43.4\%\) in Regimes I--III, respectively. This indicates that the improvement is not limited to the optimal-solution probability; the proposed method also shifts the overall sampling distribution toward lower-cost regions.

\textcolor{blue}{\textbf{Figures}}~\ref{fig:8} and~\ref{fig:9} extend the comparison to the two four-node cases. Case~I is the three-node instance and is compared with standard QAOA only; the block-wise XY-QAOA baseline is evaluated only for the two four-node instances, Case~II and Case~III. These instances are more demanding because the full computational basis has \(4096\) states. In both cases, standard QAOA has relatively low optimal-solution probability, reflecting the difficulty of sampling an optimal solution from the full Hilbert space. The block-wise XY-QAOA baseline improves substantially over standard QAOA in many probability and energy-gap comparisons, which confirms that feasibility-aware structure is useful. Since \(\lambda\) is a hyperparameter of the proposed hybrid mixer, we report full \(\lambda\)-sensitivity curves in \textcolor{blue}{\textbf{Figures}}~\ref{fig:11} and~\ref{fig:12}. For concise comparison below, we cite the best-sweep proposed results from those reported curves; these numbers should be interpreted as the performance of the tuned proposed ansatz, not as a separate post-hoc choice of \(\lambda\) for each individual metric. In Case~II, the reported best-sweep proposed results reach optimal-solution probabilities of \(0.1553\), \(0.1327\), and \(0.1141\) across the statevector, ideal finite-shot, and noisy finite-shot regimes, compared with \(0.1041\), \(0.0788\), and \(0.0483\) for block-wise XY-QAOA. The expected energy gap is also lower for the reported best-sweep proposed result in all three regimes: \(686.15\) versus \(720.62\), \(786.93\) versus \(878.08\), and \(941.03\) versus \(1106.98\), respectively. The sampling-rank metric requires a more nuanced reading: standard QAOA often attains rank one or another low rank because, once it assigns any appreciable mass to the optimum in the small benchmark instances, the optimum can become the most frequent sampled bitstring. This does not imply a better overall distribution. Standard QAOA still has much lower optimal-solution probability and much larger expected energy gap in the four-node cases, whereas the proposed method places substantially more probability mass near the optimal feasible region.

\begin{figure}[htp]
    \centering
    \begin{subfigure}[b]{0.45\linewidth}
        \centering
        \includegraphics[width=\linewidth]{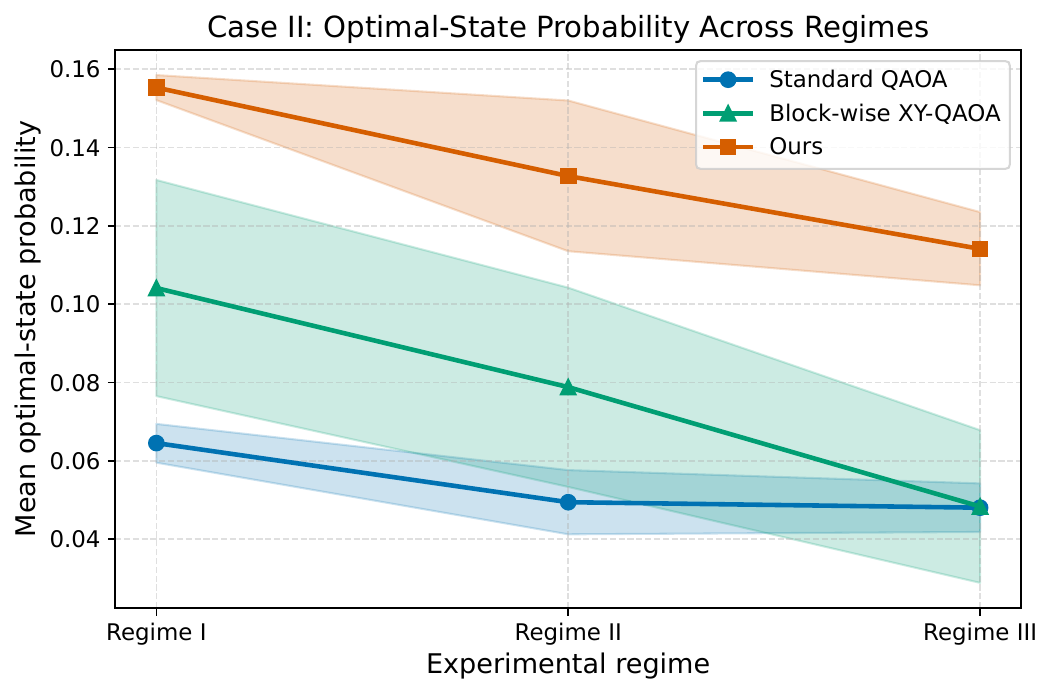}
        \caption{Optimal-solution probability across regimes}
        \label{fig:8a}
    \end{subfigure}
    \hfill
    \begin{subfigure}[b]{0.45\linewidth}
        \centering
        \includegraphics[width=\linewidth]{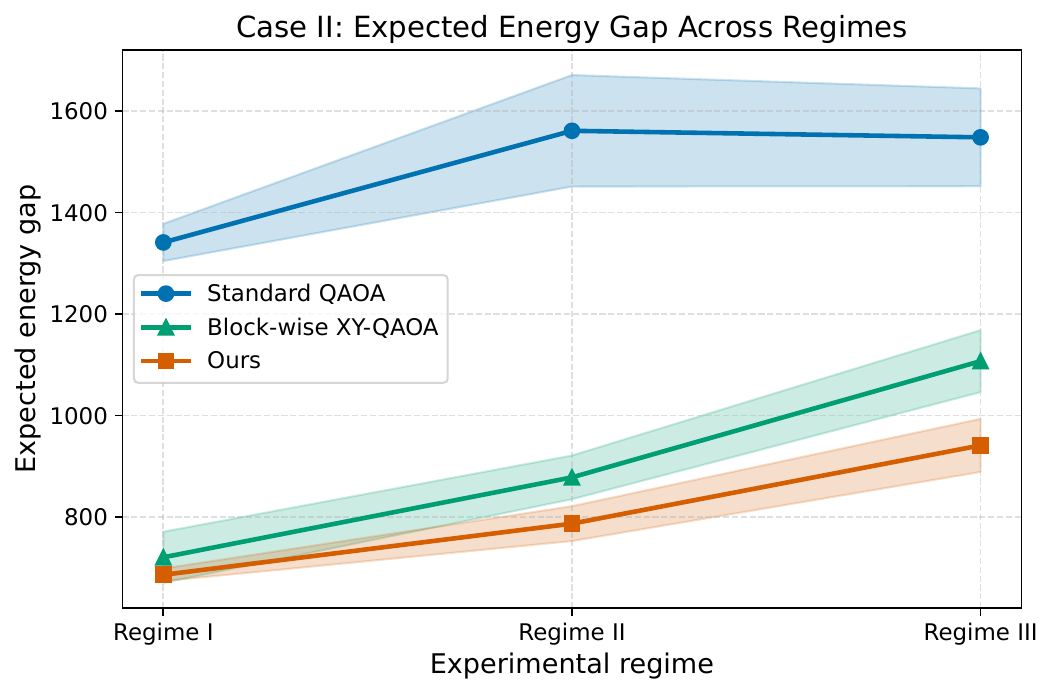}
        \caption{Expected energy gap across regimes}
        \label{fig:8b}
    \end{subfigure}

    \begin{subfigure}[b]{0.45\linewidth}
        \centering
        \includegraphics[width=\linewidth]{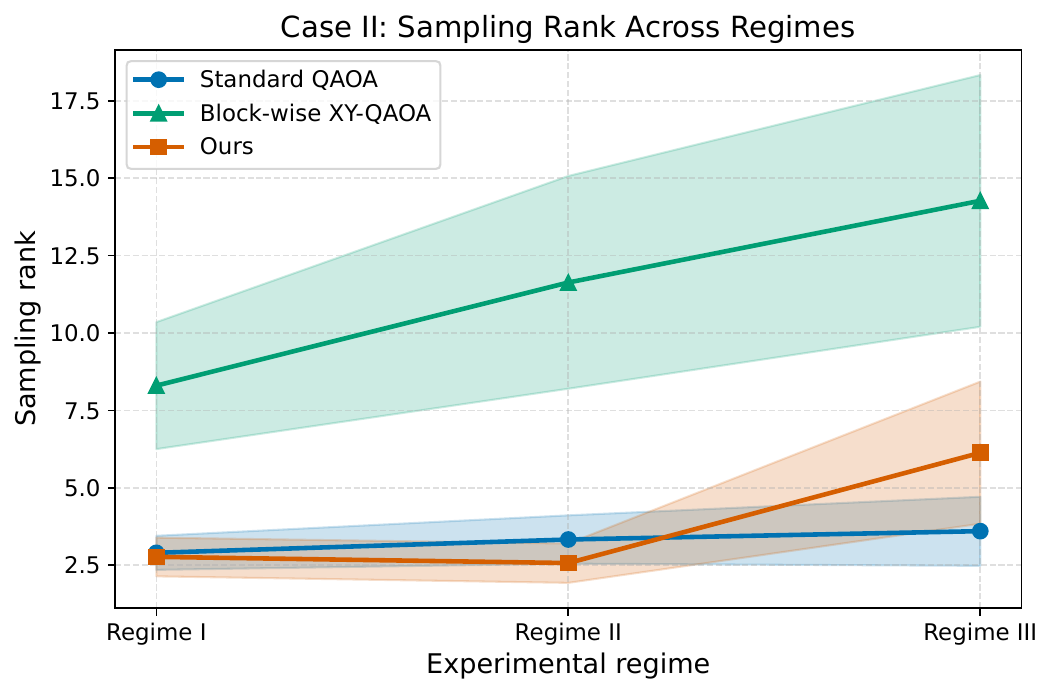}
        \caption{Sampling rank across regimes}
        \label{fig:8c}
    \end{subfigure}

    \caption{Performance comparison across the three experimental regimes. (Case II)}
    \label{fig:8}
\end{figure}

\begin{figure}[htp]
    \centering
    \begin{subfigure}[b]{0.45\linewidth}
        \centering
        \includegraphics[width=\linewidth]{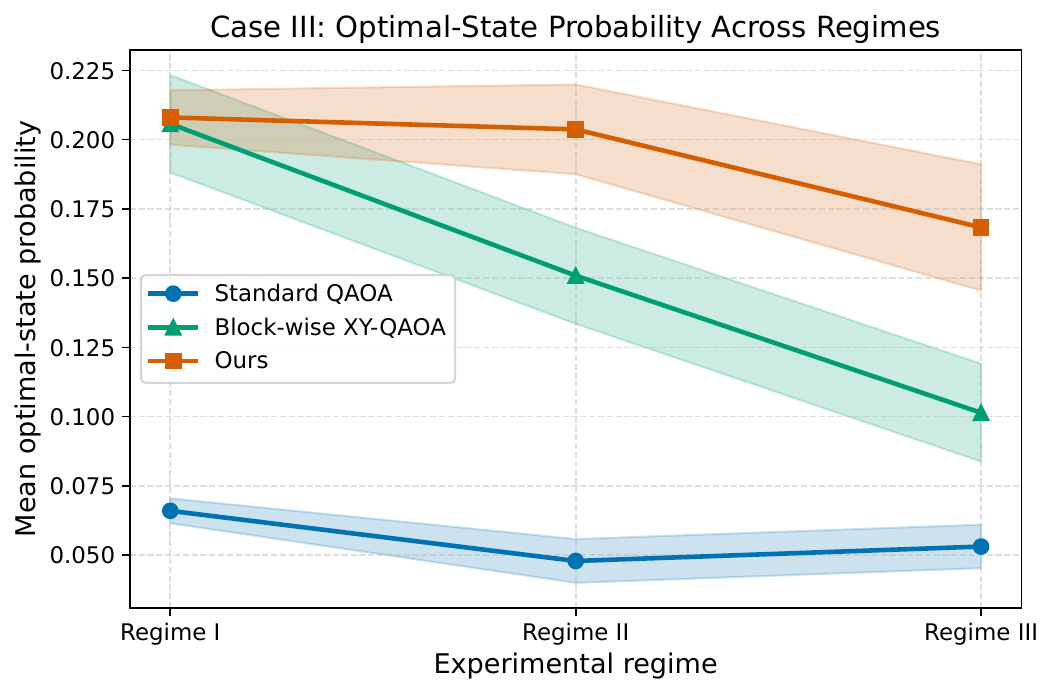}
        \caption{Optimal-solution probability across regimes}
        \label{fig:9a}
    \end{subfigure}
    \hfill
    \begin{subfigure}[b]{0.45\linewidth}
        \centering
        \includegraphics[width=\linewidth]{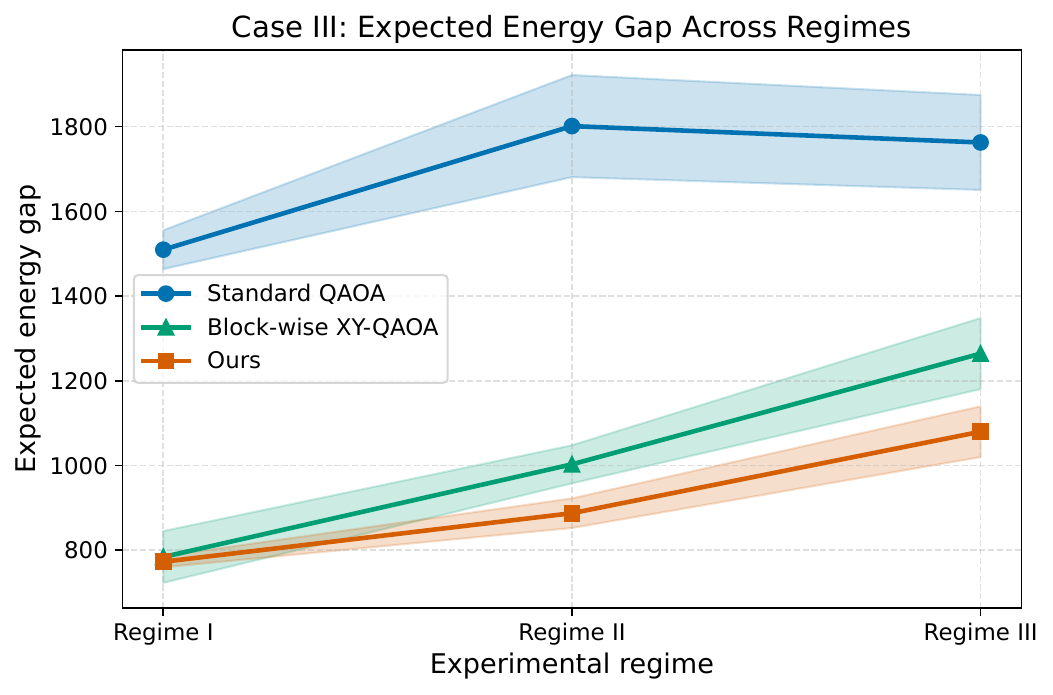}
        \caption{Expected energy gap across regimes}
        \label{fig:9b}
    \end{subfigure}

    \begin{subfigure}[b]{0.45\linewidth}
        \centering
        \includegraphics[width=\linewidth]{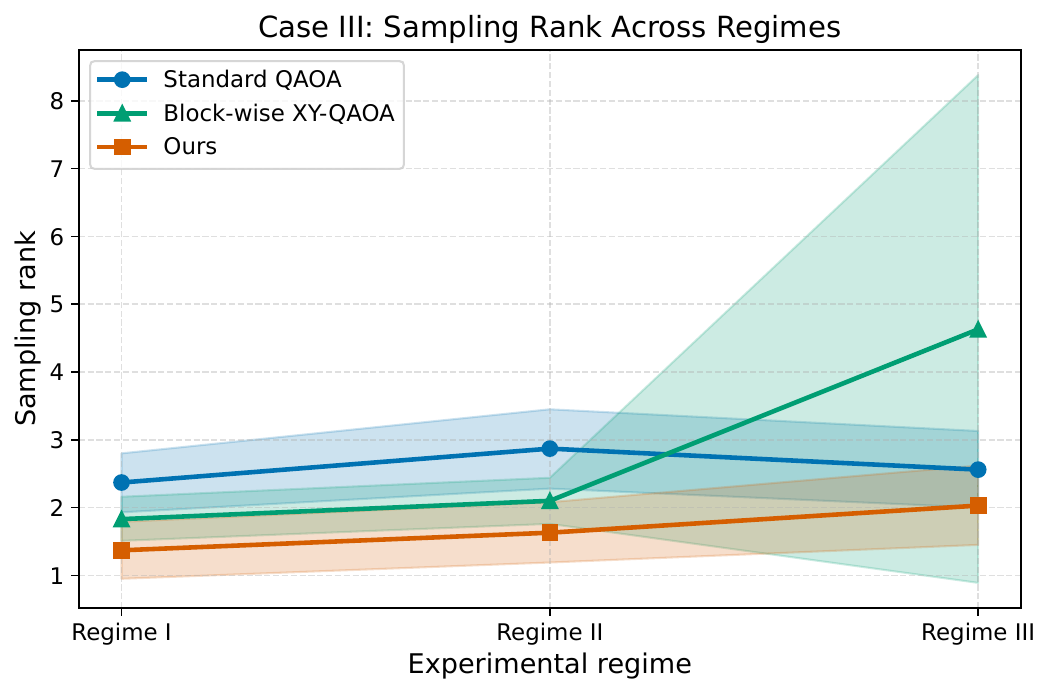}
        \caption{Sampling rank across regimes}
        \label{fig:9c}
    \end{subfigure}

    \caption{Performance comparison across the three experimental regimes. (Case III)}
    \label{fig:9}
\end{figure}
\FloatBarrier

\paragraph{Role of the block-wise XY-QAOA baseline}\label{sec:9}\label{line:9}
The block-wise XY-QAOA baseline is important because it tests whether the observed gains are simply due to using \(XY\)-type exchange gates. Standard QAOA uses a Pauli-\(X\) mixer only, while block-wise XY-QAOA uses \(XY\) exchange terms only. The proposed method combines these two mechanisms by using \(XY\) exchange terms to preserve selected overlap structures and \(X\)-type rotations to retain controlled exploration on qubits outside those selected overlap structures.

This distinction also explains why the baseline requires a structured initialization. An \(XY\)-only mixer preserves the Hamming weight inside each block, so it cannot move amplitude into a block-weight sector that is absent at initialization. The block-wise baseline therefore initializes the four-node problem in the full outgoing-degree sector with block-weight pattern \((2,1,1,1)\), using weight two for the depot-outgoing block and weight one for each customer-outgoing block. All fully feasible four-node routes lie in this sector and are included in the initial support, although the support also contains states that violate other VRP constraints. This construction gives the baseline useful feasibility-aware structure, but it also restricts the dynamics to the selected block-weight sector. In the comparisons below, the proposed model is reported at the best \(\lambda\) selected from its sensitivity curves.

The block-wise baseline is evaluated only for the two four-node instances, Case~II and Case~III; no block-wise comparison is made for the three-node Case~I. Across all six executed block-wise comparisons, the proposed method has a higher mean optimal-solution probability and a lower expected energy gap than block-wise XY-QAOA. In Case~II, the proposed method improves the optimal-solution probability from \(0.1041\), \(0.0788\), and \(0.0483\) under block-wise XY-QAOA to \(0.1553\), \(0.1327\), and \(0.1141\) across the statevector, ideal finite-shot, and noisy finite-shot regimes. The expected energy gap is also lower, decreasing from \(720.62\), \(878.08\), and \(1106.98\) to \(686.15\), \(786.93\), and \(941.03\), respectively. In Case~III, the statevector margin is smaller but remains favorable to the proposed method, while the finite-shot and noisy finite-shot advantages are larger. The optimal-solution probability increases from \(0.1509\) to \(0.2037\) and from \(0.1014\) to \(0.1683\), while the expected energy gap decreases from \(1002.56\) to \(887.14\) and from \(1264.00\) to \(1079.60\). These results indicate that the gain is not merely an \(XY\)-mixer effect; it comes from pairing overlap-aware initialization, selected \(XY\)-based preservation, and limited \(X\)-type exploration.

\paragraph{Effect of QAOA depth}\label{sec:12}\label{line:12}
The \(p\)-depth study evaluates how the number of QAOA layers affects solution quality and implementation cost. The appendix depth tables (\textcolor{blue}{\textbf{Tables}}~\ref{table:23}--\ref{table:46}) show that increasing \(p\) from \(1\) to \(3\) generally improves the proposed method, especially in optimal-solution probability and expected energy gap. For Case~I, the proposed method in the statevector regime increases the mean optimal-solution probability from \(0.2695\) at \(p=1\) to \(0.8578\) at \(p=3\), which is an approximately \(218\%\) relative increase. Over the same depth range, the expected energy gap decreases from \(1376.0\) to \(184.8\), an approximately \(86.6\%\) reduction. Similar improvements appear in the finite-shot and noisy regimes: the optimal-solution probability increases from \(0.2487\) to \(0.7846\) in Regime II and from \(0.2450\) to \(0.7241\) in Regime III, while the corresponding expected gaps decrease by approximately \(79.9\%\) and \(75.1\%\).

The four-node cases show the same pattern, but also make the depth trade-off more visible. In Case~II, the proposed statevector optimal-solution probability improves from \(0.0234\) at \(p=1\) to \(0.1944\) at \(p=3\), corresponding to roughly a \(730\%\) relative increase, while the expected energy gap drops from \(2403.7\) to \(521.9\), a \(78.3\%\) reduction. In the ideal finite-shot and noisy finite-shot regimes, Case~II also improves from \(0.0953\) to \(0.1551\) and from \(0.0914\) to \(0.1365\), respectively. For Case~III, the statevector optimal-solution probability increases from \(0.0989\) at \(p=1\) to \(0.2367\) at \(p=3\), while the expected gap decreases from \(1160.6\) to \(581.4\). The noisy Case~III setting also improves from \(0.0964\) to \(0.1923\), with the expected gap decreasing from \(1216.9\) to \(938.3\). These concrete appendix results indicate that \(p=1\) alone does not adequately characterize the behavior of the proposed ansatz.

At the same time, the \(p\)-depth results also show the expected cost of increasing depth. A larger \(p\) increases the number of trainable parameters from \(2\) at \(p=1\) to \(6\) at \(p=3\), repeats the cost and mixer layers, and increases the number of two-qubit operations. The finite-shot and noisy results are therefore less uniformly monotonic than the statevector results. For example, \textcolor{blue}{\textbf{Table}}~\ref{table:42} shows that, for the block-wise XY-QAOA baseline in Case~II under ideal finite-shot sampling, the optimal-solution probability decreases from \(0.1089\) at \(p=1\) to \(0.0536\) at \(p=3\), while the expected energy gap still decreases from \(902.7\) to \(870.3\) and the sampling rank worsens from \(4.90\) to \(14.00\). This is expected because deeper circuits can be more expressive, but they are also harder to optimize and more vulnerable to sampling and noise effects. The \(p\)-depth study therefore supports a balanced interpretation: larger \(p\) can improve solution quality, but the useful depth is constrained by sampling noise, optimizer budget, and hardware noise.

\paragraph{Effect of noise}
The noisy finite-shot results show that the proposed method remains beneficial under the optimistic hardware-inspired noise model, but the advantage is attenuated relative to the ideal statevector regime. This is especially visible when comparing the statevector and noisy panels across \textcolor{blue}{\textbf{Figures}}~\ref{fig:7}, \ref{fig:8}, and~\ref{fig:9}. The reduction is expected because the proposed and block-wise XY-QAOA circuits use additional two-qubit \(R_{XX}\) and \(R_{YY}\) operations, and two-qubit gates are more noise-sensitive than single-qubit rotations.\label{sec:19}\label{line:19}

Nevertheless, the noisy results are still favorable. In the three-node case, the proposed method remains clearly better than standard QAOA. In the four-node cases, the proposed method also remains competitive with or better than both baselines. This suggests that the structural benefit of the proposed ansatz can survive the modeled finite-shot and gate-noise conditions, although the full advantage would require sufficiently high-fidelity hardware.

\paragraph{Sensitivity to the mixer parameter \(\lambda\)}
\textcolor{blue}{\textbf{Figures}}~\ref{fig:10}, \ref{fig:11}, and~\ref{fig:12} show that the hybrid mixer has a nontrivial dependence on the weighting parameter \(\lambda\). This parameter controls the strength of the \(X\)-mixer component relative to the \(XY\)-exchange component. When \(\lambda\) is too small, the dynamics are too close to a restrictive \(XY\)-only evolution. This can limit the ability to move probability mass toward lower-cost configurations when useful transitions require changing qubits outside the protected structures. When \(\lambda\) is too large, the \(X\) component can dominate and weaken the structural advantage introduced by the constraint-aware initialization.

The sensitivity curves therefore support the main design intuition of the hybrid mixer. The \(XY\) part preserves selected one-hot or overlap structures, while the \(X\) part introduces controlled flexibility through qubits outside the protected structures. The best performance is obtained when these two effects are balanced. In the reported experiments, the selected \(\lambda\) for Case~I is around the middle of the enlarged search range, while the selected value for both four-node cases is \(\lambda=2.0\) based on the full-method sweep.

\begin{figure}[H]
    \centering
    \begin{subfigure}[b]{0.45\linewidth}
        \centering
        \includegraphics[width=\linewidth]{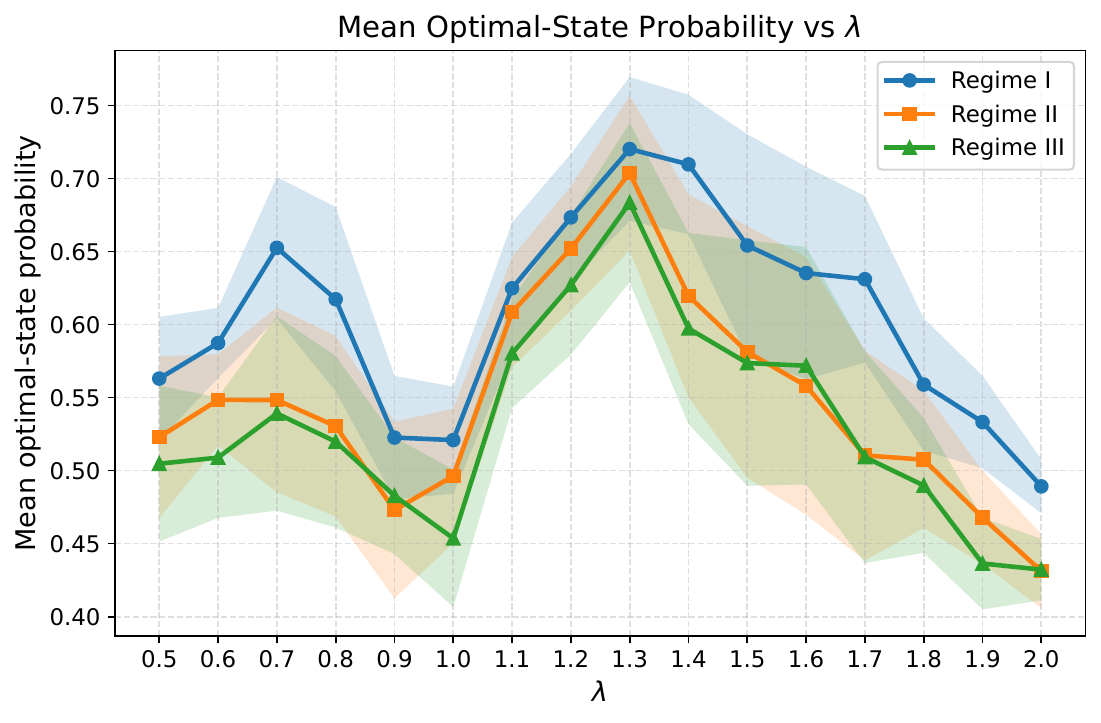}
        \caption{Optimal-solution probability under different $\lambda$}
        \label{fig:10a}
    \end{subfigure}
    \hfill
    \begin{subfigure}[b]{0.45\linewidth}
        \centering
        \includegraphics[width=\linewidth]{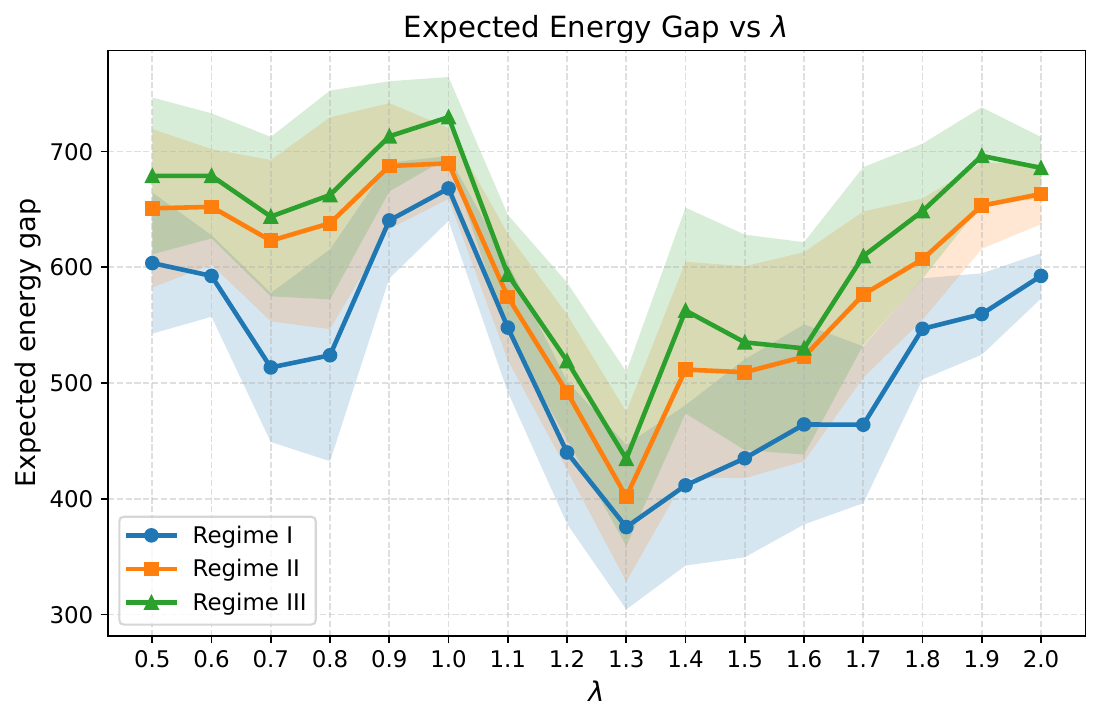}
        \caption{Expected energy gap under different $\lambda$}
        \label{fig:10b}
    \end{subfigure}

    \begin{subfigure}[b]{0.45\linewidth}
        \centering
        \includegraphics[width=\linewidth]{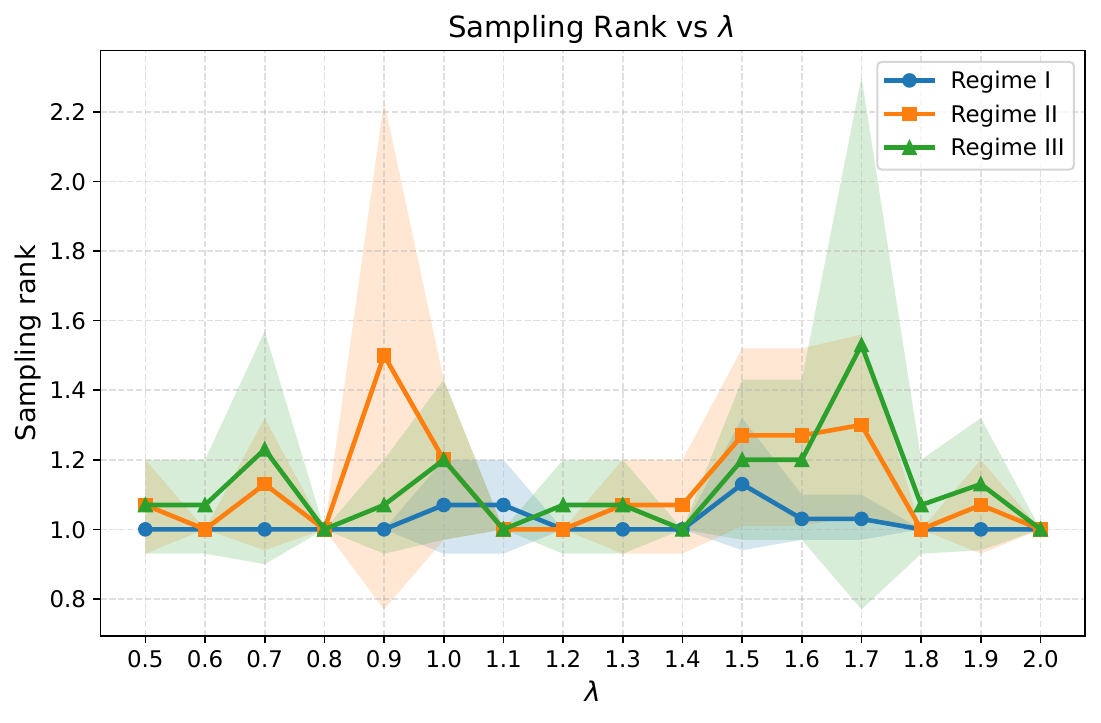}
        \caption{Sampling rank under different $\lambda$}
        \label{fig:10c}
    \end{subfigure}

    \caption{Sensitivity of the proposed method to the mixing parameter $\lambda$ across the three experimental regimes (Case I)}
    \label{fig:10}
\end{figure}

\begin{figure}[H]
    \centering
    \begin{subfigure}[b]{0.45\linewidth}
        \centering
        \includegraphics[width=\linewidth]{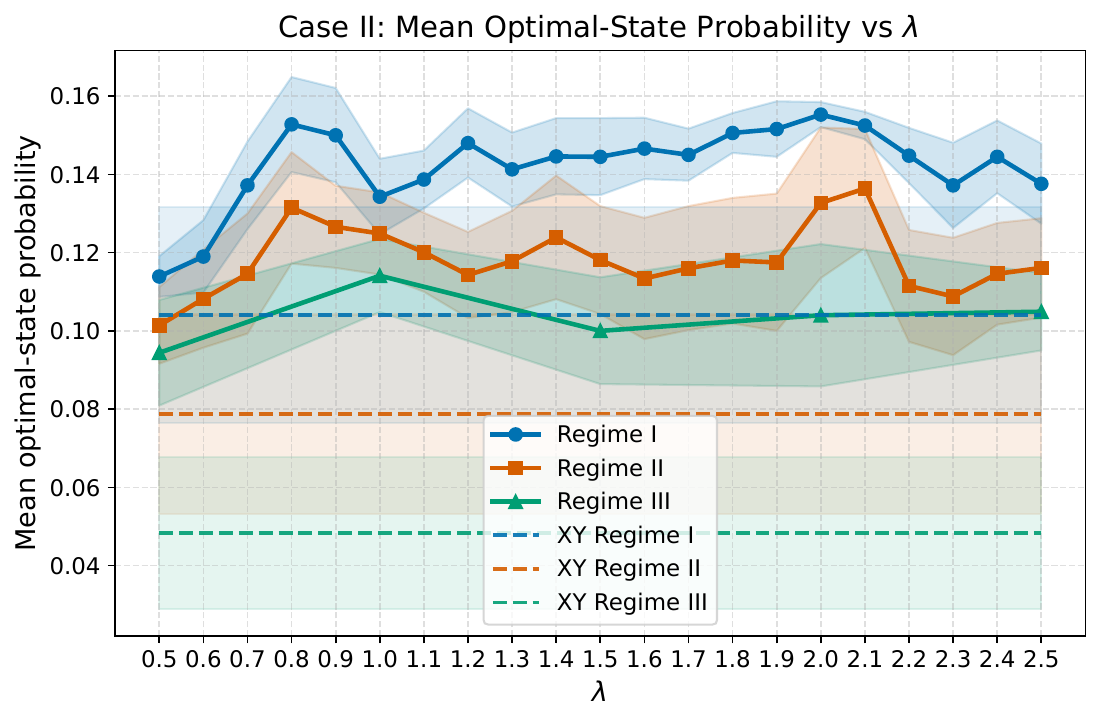}
        \caption{Optimal-solution probability under different $\lambda$}
        \label{fig:11a}
    \end{subfigure}
    \hfill
    \begin{subfigure}[b]{0.45\linewidth}
        \centering
        \includegraphics[width=\linewidth]{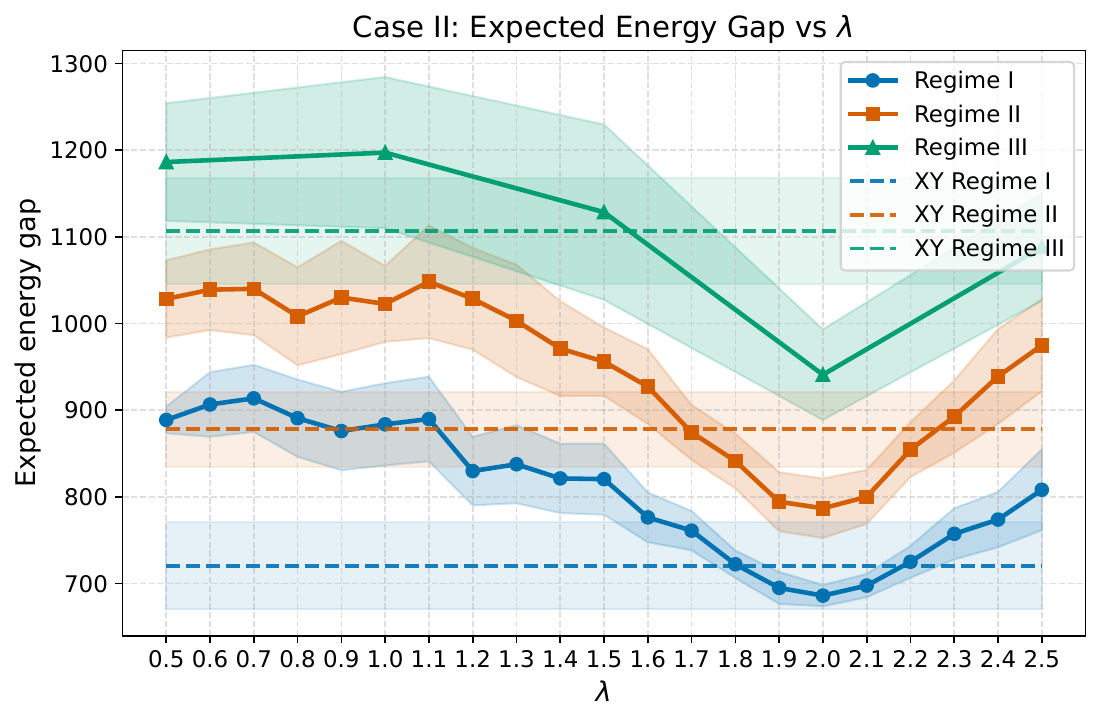}
        \caption{Expected energy gap under different $\lambda$}
        \label{fig:11b}
    \end{subfigure}

    \begin{subfigure}[b]{0.45\linewidth}
        \centering
        \includegraphics[width=\linewidth]{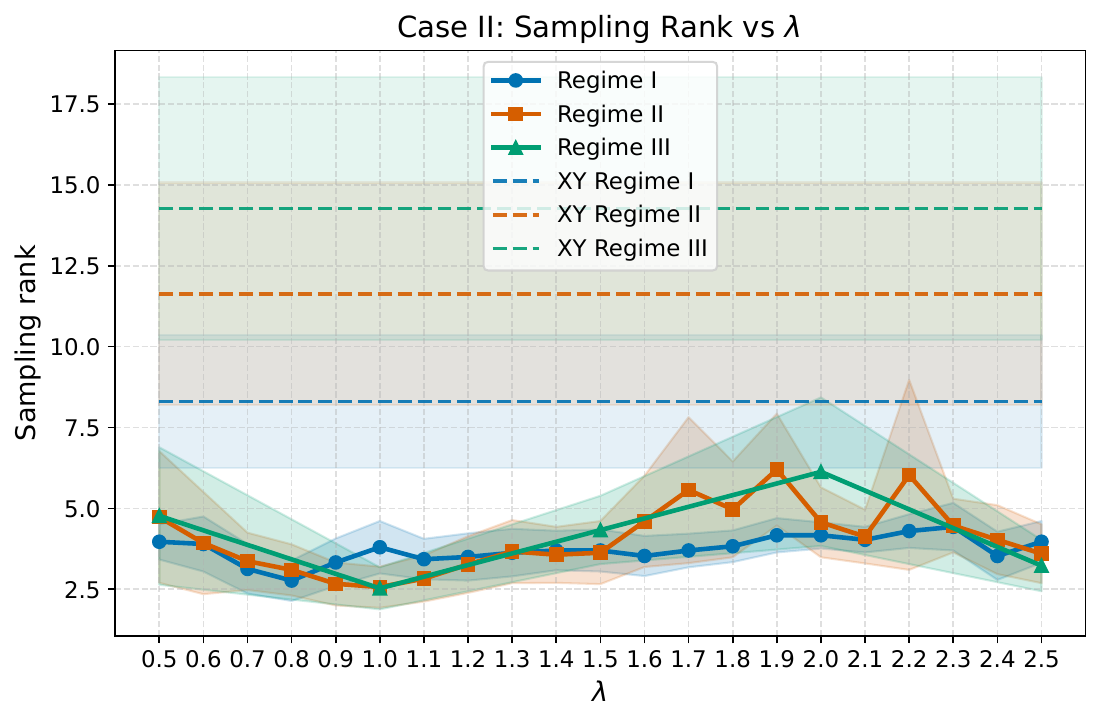}
        \caption{Sampling rank under different $\lambda$}
        \label{fig:11c}
    \end{subfigure}

    \caption{Sensitivity of the proposed method to the mixing parameter $\lambda$ across the three experimental regimes (Case II)}
    \label{fig:11}
\end{figure}

\begin{figure}[H]
    \centering
    \begin{subfigure}[b]{0.45\linewidth}
        \centering
        \includegraphics[width=\linewidth]{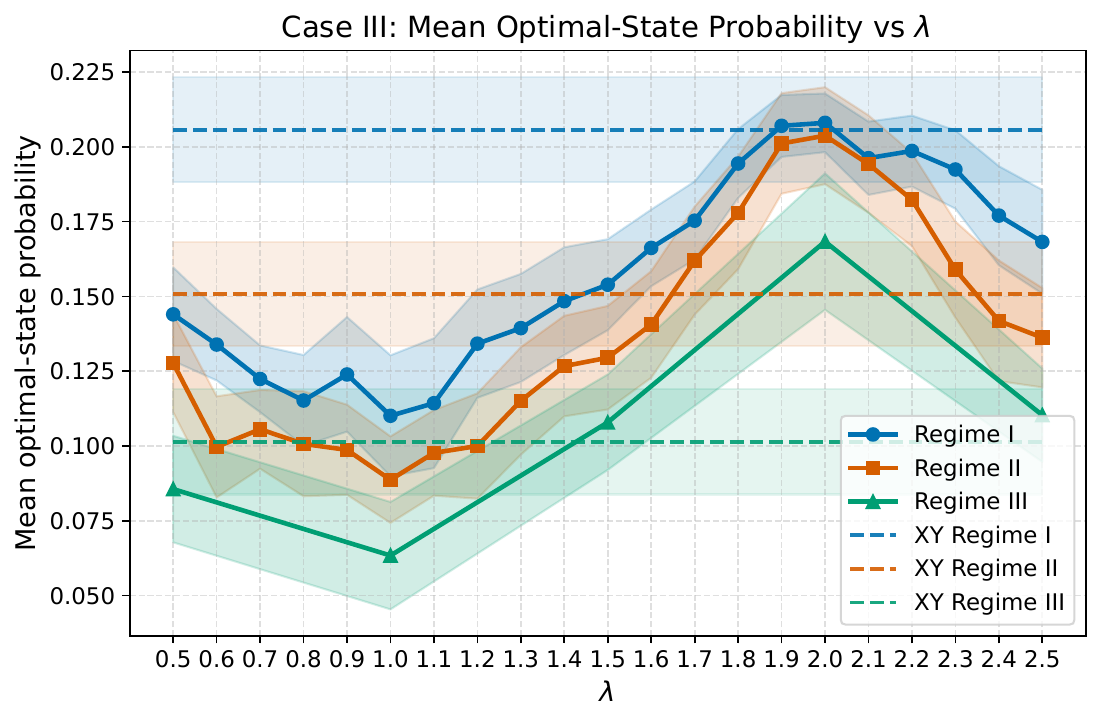}
        \caption{Optimal-solution probability under different $\lambda$}
        \label{fig:12a}
    \end{subfigure}
    \hfill
    \begin{subfigure}[b]{0.45\linewidth}
        \centering
        \includegraphics[width=\linewidth]{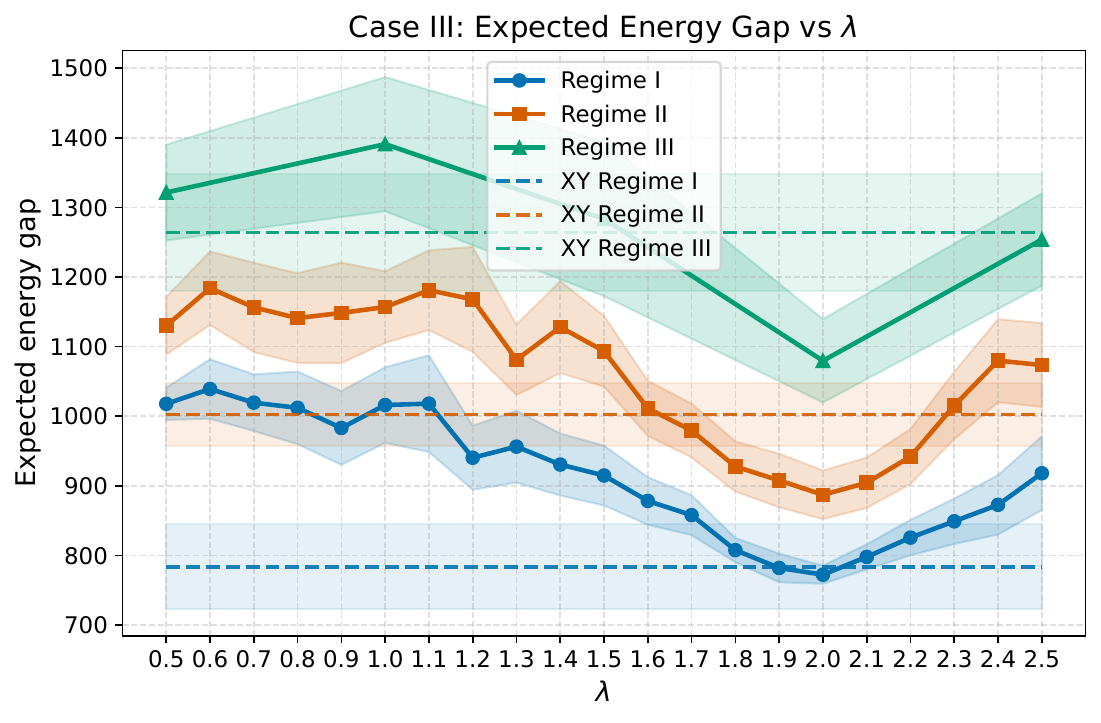}
        \caption{Expected energy gap under different $\lambda$}
        \label{fig:12b}
    \end{subfigure}

    \begin{subfigure}[b]{0.45\linewidth}
        \centering
        \includegraphics[width=\linewidth]{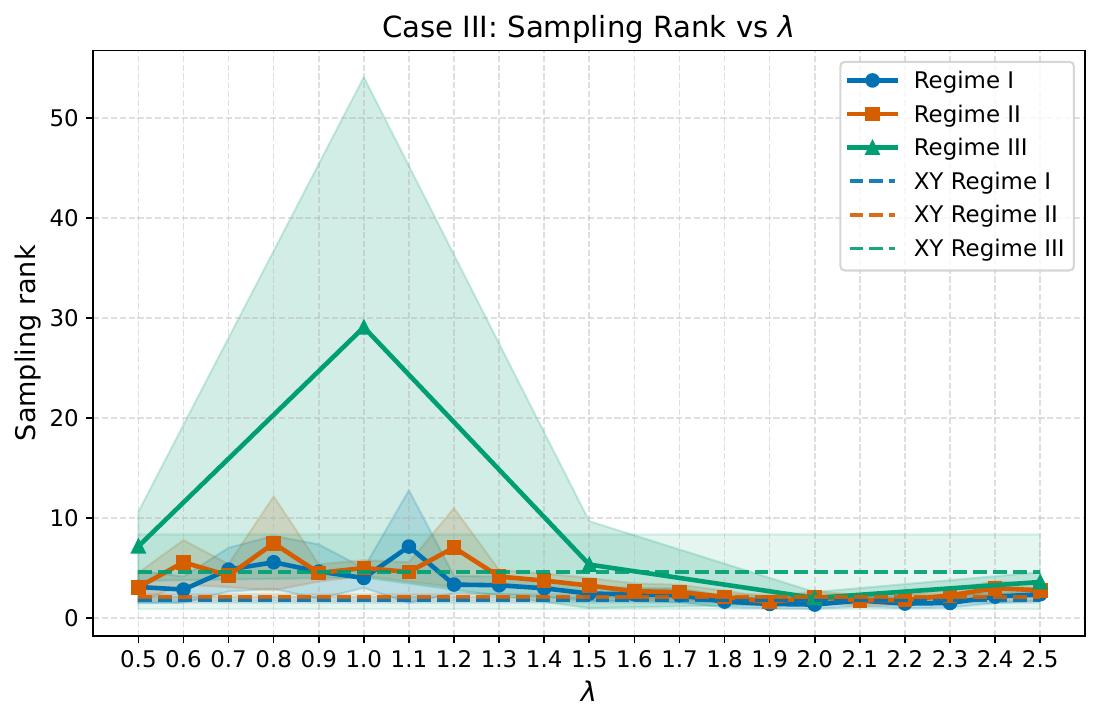}
        \caption{Sampling rank under different $\lambda$}
        \label{fig:12c}
    \end{subfigure}

    \caption{Sensitivity of the proposed method to the mixing parameter $\lambda$ across the three experimental regimes (Case III)}
    \label{fig:12}
\end{figure}
\FloatBarrier

\paragraph{Ablation study}\label{sec:6}\label{line:6}
The ablation results in the appendix (\textcolor{blue}{\textbf{Tables}}~\ref{table:14}--\ref{table:22}) and the summary bar charts in \textcolor{blue}{\textbf{Figures}}~\ref{fig:13}--\ref{fig:15} directly disentangle the two proposed ingredients. The Init-only variant combines the proposed constraint-aware initialization with the conventional transverse-field Pauli-\(X\) mixer, implemented by standard \(R_X\) rotations on all qubits. The Mixer-only variant combines the standard uniform initialization \(\ket{+}^{\bigotimes n}\) with the proposed hybrid \(XY\)--\(X\) mixer. The results show a strong interaction effect between the two components. The full proposed method is consistently stronger than both ablated variants, while both ablated variants are also weaker than standard QAOA in optimal-solution probability across all tested cases and regimes.

\begin{figure}[htbp]
\centering

\begin{subfigure}[t]{0.45\textwidth}
    \centering
    \includegraphics[width=\linewidth]{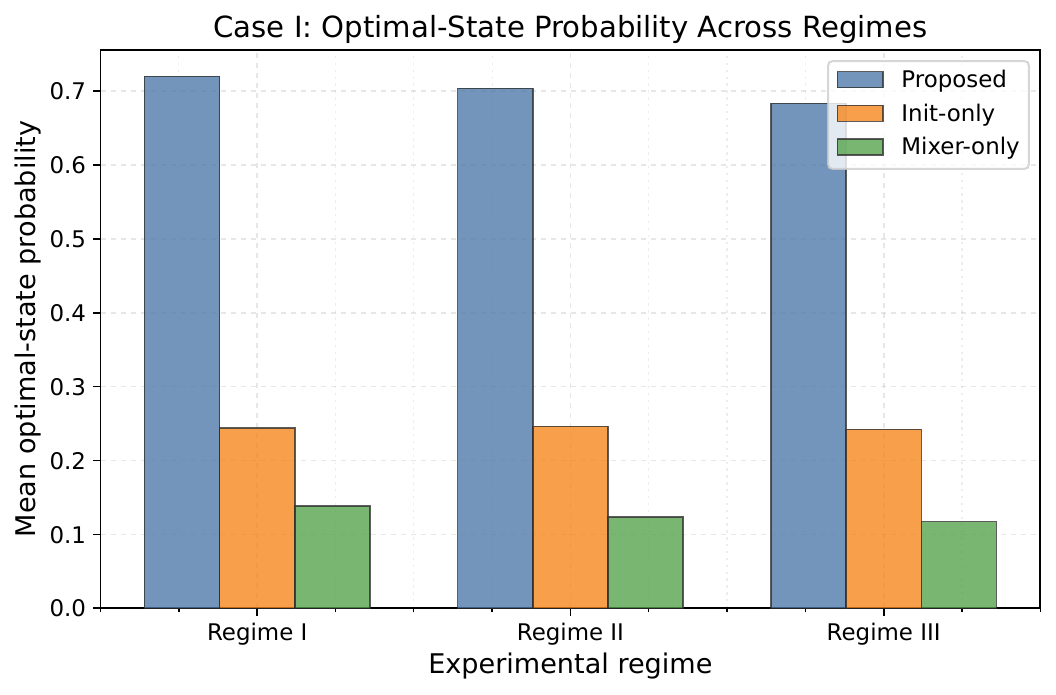}
    \caption{Optimal-solution probability}
    \label{fig:13a}
\end{subfigure}
\hfill
\begin{subfigure}[t]{0.45\textwidth}
    \centering
    \includegraphics[width=\linewidth]{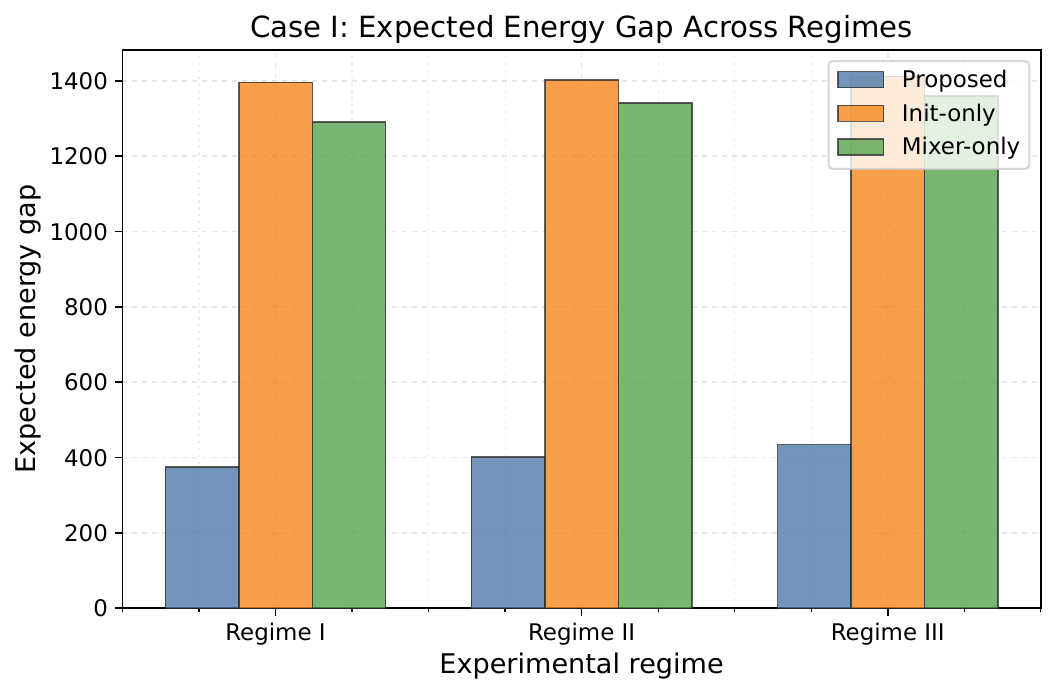}
    \caption{Expected energy gap}
    \label{fig:13b}
\end{subfigure}

\begin{subfigure}[t]{0.45\textwidth}
    \centering
    \includegraphics[width=\linewidth]{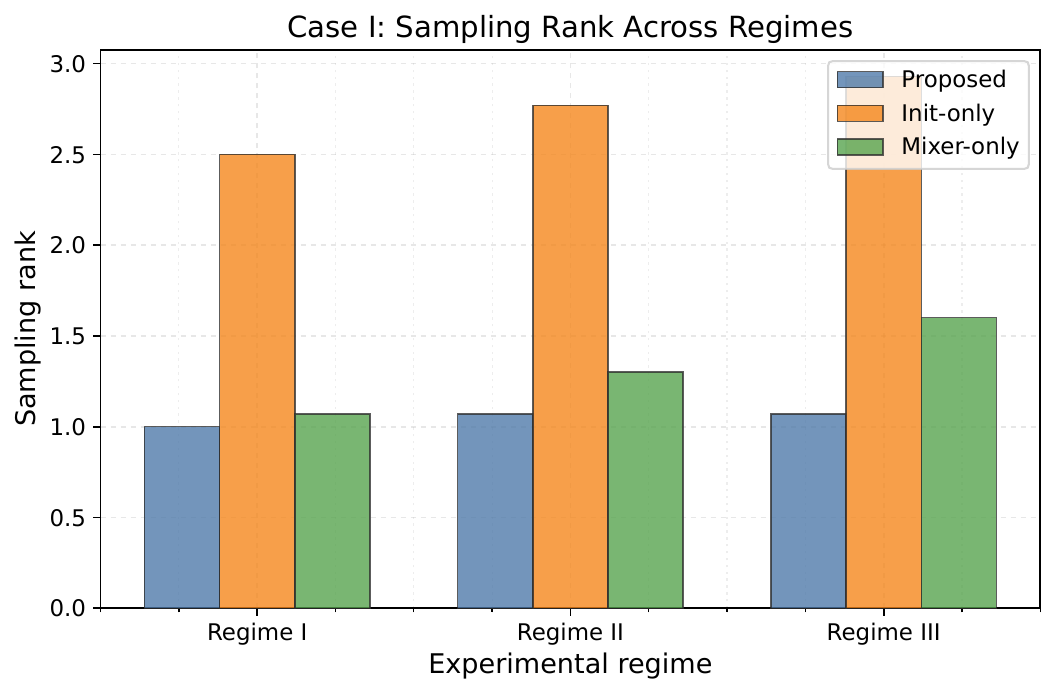}
    \caption{Sampling rank}
    \label{fig:13c}
\end{subfigure}

\caption{Ablation study for Case I (3-node).}
\label{fig:13}
\end{figure}

For Case~I, the full proposed method reaches \(p^{*}=0.7201\), \(0.7036\), and \(0.6834\) in the statevector, ideal finite-shot, and noisy finite-shot regimes, respectively. The corresponding Init-only probabilities are \(0.2439\), \(0.2462\), and \(0.2420\), while the Mixer-only probabilities are \(0.1381\), \(0.1236\), and \(0.1173\). The full method therefore increases \(p^{*}\) by \(182.4\%\)--\(195.2\%\) over Init-only and by \(421.4\%\)--\(482.6\%\) over Mixer-only. It also reduces the expected energy gap by \(69.2\%\)--\(73.1\%\) relative to Init-only and by \(68.0\%\)--\(70.9\%\) relative to Mixer-only. Standard QAOA gives \(p^{*}=0.4994\), \(0.4341\), and \(0.4170\) in the same three regimes, which means that both ablated variants fall below even the standard baseline.

\begin{figure}[htbp]
\centering
\FloatBarrier

\begin{subfigure}[t]{0.45\textwidth}
    \centering
    \includegraphics[width=\linewidth]{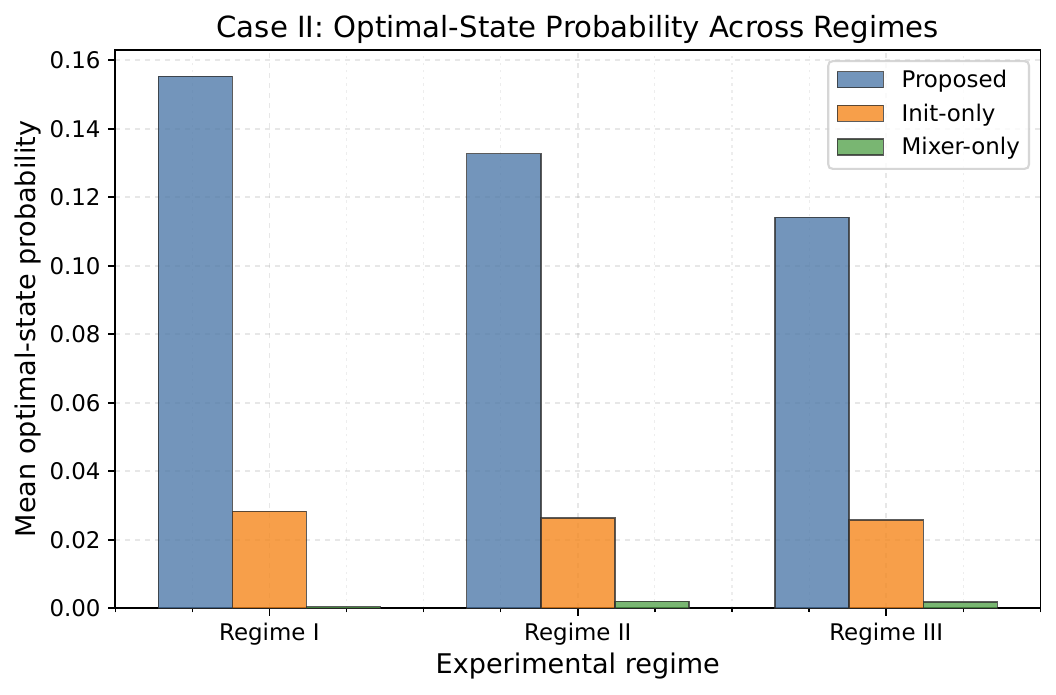}
    \caption{Optimal-solution probability}
    \label{fig:14a}
\end{subfigure}
\hfill
\begin{subfigure}[t]{0.45\textwidth}
    \centering
    \includegraphics[width=\linewidth]{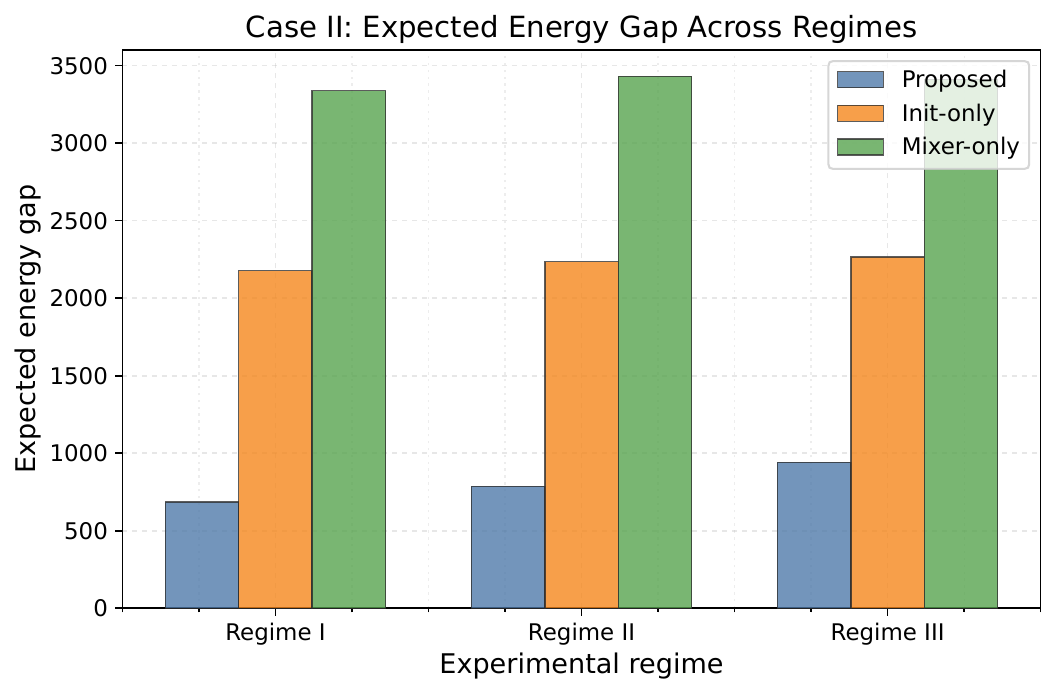}
    \caption{Expected energy gap}
    \label{fig:14b}
\end{subfigure}

\begin{subfigure}[t]{0.45\textwidth}
    \centering
    \includegraphics[width=\linewidth]{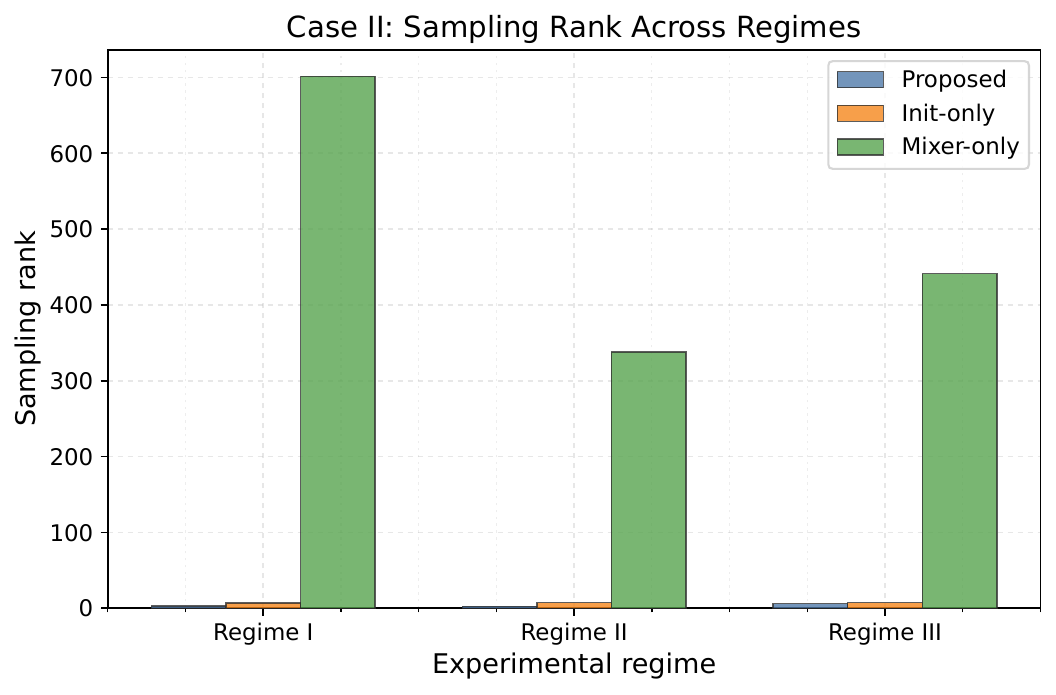}
    \caption{Sampling rank}
    \label{fig:14c}
\end{subfigure}

\caption{Ablation study for Case II (4-node balanced symmetric case).}
\label{fig:14}
\end{figure}

\begin{figure}[htbp]
\centering

\begin{subfigure}[t]{0.45\textwidth}
    \centering
    \includegraphics[width=\linewidth]{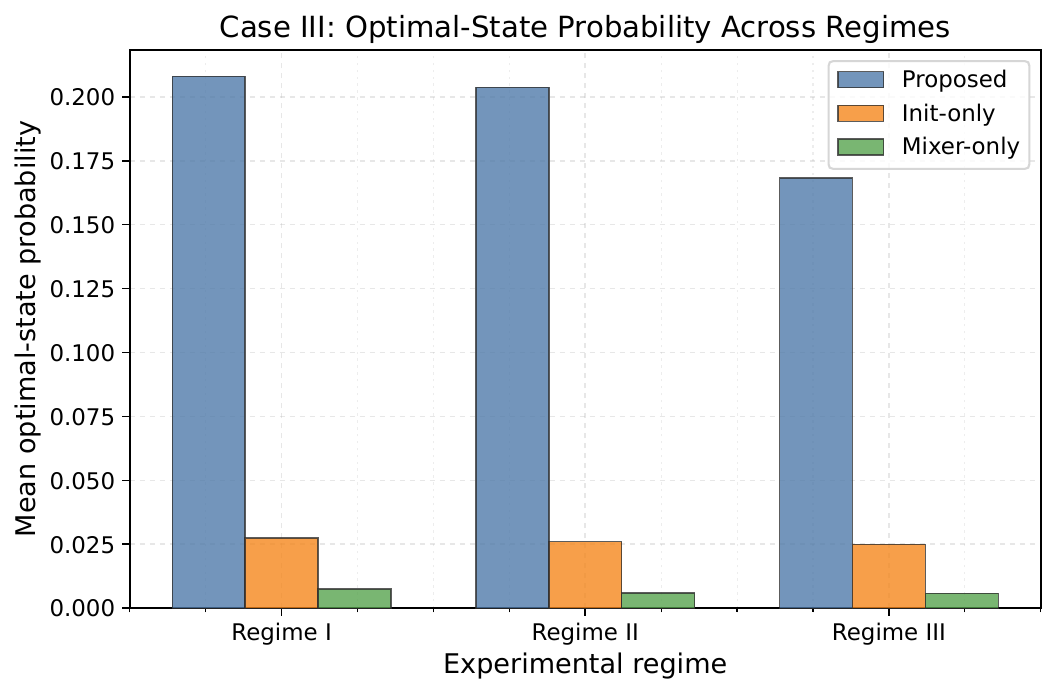}
    \caption{Optimal-solution probability}
    \label{fig:15a}
\end{subfigure}
\hfill
\begin{subfigure}[t]{0.45\textwidth}
    \centering
    \includegraphics[width=\linewidth]{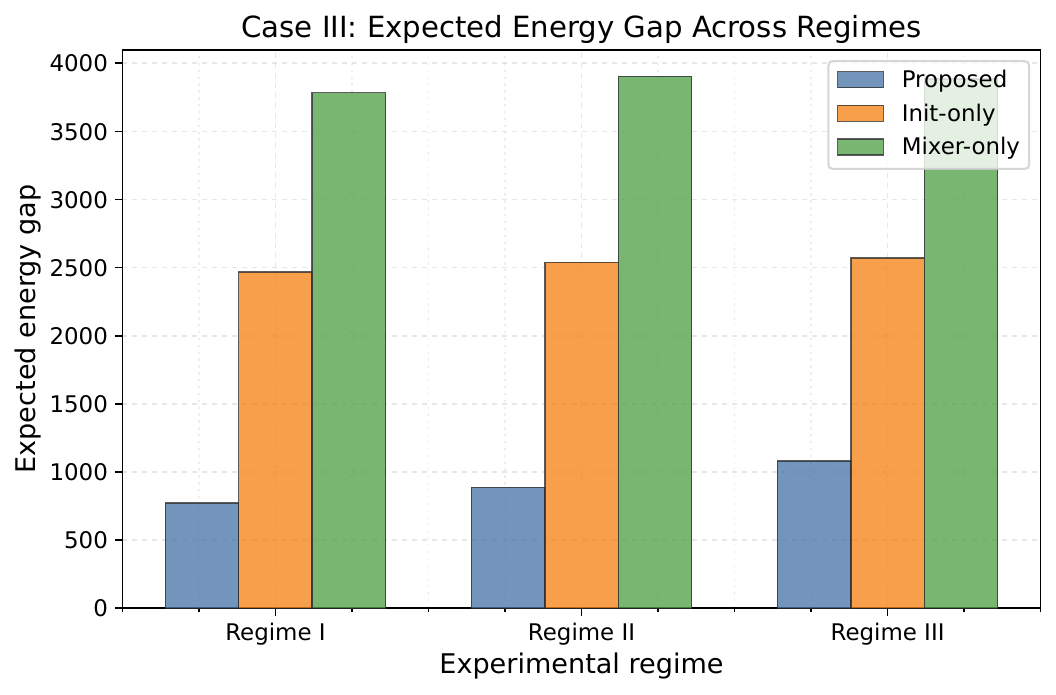}
    \caption{Expected energy gap}
    \label{fig:15b}
\end{subfigure}

\begin{subfigure}[t]{0.45\textwidth}
    \centering
    \includegraphics[width=\linewidth]{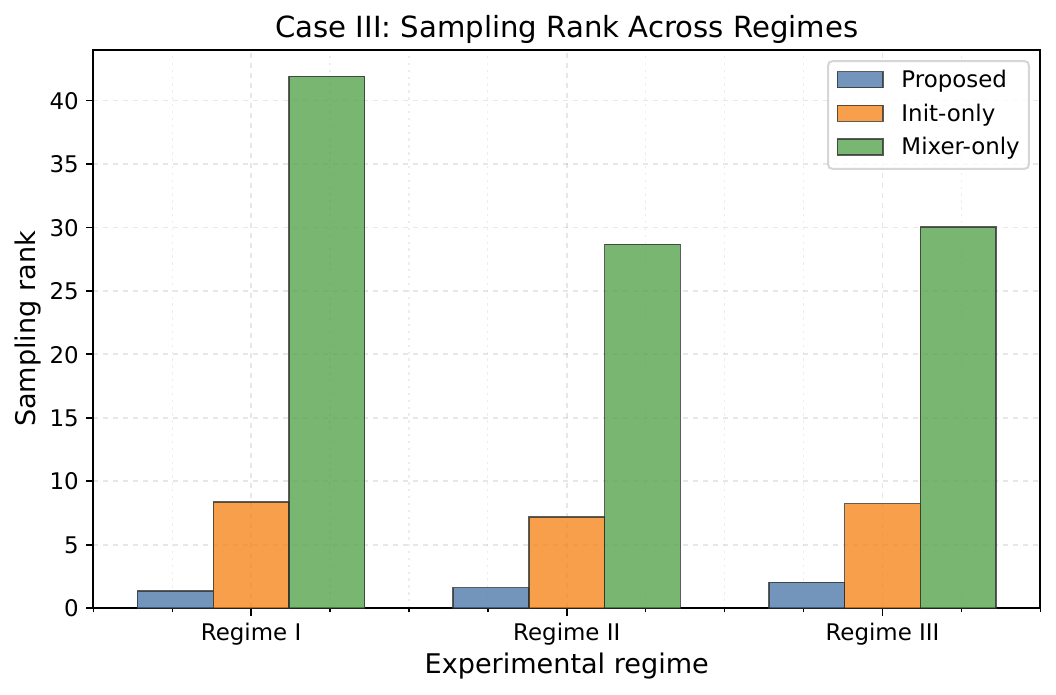}
    \caption{Sampling rank}
    \label{fig:15c}
\end{subfigure}

\caption{Ablation study for Case III (4-node customer cluster case).}
\label{fig:15}
\end{figure}
\FloatBarrier

The four-node cases show an even stronger separation. In the matched Case~II statevector ablation with \(p=2\) and a maximum of \(2000\) COBYLA iterations per restart, the full method gives \(p^{*}=0.1553\) and expected gap \(686.24\). Init-only gives \(p^{*}=0.0283\) and gap \(2177.12\), while Mixer-only gives \(p^{*}=0.0005\) and gap \(3338.04\). The full method therefore increases \(p^{*}\) by \(448.8\%\) over Init-only and by a factor of \(310.6\) over Mixer-only, while reducing the expected gap by \(68.5\%\) and \(79.4\%\), respectively. In Case~III under ideal finite-shot sampling, the full method gives \(p^{*}=0.2037\) and gap \(887.14\), compared with \(p^{*}=0.0260\) and gap \(2538.54\) for Init-only and \(p^{*}=0.0060\) and gap \(3904.48\) for Mixer-only. These values correspond to a \(683.5\%\) increase over Init-only, a \(33.9\)-fold increase over Mixer-only, and expected-gap reductions of \(65.1\%\) and \(77.3\%\). Across the six four-node case-regime combinations, Init-only reaches only \(41.5\%\)--\(54.3\%\) of the standard-QAOA optimal-solution probability, and Mixer-only reaches only \(0.8\%\)--\(12.5\%\) of it.

These results indicate that neither component works as a competitive standalone substitute for the full ansatz. The initialization alone places amplitude on a more structured support, but without the matching hybrid mixer it cannot effectively exploit that support during optimization. The mixer alone starts from an unstructured uniform state, so its exchange and \(X\)-type moves have no overlap-aware probability concentration to preserve or refine. The proposed method obtains its advantage only when the two components are paired. The initialization supplies a structured overlap-aware support, and the hybrid mixer explores this support while preserving the selected encoded constraints and allowing controlled transitions through qubits left outside the selected overlap structures.

\paragraph{Interpretation of the three methods}
Taken together, the results suggest a clear hierarchy of structural information. Standard QAOA is the most flexible but least informed because it explores the full basis and does not preserve VRP structure. Block-wise XY-QAOA is more informed but more restrictive because it preserves block-wise Hamming weights and improves over standard QAOA in many four-node settings, but it cannot leave its fixed block sectors. The proposed method lies between these two extremes. It encodes selected constraints at initialization, preserves those selected structures through \(XY\) exchange terms, and uses weighted \(X\) rotations on qubits left outside the selected overlap structures to maintain exploration. The empirical results indicate that this intermediate design gives the best overall balance between feasibility guidance and search flexibility.
\par Overall, the results support the following conclusions. First, the proposed method improves over standard QAOA on the three-node toy instance and remains effective on the two four-node benchmark instances. Second, the block-wise XY-QAOA baseline confirms that feasibility-aware structure is important, while the proposed overlap-aware design shows a stronger balance between structure preservation and exploration under the reported tuned settings. Third, increasing \(p\) improves performance in many settings, but it also increases trainable parameters, circuit depth, two-qubit gate count, and optimizer cost. Fourth, the proposed method remains beneficial under finite-shot and noisy simulations, although its relative advantage is reduced by noise. Finally, the ablation study confirms that the constraint-aware initialization and the hybrid mixer must be used together, since Init-only and Mixer-only are both weaker than the full method and fall below standard QAOA in optimal-solution probability across all tested regimes.
\textcolor{black}{The explicit circuit realizations used for Cases~I--III are collected in \textcolor{blue}{\textbf{Appendix}}~\ref{app:circuit_realizations}. \textcolor{blue}{\textbf{Figures}}~\ref{fig:16}--\ref{fig:21} show the corresponding standard and proposed circuits.}
}

\newpage
\section{Conclusions} \label{section:6}

{\color{black}
\par This study investigates how QAOA can be made more effective for the vehicle routing problem by explicitly addressing the feasibility issue inherent in constrained combinatorial optimization. Under standard QAOA, the conventional uniform-superposition initialization distributes probability mass over the entire Hilbert space, while the standard Pauli-$X$ mixer explores this space without explicitly respecting problem constraints. For VRP instances, where feasible solutions typically occupy only a very small portion of the full binary space, this leads to an inefficient search process in which substantial probability mass can remain assigned to infeasible or high-cost states.
\par To mitigate this issue, we proposed a feasibility-aware QAOA framework built on two main ideas. First, we introduced a constraint-aware initialization that incorporates selected local VRP constraints directly into the initial state, thereby reducing the number of superposed basis states and increasing the initial probability mass assigned to structurally admissible configurations. Second, we proposed a hybrid \(XY\)--\(X\) mixer that preserves the selected one-hot or overlap structures through \(XY\)-exchange terms while allowing controlled exploration through \(X\)-rotations on qubits left outside the protected structures. The resulting design is intentionally partial rather than fully feasibility-preserving because it encodes an informative subset of constraints and leaves the remaining VRP constraints, including remaining customer-degree and depot departure/return constraints, to be handled through the QUBO penalty Hamiltonian and the subsequent variational search.
\par The numerical study evaluates this framework on one three-node toy instance and two four-node benchmark instances under ideal statevector, ideal finite-shot, and noisy finite-shot regimes. Relative to standard QAOA, the proposed method improves optimal-solution probability and reduces expected energy gap across the tested cases and regimes, indicating both a higher probability of sampling an optimal solution and a better overall concentration of the final sampling distribution around low-cost solutions. The block-wise XY-QAOA baseline further shows that the improvement is not simply a consequence of using an \(XY\)-type mixer. That baseline often improves over standard QAOA by preserving fixed Hamming-weight block structure, but the proposed overlap-aware initialization and hybrid mixer generally provide a stronger balance between constraint preservation and search flexibility under the reported tuned settings.
\par The ablation and depth-sensitivity studies clarify the source and limitations of the observed gains. The ablation results show that the constraint-aware initialization and the hybrid mixer are complementary. The initialization concentrates probability mass on a more relevant support, while the mixer explores that support without fully destroying the selected encoded constraints. The Init-only and Mixer-only variants both fall below standard QAOA in optimal-solution probability across all tested regimes, which further indicates that the two proposed components must be used together. The \(p\)-depth study shows that increasing the number of QAOA layers can improve optimal-solution probability and expected energy gap, but this comes with more trainable parameters, larger circuit depth, additional two-qubit gates, and higher optimization cost. Thus, the useful depth is constrained by the available optimizer budget, sampling noise, and hardware noise.
\par The noisy finite-shot experiments highlight an important practical limitation. Although the proposed ansatz remains beneficial under the optimistic hardware-inspired noise model used in this study, its relative advantage is attenuated compared with the ideal statevector regime. This is expected because the more structured mixer introduces additional two-qubit \(R_{XX}\) and \(R_{YY}\) operations, which are more sensitive to gate noise than single-qubit rotations. The prescribed initial superposition is injected directly in the simulations, so the reported noise model does not include the native-gate decomposition or noise of a hardware state-preparation circuit. Therefore, the algorithmic benefit of constraint-aware initialization and hybrid mixing must be considered together with state-preparation cost, circuit depth, two-qubit gate overhead, readout error, and hardware connectivity.
\par More broadly, the transition from promising small-scale demonstrations to practically useful transportation applications will require substantial progress beyond the present setting. As emphasized in recent discussions on quantum optimization for transportation systems, future applicability to larger and more realistic problems will depend not only on lower-noise gates and better readout, but also on the ability to control more qubits, manage gate complexity, and support effective error mitigation or correction (\citep{du2026overcoming,massimiliano2026quantum}). Thus, while the present work shows that feasibility-aware QAOA design can improve performance on several small VRP instances, solving larger and more realistic routing problems will ultimately require simultaneous advances in both quantum algorithms and quantum hardware.
\par Several directions for future research follow naturally from this work. First, the proposed framework should be extended to richer VRP variants, such as capacitated, time-window-constrained, or multi-depot settings, where the feasible subspace becomes substantially more complex. Second, a more systematic analysis of circuit complexity should quantify the trade-off between feasibility promotion and implementation overhead under different mixer constructions and should include native-gate synthesis of the constraint-aware initial state. Third, it will be important to test the proposed design under more realistic hardware constraints, including native coupling maps, hardware-specific gate decompositions, state-preparation noise, and non-depolarizing noise models. Finally, it would be worthwhile to investigate how the present feasibility-aware design can be combined with other QAOA enhancement strategies, such as warm-start methods or decomposition-based approaches, to further improve scalability and robustness.
\par In summary, this work suggests that for constrained routing problems such as the VRP, the treatment of feasibility is a central algorithmic issue rather than a secondary modeling detail. By incorporating constraint structure directly into both the initialization and the mixer, the proposed framework achieves a more targeted search process than standard QAOA on the tested small-scale instances. At the same time, the study also makes clear that such algorithmic improvements alone are not sufficient for large-scale practical deployment. Their ultimate value will depend on continued progress in quantum hardware, especially in terms of lower noise, better readout, and the ability to reliably operate on larger and deeper circuits.
}

%\newpage
\section{Appendix}

\textcolor{black}{Throughout this appendix, Case~I denotes the three-node VRP instance, Case~II denotes the balanced symmetric four-node VRP instance, and Case~III denotes the customer-cluster four-node VRP instance.}

\textcolor{black}{This appendix collects the detailed QUBO and Ising derivations, ansatz specifications, complete numerical tables, circuit realizations, and optimization procedures that support the main text.}

\subsection{\textcolor{black}{Detailed QUBO and Ising Derivation for the Three-Node VRP Instance}}
\label{app:qubo_ising_derivation} \label{sec:2} \label{line:2}

\textcolor{black}{This section provides the full QUBO construction and QUBO-to-Ising conversion for the three-node VRP instance used in \textcolor{blue}{\textbf{Section}}~\ref{subsec:problem_setup}. The derivation is included for reproducibility, while the main text reports only the final Hamiltonian used in the QAOA circuit.}

\textcolor{black}{The constrained three-node VRP formulation in \eqref{eq:34} is converted into an unconstrained binary objective by adding quadratic penalty terms. For binary variables, an exactly-one constraint $x+y=1$ can be penalized as}
\textcolor{black}{
\begin{equation}
P(x+y-1)^2
=
P(1-x-y+2xy)
\label{eq:57}
\end{equation}}
\textcolor{black}{because $x^2=x$ and $y^2=y$ for $x,y\in\{0,1\}$. Similarly, an exactly-two constraint $x+y=2$ can be penalized as}
\textcolor{black}{
\begin{equation}
P(x+y-2)^2
=
P(4-3x-3y+2xy)
\label{eq:58}
\end{equation}}
\textcolor{black}{and an at-least-one constraint $x+y\geq 1$ can be penalized as}
\textcolor{black}{\begin{equation}
P(1-x-y+xy)
=
P(1-x)(1-y)
\label{eq:59}
\end{equation}}

\textcolor{black}{The penalty coefficient is set as}
\textcolor{black}{
\begin{equation}
\begin{aligned}
P
&=
2\left(
|w_{0,1}|
+
|w_{0,2}|
+
|w_{1,0}|
+
|w_{1,2}|
+
|w_{2,0}|
+
|w_{2,1}|
\right)
\\
&=
435.6
\end{aligned}
\label{eq:60}
\end{equation}}

\textcolor{black}{As an example, the depot-return constraint $x_{1,0}+x_{2,0}=2$ yields}
\textcolor{black}{\begin{equation}
P(x_{1,0}+x_{2,0}-2)^2
=
P\left(
4-3x_{1,0}-3x_{2,0}+2x_{1,0}x_{2,0}
\right)
\label{eq:61}
\end{equation}}

\textcolor{black}{Substituting $P=435.6$ gives}
\textcolor{black}{\begin{equation}
435.6
\left(
4-3x_{1,0}-3x_{2,0}+2x_{1,0}x_{2,0}
\right)
\label{eq:62}
\end{equation}}

\textcolor{black}{Combining the travel-cost objective with all penalty terms gives}
\textcolor{black}{\begin{equation}
\label{eq:63}
\begin{aligned}
f_{\mathrm{QUBO}}(\mathbf{x})
&=
61.3x_{0,1}
+4.7x_{0,2}
+61.3x_{1,0}
+42.9x_{1,2}
+4.7x_{2,0}
+42.9x_{2,1}
\\
&\quad
+435.6
\left(
4-3x_{1,0}-3x_{2,0}
+2x_{1,0}x_{2,0}
\right)
\\
&\quad
+435.6
\left(
4-3x_{0,1}-3x_{0,2}
+2x_{0,1}x_{0,2}
\right)
\\
&\quad
+435.6
\left(
1-x_{1,0}-x_{1,2}
+2x_{1,0}x_{1,2}
\right)
\\
&\quad
+435.6
\left(
1-x_{0,1}-x_{2,1}
+2x_{0,1}x_{2,1}
\right)
\\
&\quad
+435.6
\left(
1-x_{2,0}-x_{2,1}
+2x_{2,0}x_{2,1}
\right)
\\
&\quad
+435.6
\left(
1-x_{0,2}-x_{1,2}
+2x_{0,2}x_{1,2}
\right)
\\
&\quad
+435.6
\left(
1-x_{1,0}-x_{2,0}
+x_{1,0}x_{2,0}
\right)
\end{aligned}
\end{equation}}

\textcolor{black}{After collecting like terms, the QUBO becomes}
\textcolor{black}{\begin{equation}
\label{eq:64}
\begin{aligned}
f_{\mathrm{QUBO}}(\mathbf{x})
&=
1306.8x_{1,0}x_{2,0}
+871.2x_{0,1}x_{0,2}
+871.2x_{1,0}x_{1,2}
\\
&\quad
+871.2x_{0,1}x_{2,1}
+871.2x_{2,0}x_{2,1}
+871.2x_{0,2}x_{1,2}
\\
&\quad
-2116.7x_{1,0}
-2173.3x_{2,0}
-1681.1x_{0,1}
\\
&\quad
-1737.7x_{0,2}
-828.3x_{1,2}
-828.3x_{2,1}
+5662.8
\end{aligned}
\end{equation}}

\textcolor{black}{We use the binary-to-Ising convention}
\textcolor{black}{\begin{equation}
x_i=\frac{I-Z_i}{2}
\label{eq:65}
\end{equation}
where $Z_i$ is the Pauli-$Z$ operator acting on qubit $q_i$. The qubit-to-variable correspondence is
\begin{equation}
\left[
q_0,q_1,q_2,q_3,q_4,q_5
\right]
\longleftrightarrow
\left[
x_{0,1}
\quad
x_{0,2}
\quad
x_{1,0}
\quad
x_{1,2}
\quad
x_{2,0}
\quad
x_{2,1}
\right]
\label{eq:66}
\end{equation}}

\textcolor{black}{Accordingly,}
\textcolor{black}{\begin{equation}
\label{eq:67}
\begin{aligned}
x_{0,1}
&\mapsto
\frac{I-Z_0}{2}
&
x_{0,2}
&\mapsto
\frac{I-Z_1}{2}
&
x_{1,0}
&\mapsto
\frac{I-Z_2}{2}
\\
x_{1,2}
&\mapsto
\frac{I-Z_3}{2}
&
x_{2,0}
&\mapsto
\frac{I-Z_4}{2}
&
x_{2,1}
&\mapsto
\frac{I-Z_5}{2}
\end{aligned}
\end{equation}}

\textcolor{black}{Substituting \eqref{eq:67} into \eqref{eq:64} gives}
\textcolor{black}{\begin{equation}
\label{eq:68}
\begin{aligned}
\widehat{C}
&=
1306.8
\left(\frac{I-Z_2}{2}\right)
\left(\frac{I-Z_4}{2}\right)
+871.2
\left(\frac{I-Z_0}{2}\right)
\left(\frac{I-Z_1}{2}\right)
\\
&\quad
+871.2
\left(\frac{I-Z_2}{2}\right)
\left(\frac{I-Z_3}{2}\right)
+871.2
\left(\frac{I-Z_0}{2}\right)
\left(\frac{I-Z_5}{2}\right)
\\
&\quad
+871.2
\left(\frac{I-Z_4}{2}\right)
\left(\frac{I-Z_5}{2}\right)
+871.2
\left(\frac{I-Z_1}{2}\right)
\left(\frac{I-Z_3}{2}\right)
\\
&\quad
-2116.7
\left(\frac{I-Z_2}{2}\right)
-2173.3
\left(\frac{I-Z_4}{2}\right)
-1681.1
\left(\frac{I-Z_0}{2}\right)
\\
&\quad
-1737.7
\left(\frac{I-Z_1}{2}\right)
-828.3
\left(\frac{I-Z_3}{2}\right)
-828.3
\left(\frac{I-Z_5}{2}\right)
+5662.8I
\end{aligned}
\end{equation}}

\textcolor{black}{Expanding and collecting like terms yields}
\textcolor{black}{\begin{equation}
\label{eq:69}
\begin{aligned}
\widehat{C}
&=
326.7Z_2Z_4
+217.8Z_0Z_1
+217.8Z_2Z_3
+217.8Z_0Z_5
+217.8Z_4Z_5
+217.8Z_1Z_3
\\
&\quad
+404.95Z_0
+433.25Z_1
+513.85Z_2
-21.45Z_3
+542.15Z_4
-21.45Z_5
+2395.8I
\end{aligned}
\end{equation}}

\textcolor{black}{The constant term $2395.8I$ shifts all computational-basis energies by the same amount and contributes only a global phase to the cost unitary. Therefore, it can be omitted without changing the minimizing bitstrings. The unscaled cost Hamiltonian is}
\textcolor{black}{\begin{equation}
\label{eq:70}
\begin{aligned}
H_C
&=
326.7Z_2Z_4
+217.8Z_0Z_1
+217.8Z_2Z_3
+217.8Z_0Z_5
+217.8Z_4Z_5
+217.8Z_1Z_3
\\
&\quad
+404.95Z_0
+433.25Z_1
+513.85Z_2
-21.45Z_3
+542.15Z_4
-21.45Z_5
\end{aligned}
\end{equation}}

\textcolor{black}{Finally, the circuit implementation uses the positively scaled Hamiltonian}
\textcolor{black}{\begin{equation}
\label{eq:71}
\widetilde{H}_C=\frac{H_C}{435.6}
\end{equation}}
\textcolor{black}{This scaling preserves the ordering of all computational-basis energies and only rescales the optimized QAOA cost angles.}

{\color{black}
\subsection{Detailed QUBO and Ising Derivation for the Four-Node Benchmark Instances}
\label{app:fournode_qubo_ising_derivation} \label{sec:4} \label{line:4}

This appendix provides the full QUBO construction and QUBO-to-Ising conversion for the two four-node benchmark instances used in \textcolor{blue}{\textbf{Section}}~\ref{subsubsec:four_node_benchmark}. The derivation follows the same penalty template as \textcolor{blue}{\textbf{Appendix}}~\ref{app:qubo_ising_derivation} for the three-node toy instance. The overlap-4 initialization and mixer used in the numerical experiments are specified separately in \textcolor{blue}{\textbf{Appendix}}~\ref{app:overlap4_init_mixer}.

\subsubsection{Quadratic penalty construction}
\label{app:fournode_penalty_construction}

\textcolor{black}{The four-node constrained VRP formulation is converted into an unconstrained binary objective by adding quadratic penalty terms for the eight degree-balance constraints listed in \eqref{eq:45a}--\eqref{eq:45h}.} We use the same penalty templates as in \textcolor{blue}{\textbf{Appendix}}~\ref{app:qubo_ising_derivation}:
\begin{subequations}
\label{eq:72}
\begin{align}
P(x+y-1)^2 &= P(1-x-y+2xy)
\label{eq:72a}
\\
P(x+y+z-1)^2 &= P(1-x-y-z+2xy+2xz+2yz)
\label{eq:72b}
\\
P(x+y+z-2)^2 &= P(4-3x-3y-3z+2xy+2xz+2yz)
\label{eq:72c}
\end{align}
\end{subequations}
where each identity follows from $x_i^2=x_i$ for binary variables.

\textcolor{black}{For a customer subset \(\mathcal{S}\subseteq\{1,2,3\}\) with \(|\mathcal{S}|\geq 2\), let \(A_{\mathcal{S}}=\sum_{i\in\mathcal{S}}\sum_{j\notin\mathcal{S}}x_{i,j}\) denote the number of selected arcs leaving \(\mathcal{S}\). The inequality \(A_{\mathcal{S}}\geq 1\) can be converted to an equality by introducing a nonnegative integer slack variable \(u_{\mathcal{S}}=A_{\mathcal{S}}-1\). Its corresponding penalty is}
\begin{equation}
P\left(
\sum_{i\in\mathcal{S}}\sum_{j\notin\mathcal{S}}x_{i,j}
-1-u_{\mathcal{S}}
\right)^2
\label{eq:73}
\end{equation}
\textcolor{black}{where \(u_{\mathcal{S}}\in\{0,\ldots,M_{\mathcal{S}}-1\}\), with \(M_{\mathcal{S}}=|\mathcal{S}|(4-|\mathcal{S}|)\), and an explicit QUBO implementation would encode \(u_{\mathcal{S}}\) using auxiliary binary variables. This construction assigns zero penalty to every bitstring satisfying \(A_{\mathcal{S}}\geq 1\), rather than only to the special case \(A_{\mathcal{S}}=1\).}\footnote{ \textcolor{black}{In the two four-node benchmark instances studied numerically, the QUBO includes only the travel-cost objective and the eight degree-balance penalties in \eqref{eq:45a}--\eqref{eq:45h}. Explicit subtour-elimination penalties and their auxiliary slack bits are omitted because exhaustive enumeration shows that exactly six bitstrings satisfy the degree-balance constraints, and each corresponds to a valid two-route solution with no disconnected subtour. Adding \eqref{eq:73} would therefore be redundant on the feasible set of interest, while increasing both the qubit requirement and the number of penalty couplings. The subtour inequalities are nevertheless stated in the main text to keep the VRP formulation complete.}
}

The penalty coefficient is set as
\begin{equation}
P=2\sum_{i\neq j}|w_{i,j}|
\label{eq:74}
\end{equation}
which gives $P^{(1)}=586.4$ for instance~1 and $P^{(2)}=667.6$ for instance~2.

\subsubsection{Full QUBO objective for instance~1}
\label{app:fournode_case1_qubo}

Combining the travel-cost objective with all degree-balance penalty terms and collecting like terms yields
\begin{equation}
\label{eq:75}
\begin{aligned}
f_{\mathrm{QUBO}}^{(1)}(\mathbf{x})
&=
1172.8\,x_{0,1}x_{0,2}
+1172.8\,x_{0,1}x_{0,3}
+1172.8\,x_{0,1}x_{2,1}
+1172.8\,x_{0,1}x_{3,1}
\\
&\quad
+1172.8\,x_{0,2}x_{0,3}
+1172.8\,x_{0,2}x_{1,2}
+1172.8\,x_{0,2}x_{3,2}
+1172.8\,x_{1,0}x_{1,2}
\\
&\quad
+1172.8\,x_{1,0}x_{1,3}
+1172.8\,x_{1,0}x_{2,0}
+1172.8\,x_{1,0}x_{3,0}
+1172.8\,x_{1,2}x_{1,3}
\\
&\quad
+1172.8\,x_{1,2}x_{3,2}
+1172.8\,x_{1,3}x_{2,3}
+1172.8\,x_{2,0}x_{2,1}
+1172.8\,x_{2,0}x_{2,3}
\\
&\quad
+1172.8\,x_{2,0}x_{3,0}
+1172.8\,x_{2,1}x_{2,3}
+1172.8\,x_{2,1}x_{3,1}
+1172.8\,x_{2,3}x_{3,2}
\\
&\quad
+1172.8\,x_{3,0}x_{3,1}
+1172.8\,x_{3,0}x_{3,2}
+1172.8\,x_{3,1}x_{3,2}
\\
&\quad
-2323.9\,x_{0,1}
-2311.4\,x_{0,2}
-2317.0\,x_{0,3}
-2323.9\,x_{1,0}
\\
&\quad
-1155.4\,x_{1,2}
-1147.9\,x_{1,3}
-2311.4\,x_{2,0}
-1155.4\,x_{2,1}
\\
&\quad
-1153.0\,x_{2,3}
-2317.0\,x_{3,0}
-1147.9\,x_{3,1}
-1153.0\,x_{3,2}
+8209.6
\end{aligned}
\end{equation}

\subsubsection{Full QUBO objective for instance~2}
\label{app:fournode_case2_qubo}

The same penalty template applied to the instance~2 distance matrix gives
\begin{equation}
\label{eq:76}
\begin{aligned}
f_{\mathrm{QUBO}}^{(2)}(\mathbf{x})
&=
1335.2\,x_{0,1}x_{0,2}
+1335.2\,x_{0,1}x_{0,3}
+1335.2\,x_{0,1}x_{2,1}
+1335.2\,x_{0,1}x_{3,1}
\\
&\quad
+1335.2\,x_{0,2}x_{0,3}
+1335.2\,x_{0,2}x_{1,2}
+1335.2\,x_{0,2}x_{3,2}
+1335.2\,x_{1,0}x_{1,2}
\\
&\quad
+1335.2\,x_{1,0}x_{1,3}
+1335.2\,x_{1,0}x_{2,0}
+1335.2\,x_{1,0}x_{3,0}
+1335.2\,x_{1,2}x_{1,3}
\\
&\quad
+1335.2\,x_{1,2}x_{3,2}
+1335.2\,x_{1,3}x_{2,3}
+1335.2\,x_{2,0}x_{2,1}
+1335.2\,x_{2,0}x_{2,3}
\\
&\quad
+1335.2\,x_{2,0}x_{3,0}
+1335.2\,x_{2,1}x_{2,3}
+1335.2\,x_{2,1}x_{3,1}
+1335.2\,x_{2,3}x_{3,2}
\\
&\quad
+1335.2\,x_{3,0}x_{3,1}
+1335.2\,x_{3,0}x_{3,2}
+1335.2\,x_{3,1}x_{3,2}
\\
&\quad
-2628.1\,x_{0,1}
-2643.6\,x_{0,2}
-2631.9\,x_{0,3}
-2628.1\,x_{1,0}
\\
&\quad
-1321.5\,x_{1,2}
-1305.8\,x_{1,3}
-2643.6\,x_{2,0}
-1321.5\,x_{2,1}
\\
&\quad
-1319.0\,x_{2,3}
-2631.9\,x_{3,0}
-1305.8\,x_{3,1}
-1319.0\,x_{3,2}
+9346.4
\end{aligned}
\end{equation}

\subsubsection{QUBO-to-Ising conversion}
\label{app:fournode_qubo_to_ising}

\textcolor{black}{We use the binary-to-Ising convention in \eqref{eq:41} and the qubit-to-variable mapping defined in the main text. Substituting \eqref{eq:75} and \eqref{eq:76} into this convention and collecting like terms gives the complete diagonal cost operators and their constant-free parts as follows:}
\begin{equation}
\label{eq:77}
\begin{aligned}
\widehat{C}^{(k)}
&=C_{\mathrm{I}}^{(k)}I+H_C^{(k)}
\qquad k\in\{1,2\}
\\
H_C^{(k)}
&=
\frac{P^{(k)}}{2}
\sum_{(i,j)\in\mathcal{E}_{ZZ}}Z_iZ_j
+
\sum_{m=0}^{11}h_m^{(k)}Z_m
\\
C_{\mathrm{I}}^{(1)}&=4837.8
\qquad
C_{\mathrm{I}}^{(2)}=5507.7
\\
\mathbf{h}^{(1)}
&=
\bigl(
-10.85\ {-17.10}\ {-14.30}\ {-10.85}\ {-595.10}\ {-598.85}
\\
&\qquad
{-17.10}\ {-595.10}\ {-596.30}\ {-14.30}\ {-598.85}\ {-596.30}
\bigr)^{\mathsf T}
\\
\mathbf{h}^{(2)}
&=
\bigl(
-21.15\ {-13.40}\ {-19.25}\ {-21.15}\ {-674.45}\ {-682.30}
\\
&\qquad
{-13.40}\ {-674.45}\ {-675.70}\ {-19.25}\ {-682.30}\ {-675.70}
\bigr)^{\mathsf T}
\\
\mathbf{h}^{(k)}
&=
\bigl(h_0^{(k)}\ h_1^{(k)}\ \cdots\ h_{11}^{(k)}\bigr)^{\mathsf T}
\end{aligned}
\end{equation}
\textcolor{black}{Here \(\widehat{C}^{(k)}\) reproduces the full QUBO value on every computational-basis state, whereas \(H_C^{(k)}\) omits only the global constant \(C_{\mathrm{I}}^{(k)}I\). Every quadratic QUBO coefficient \(2P^{(k)}\) contributes a \(ZZ\) coefficient \(P^{(k)}/2\). The shared interaction-edge set is}
\begin{equation}
\label{eq:78}
\begin{aligned}
\mathcal{E}_{ZZ}
=\{
&(0,1)\quad (0,2)\quad (0,7)\quad (0,10)\\
&(1,2)\quad (1,4)\quad (1,11)\\
&(2,5)\quad (2,8)\\
&(3,4)\quad (3,5)\quad (3,6)\quad (3,9)\\
&(4,5)\quad (4,11)\\
&(5,8)\\
&(6,7)\quad (6,8)\quad (6,9)\\
&(7,8)\quad (7,10)\\
&(9,10)\quad (9,11)\\
&(10,11)
\}
\end{aligned}
\end{equation}
\textcolor{black}{with each pair \((q_i,q_j)\) corresponding to one of the QUBO monomials listed in \eqref{eq:75} and \eqref{eq:76}. Because a global identity term contributes only an unobservable phase, the circuit uses the positively scaled constant-free Hamiltonians}
\begin{equation}
\label{eq:79}
\widetilde{H}_C^{(k)}=\frac{H_C^{(k)}}{P^{(k)}}
\qquad k\in\{1,2\}
\end{equation}
\textcolor{black}{Since \(P^{(k)}>0\), this scaling preserves the ordering and minimizers of all computational-basis energies. It changes only the energy scale and the corresponding parameterization of the QAOA cost angles. After scaling, every \(ZZ\) coefficient equals \(0.5\), and each single-qubit coefficient is \(h_m^{(k)}/P^{(k)}\).}

\subsection{\textcolor{black}{Initialization and Mixer for the Four-Node Benchmarks}}
\label{app:overlap4_init_mixer} \label{sec:5} \label{line:5}

\textcolor{black}{This appendix gives a detailed explanation of the constraint-aware initialization and the hybrid XY--X mixer used in the four-node numerical experiments. The same ansatz is used for the two four-node benchmark instances; only the travel-cost coefficients, and therefore the cost Hamiltonian, are different.}

\subsubsection{\textcolor{black}{Initialization support}}

\textcolor{black}{This subsubsection specifies the selected overlap constraints, their joint support, and the resulting initial-state probability assigned to globally optimal solutions.}

\textcolor{black}{Recall that the twelve qubits are ordered as}
\begin{equation}
\label{eq:80}
\textcolor{black}{
(q_0,\ldots,q_{11})
\longleftrightarrow
\bigl(
x_{0,1}\quad x_{0,2}\quad x_{0,3}
\quad
x_{1,0}\quad x_{1,2}\quad x_{1,3}
\quad
x_{2,0}\quad x_{2,1}\quad x_{2,3}
\quad
x_{3,0}\quad x_{3,1}\quad x_{3,2}
\bigr)
}
\end{equation}

\textcolor{black}{The proposed four-node initialization does not encode all VRP constraints. Instead, it encodes the following four selected overlapping degree constraints:}
\begin{subequations}
\label{eq:81}
\begin{align}
\textcolor{black}{
I_1:\quad q_0+q_7+q_{10}
      = x_{0,1}+x_{2,1}+x_{3,1}=1
}
\label{eq:81a}
\\
\textcolor{black}{
I_2:\quad q_1+q_4+q_{11}
      = x_{0,2}+x_{1,2}+x_{3,2}=1
}
\label{eq:81b}
\\
\textcolor{black}{
O_1:\quad q_3+q_4+q_5
      = x_{1,0}+x_{1,2}+x_{1,3}=1
}
\label{eq:81c}
\\
\textcolor{black}{
O_2:\quad q_6+q_7+q_8
      = x_{2,0}+x_{2,1}+x_{2,3}=1
}
\label{eq:81d}
\end{align}
\end{subequations}

\textcolor{black}{Here \(I_1\) and \(I_2\) are incoming-degree constraints for customers 1 and 2, while \(O_1\) and \(O_2\) are outgoing-degree constraints for customers 1 and 2. The variables \(q_4=x_{1,2}\) and \(q_7=x_{2,1}\) are called overlap qubits because each of them appears in two of the selected constraints: \(q_4\) appears in both \(I_2\) and \(O_1\), and \(q_7\) appears in both \(I_1\) and \(O_2\).}

\textcolor{black}{Let \(\mathcal{C}_{\mathrm{init}}\) denote the set of computational-basis states satisfying these four constraints:}
\begin{equation}
\label{eq:82}
\textcolor{black}{
\mathcal{C}_{\mathrm{init}}
=
\left\{
\mathbf{x}\in\{0,1\}^{12}:
I_1(\mathbf{x})=I_2(\mathbf{x})=O_1(\mathbf{x})=O_2(\mathbf{x})=1
\right\}
}
\end{equation}

\textcolor{black}{Exhaustive enumeration gives}
\begin{equation}
\label{eq:83}
\textcolor{black}{
|\mathcal{C}_{\mathrm{init}}|=100
}
\end{equation}

\textcolor{black}{The proposed initial state is the uniform superposition over these 100 basis states:}
\begin{equation}
\label{eq:84}
\textcolor{black}{
|\psi(0)\rangle
=
\frac{1}{\sqrt{100}}
\sum_{\mathbf{x}\in\mathcal{C}_{\mathrm{init}}}
|\mathbf{x}\rangle
}
\end{equation}

\begin{equation}
\label{eq:85}
\begin{aligned}
|\psi(0)\rangle
&=
\frac{1}{\sqrt{100}}
\Bigg\{
\Big[
\left(|10\rangle+|01\rangle\right)_{(q_1,q_{11})}
\bigotimes
\left(|10\rangle+|01\rangle\right)_{(q_3,q_5)}
\bigotimes
|0\rangle_{q_4}
\Big]
\\
&\quad\bigotimes
\Big[
\left(|10\rangle+|01\rangle\right)_{(q_0,q_{10})}
\bigotimes
\left(|10\rangle+|01\rangle\right)_{(q_6,q_8)}
\bigotimes
|0\rangle_{q_7}
\Big]
\bigotimes
\left(|00\rangle+|01\rangle+|10\rangle+|11\rangle\right)_{(q_2,q_9)}
\\
&\quad+
\Big[
\left(|10\rangle+|01\rangle\right)_{(q_1,q_{11})}
\bigotimes
\left(|10\rangle+|01\rangle\right)_{(q_3,q_5)}
\bigotimes
|0\rangle_{q_4}
\Big]
\\
&\quad\bigotimes
\Big[
|00\rangle_{(q_0,q_{10})}
\bigotimes
|00\rangle_{(q_6,q_8)}
\bigotimes
|1\rangle_{q_7}
\Big]
\bigotimes
\left(|00\rangle+|01\rangle+|10\rangle+|11\rangle\right)_{(q_2,q_9)}
\\
&\quad+
\Big[
|00\rangle_{(q_1,q_{11})}
\bigotimes
|00\rangle_{(q_3,q_5)}
\bigotimes
|1\rangle_{q_4}
\Big]
\\
&\quad\bigotimes
\Big[
\left(|10\rangle+|01\rangle\right)_{(q_0,q_{10})}
\bigotimes
\left(|10\rangle+|01\rangle\right)_{(q_6,q_8)}
\bigotimes
|0\rangle_{q_7}
\Big]
\bigotimes
\left(|00\rangle+|01\rangle+|10\rangle+|11\rangle\right)_{(q_2,q_9)}
\\
&\quad+
\Big[
|00\rangle_{(q_1,q_{11})}
\bigotimes
|00\rangle_{(q_3,q_5)}
\bigotimes
|1\rangle_{q_4}
\Big]
\bigotimes
\Big[
|00\rangle_{(q_0,q_{10})}
\bigotimes
|00\rangle_{(q_6,q_8)}
\bigotimes
|1\rangle_{q_7}
\Big]
\\
&\quad\bigotimes
\left(|00\rangle+|01\rangle+|10\rangle+|11\rangle\right)_{(q_2,q_9)}
\Bigg\}
\end{aligned}
\end{equation}

\textcolor{black}{In \eqref{eq:85}, every subscript identifies the named qubit register on which that ket factor is supported, and the resulting state is embedded in the global qubit order \((q_0,\ldots,q_{11})\) given in \eqref{eq:80}. Because the four selected constraints depend only on the structure of the VRP and not on the numerical travel distances, the same support \(\mathcal{C}_{\mathrm{init}}\) is used for both four-node benchmark instances.}

\textcolor{black}{It is important to note that the four selected constraints do not force the overlap qubits \(q_4\) and \(q_7\) to take any particular fixed value. In particular, they do not imply \((q_4,q_7)=(0,0)\). Rather, all four sectors}
\begin{equation}
\label{eq:86}
\textcolor{black}{
(q_4,q_7)\in\{(0,0)\quad (0,1)\quad (1,0)\quad (1,1)\}
}
\end{equation}
\textcolor{black}{appear in the initialization support, but with different multiplicities. The word ``frozen'' refers only to the mixer: the mixer does not act on \(q_4\) and \(q_7\). It does not mean that these qubits are fixed to zero in the initial state.}

\textcolor{black}{The following sector counts state how many computational-basis states in \(\mathcal{C}_{\mathrm{init}}\) have each fixed value of the two overlap qubits:}
\begin{equation}
\label{eq:87}
\textcolor{black}{
\begin{aligned}
\#\{\mathbf{x}\in\mathcal{C}_{\mathrm{init}}:(q_4,q_7)=(0,0)\}&=64\\
\#\{\mathbf{x}\in\mathcal{C}_{\mathrm{init}}:(q_4,q_7)=(0,1)\}&=16\\
\#\{\mathbf{x}\in\mathcal{C}_{\mathrm{init}}:(q_4,q_7)=(1,0)\}&=16\\
\#\{\mathbf{x}\in\mathcal{C}_{\mathrm{init}}:(q_4,q_7)=(1,1)\}&=4
\end{aligned}
}
\end{equation}

\textcolor{black}{These numbers can be understood directly. Consider first \(q_4\), which appears in \(I_2\) and \(O_1\). If \(q_4=0\), then the two constraints reduce to}
\begin{equation}
\label{eq:88}
\textcolor{black}{
q_1+q_{11}=1
\qquad
q_3+q_5=1
}
\end{equation}
\textcolor{black}{which gives two choices for \((q_1,q_{11})\) and two choices for \((q_3,q_5)\), hence \(2\times 2=4\) choices. If \(q_4=1\), then both constraints force}
\begin{equation}
\label{eq:89}
\textcolor{black}{
q_1=q_{11}=q_3=q_5=0
}
\end{equation}
\textcolor{black}{which gives only one choice. Thus setting an overlap qubit to zero leaves four local choices, whereas setting it to one leaves one local choice. The same argument applies to \(q_7\), which appears in \(I_1\) and \(O_2\). Finally, the two qubits \(q_2=x_{0,3}\) and \(q_9=x_{3,0}\) are not included in the four selected constraints, so they remain free and contribute a factor \(2^2=4\) to every sector. Therefore,}
\begin{subequations}
\label{eq:90}
\begin{align}
\textcolor{black}{
(q_4,q_7)=(0,0):\quad 4\times 4\times 4 = 64
}
\label{eq:90a}
\\
\textcolor{black}{
(q_4,q_7)=(0,1):\quad 4\times 1\times 4 = 16
}
\label{eq:90b}
\\
\textcolor{black}{
(q_4,q_7)=(1,0):\quad 1\times 4\times 4 = 16
}
\label{eq:90c}
\\
\textcolor{black}{
(q_4,q_7)=(1,1):\quad 1\times 1\times 4 = 4
}
\label{eq:90d}
\end{align}
\end{subequations}

\textcolor{black}{These four sector counts sum to}
\begin{equation}
\label{eq:91}
\textcolor{black}{
64+16+16+4=100
}
\end{equation}
\textcolor{black}{which agrees with the exhaustive enumeration of \(\mathcal{C}_{\mathrm{init}}\).}

\textcolor{black}{Within this 100-state support, exactly six basis states satisfy all eight degree-balance constraints and encode valid two-route solutions. Moreover, exactly two of these feasible states are globally optimal for each benchmark instance, as verified by exhaustive enumeration. Hence the initial feasible-solution probability is}
\begin{equation}
\label{eq:92}
\textcolor{black}{
p_{\mathrm{feas}}(0)=\frac{6}{100}=0.06
}
\end{equation}
\textcolor{black}{and the initial optimal-solution probability defined in \eqref{eq:54} is}
\begin{equation}
\label{eq:93}
\textcolor{black}{
p^*(0)=\frac{2}{100}=0.02
}
\end{equation}

\subsubsection{\textcolor{black}{Hybrid mixer and preserved structures}}

\textcolor{black}{The hybrid mixer used with the overlap-4 initialization is}
\begin{equation}
\label{eq:94}
\textcolor{black}{
H_M
=
(X_0X_{10}+Y_0Y_{10})
+
(X_1X_{11}+Y_1Y_{11})
+
(X_3X_5+Y_3Y_5)
+
(X_6X_8+Y_6Y_8)
+
\lambda (X_2+X_9)
}
\end{equation}

\textcolor{black}{The first four terms are \(XY\)-exchange terms. They exchange probability amplitude between two qubits inside one selected one-hot constraint. Therefore, they preserve the corresponding selected equality constraint. The last term is a weighted transverse-field component acting on \(q_2\) and \(q_9\), which are not part of the four selected constraints.}

{\color{black}
\begin{table}[htbp]
\centering
\caption{\textcolor{black}{Overlap-4 mixer mapping for the four-node benchmarks.}}
\label{table:4}
\begin{tabular}{clcl}
\thicktoprule
Circuit term & Qubits & Arc variables & Preserved selected constraint \\
\midrule
$X_0X_{10}+Y_0Y_{10}$ & $(q_0,q_{10})$ & $(x_{0,1},x_{3,1})$ & $I_1$ \\
$X_1X_{11}+Y_1Y_{11}$ & $(q_1,q_{11})$ & $(x_{0,2},x_{3,2})$ & $I_2$ \\
$X_3X_5+Y_3Y_5$       & $(q_3,q_5)$   & $(x_{1,0},x_{1,3})$ & $O_1$ \\
$X_6X_8+Y_6Y_8$       & $(q_6,q_8)$   & $(x_{2,0},x_{2,3})$ & $O_2$ \\
$\lambda(X_2+X_9)$    & $(q_2,q_9)$   & $(x_{0,3},x_{3,0})$ & none selected \\
\midrule
Frozen in mixer       & $(q_4,q_7)$   & $(x_{1,2},x_{2,1})$ & overlap-sector qubits \\
\thickbottomrule
\end{tabular}
\end{table}
}

\textcolor{black}{For example, the constraint \(I_1\) is}
\begin{equation}
\label{eq:95}
\textcolor{black}{
q_0+q_7+q_{10}=1
}
\end{equation}
\textcolor{black}{The mixer applies an \(XY\)-exchange between \(q_0\) and \(q_{10}\), while leaving the overlap qubit \(q_7\) unchanged. If the two exchanged bits differ, the exchange simply swaps \(10\leftrightarrow 01\). Therefore, \(q_0+q_7+q_{10}\) remains equal to one. If the two bits are equal, the exchange does not change the Hamming weight of the pair. Hence the selected constraint \(I_1\) is preserved. The same reasoning applies to the other three \(XY\) pairs and their corresponding selected constraints \(I_2\), \(O_1\), and \(O_2\).}

\textcolor{black}{The qubits \(q_4\) and \(q_7\) are not mixed because each of them participates in two selected constraints. Acting directly on these overlap qubits could change two constraints at once. By excluding them from both the \(XY\)-exchange pairs and the single-qubit \(X\) terms, the mixer acts separately within each \((q_4,q_7)\) sector described in \eqref{eq:87}.}

\textcolor{black}{The single-qubit terms on \(q_2\) and \(q_9\) are included because these two qubits are not constrained by \(I_1,I_2,O_1,O_2\). They allow the ansatz to explore additional depot-related degrees of freedom:}
\begin{equation}
\label{eq:96}
\textcolor{black}{
q_2=x_{0,3}
\qquad
q_9=x_{3,0}
}
\end{equation}

\textcolor{black}{Thus, the hybrid XY--X mixer preserves the four selected overlap constraints used to define the initialization support. It should not be interpreted as preserving the full VRP feasible subspace. The remaining VRP constraints are encoded in the QUBO penalty Hamiltonian and are evaluated through the final sampling metrics.}

\textcolor{black}{In the circuit implementation, one QAOA layer applies}
\begin{equation}
\label{eq:97}
\textcolor{black}{
e^{-i\beta_\ell H_M}\,
e^{-i\gamma_\ell \widetilde{H}_C}
}
\end{equation}
\textcolor{black}{where \(\widetilde{H}_C\) is the scaled cost Hamiltonian and \(H_M\) is the hybrid mixer above. Because operators act on a state from right to left, the cost unitary is applied first and the mixer unitary second in each layer, matching the circuit implementation. Each \(XY\) exchange is implemented by}
\begin{equation}
\label{eq:98}
\textcolor{black}{
e^{-i\beta_\ell(X_iX_j+Y_iY_j)}
=
R_{XX}^{(i,j)}(2\beta_\ell)
R_{YY}^{(i,j)}(2\beta_\ell)
}
\end{equation}
\textcolor{black}{where the equality follows because \(X_iX_j\) and \(Y_iY_j\) commute. The weighted single-qubit terms are implemented by}
\begin{equation}
\label{eq:99}
\textcolor{black}{
e^{-i\beta_\ell\lambda X_k}
=
R_X^{(k)}(2\beta_\ell\lambda)
\qquad
k\in\{2,9\}
}
\end{equation}
\textcolor{black}{These identities use the Qiskit rotation convention \(R_P(\theta)=e^{-i\theta P/2}\).}

\textcolor{black}{The initial state in \eqref{eq:84} is injected directly as the prescribed logical superposition. This simulator-level operation isolates the variational dynamics associated with the intended support but is not a native-gate state-preparation circuit. In the noisy simulations, state-preparation gates and state-preparation noise are not modeled; gate and readout noise are applied only to the subsequent QAOA evolution and measurement. A hardware implementation would require a separate decomposition of the prescribed superposition into native gates, and the resulting depth and two-qubit gate cost are outside the scope of the present experiments.}
}

\subsection{\textcolor{black}{Full Baseline Comparison Tables}}

\textcolor{black}{This subsection reports the complete summary statistics for the standard, block-wise XY-QAOA, and proposed methods across the three experimental regimes. These tables complement the selected comparisons discussed in the Results Analysis section.}

\begin{table}[H]
\centering
\caption{\textcolor{black}{Performance comparison (Case I, Regime I)}}
\label{table:5}
\resizebox{\textwidth}{!}{
\begin{tabular}{cccc}
\thicktoprule
 Model & \textcolor{black}{Mean optimal-solution probability} & Sampling rank & Expected energy gap   \\
\midrule
Standard QAOA (700)& 0.4994, 0.0924, [0.4649, 0.5339]   & 1.00, 0.00, - & 638.75, 118.26, [594.60, 682.91]  \\
Ours $\lambda = 0.5$ (1200) & 0.5630, 0.1130, [0.5208, 0.6052] & 1.00, 0.00, [1.00, 1.00] & 603.63, 163.30, [542.66, 664.60] \\
Ours $\lambda = 0.6$ (1500) & 0.5873, 0.0645, [0.5633, 0.6114] & 1.00, 0.00, [1.00, 1.00] & 592.52, 94.37, [557.28, 627.75] \\
Ours $\lambda = 0.7$ (1900) & 0.6525, 0.1296, [0.6041, 0.7009] & 1.00, 0.00, [1.00, 1.00] & 513.44, 172.42, [449.07, 577.81] \\
Ours $\lambda = 0.8$ (900) & 0.6173, 0.1688, [0.5543, 0.6804] & 1.00, 0.00, [1.00, 1.00] & 523.99, 245.22, [432.43, 615.54] \\
Ours $\lambda = 0.9$ (1700) & 0.5226, 0.1131, [0.4804, 0.5648] & 1.00, 0.00, [1.00, 1.00] & 640.48, 133.86, [590.50, 690.46] \\
Ours $\lambda = 1.0$ (1500) & 0.5209, 0.0982, [0.4842, 0.5575] & 1.07, 0.37, [0.93, 1.20] & 668.12, 75.16, [640.06, 696.18] \\
Ours $\lambda = 1.1$ (1600) & 0.6248, 0.1203, [0.5799, 0.6698] & 1.07, 0.37, [0.93, 1.20] & 547.91, 149.14, [492.23, 603.60] \\
Ours $\lambda = 1.2$ (1300) & 0.6733, 0.1165, [0.6299, 0.7168] & 1.00, 0.00, [1.00, 1.00] & 439.97, 165.05, [378.34, 501.59] \\
Ours $\lambda = 1.3$ (1300) & \textbf{0.7201, 0.1319, [0.6709, 0.7694]} & \textbf{1.00, 0.00, [1.00, 1.00]} & \textbf{375.53, 190.45, [304.42, 446.63]} \\
Ours $\lambda = 1.4$ (900) & 0.7096, 0.1277, [0.6619, 0.7573] & 1.00, 0.00, [1.00, 1.00] & 411.57, 185.66, [342.26, 480.89] \\
Ours $\lambda = 1.5$ (2000) & 0.6541, 0.2038, [0.5781, 0.7302] & 1.13, 0.51, [0.94, 1.32] & 434.98, 229.01, [349.48, 520.49] \\
Ours $\lambda = 1.6$ (400) & 0.6352, 0.1947, [0.5625, 0.7080] & 1.03, 0.18, [0.97, 1.10] & 464.19, 231.37, [377.81, 550.58] \\
Ours $\lambda = 1.7$ (1000) & 0.6310, 0.1526, [0.5740, 0.6880] & 1.03, 0.18, [0.97, 1.10] & 463.96, 181.44, [396.22, 531.71] \\
Ours $\lambda = 1.8$ (1600) & 0.5589, 0.1214, [0.5136, 0.6042] & 1.00, 0.00, [1.00, 1.00] & 546.83, 116.99, [503.15, 590.51] \\
Ours $\lambda = 1.9$ (1800) & 0.5332, 0.0847, [0.5016, 0.5649] & 1.00, 0.00, [1.00, 1.00] & 559.59, 94.21, [524.42, 594.76] \\
Ours $\lambda = 2.0$ (1700) & 0.4892, 0.0494, [0.4708, 0.5077] & 1.00, 0.00, [1.00, 1.00] & 592.57, 52.40, [573.01, 612.13] \\
\thickbottomrule
\end{tabular}
}
\begin{tablenotes}
\footnotesize
\item \textcolor{black}{Note: In \textcolor{blue}{\textbf{Tables}}~\ref{table:5}--\ref{table:13}, the three statistics reported in each column except the first are the mean, standard deviation, and 95\% confidence interval, respectively. In the first column, the number in parentheses is the maximum COBYLA iteration budget used for that configuration.}
\end{tablenotes}
\end{table}

\begin{table}[H]
\centering
\caption{\textcolor{black}{Performance comparison (Case I, Regime II)}}
\label{table:6}
\resizebox{\textwidth}{!}{
\begin{tabular}{cccc}
\thicktoprule
 Model & \textcolor{black}{Mean optimal-solution probability} & Sampling rank & Expected energy gap   \\
\midrule
Standard QAOA (200)& 0.4341, 0.1311, [0.3852, 0.4831]   & 1.00, 0.00, - & 731.71, 169.26, [668.52, 794.91]  \\
Ours $\lambda = 0.5$ (200) & 0.5229, 0.1483, [0.4675, 0.5783] & 1.07, 0.37, [0.93, 1.20] & 650.96, 183.51, [582.45, 719.47] \\
Ours $\lambda = 0.6$ (200) & 0.5484, 0.0838, [0.5171, 0.5797] & 1.00, 0.00, [1.00, 1.00] & 652.18, 133.80, [602.23, 702.14] \\
Ours $\lambda = 0.7$ (200) & 0.5483, 0.1694, [0.4851, 0.6116] & 1.13, 0.51, [0.94, 1.32] & 622.98, 186.65, [553.29, 692.67] \\
Ours $\lambda = 0.8$ (200) & 0.5304, 0.1655, [0.4686, 0.5922] & 1.00, 0.00, [1.00, 1.00] & 638.07, 245.15, [546.54, 729.60] \\
Ours $\lambda = 0.9$ (200) & 0.4730, 0.1617, [0.4126, 0.5334] & 1.50, 1.94, [0.77, 2.23] & 687.52, 145.51, [633.19, 741.85] \\
Ours $\lambda = 1.0$ (200) & 0.4963, 0.1233, [0.4503, 0.5424] & 1.20, 0.61, [0.97, 1.43] & 689.78, 82.10, [659.12, 720.43] \\
\end{tabular}
}
\end{table}

\begin{table}[H]
\centering
\caption{\textcolor{black}{Performance comparison (Case I, Regime III)}}
\label{table:7}
\resizebox{\textwidth}{!}{
\begin{tabular}{cccc}
\thicktoprule
 Model & \textcolor{black}{Mean optimal-solution probability} & Sampling rank & Expected energy gap   \\
\midrule
Standard QAOA (200) & 0.4170, 0.1246, [0.3705, 0.4635] & 1.00, 0.00, - & 767.92, 172.48, [703.53, 832.32] \\
Ours $\lambda = 0.5$ (200) & 0.5047, 0.1430, [0.4514, 0.5581] & 1.07, 0.37, [0.93, 1.20] & 678.95, 181.39, [611.23, 746.68] \\
Ours $\lambda = 0.6$ (200) & 0.5089, 0.1106, [0.4676, 0.5502] & 1.07, 0.37, [0.93, 1.20] & 678.98, 144.76, [624.93, 733.03] \\
Ours $\lambda = 0.7$ (200) & 0.5391, 0.1781, [0.4726, 0.6056] & 1.23, 0.90, [0.90, 1.57] & 643.79, 184.50, [574.91, 712.68] \\
Ours $\lambda = 0.8$ (200) & 0.5198, 0.1574, [0.4610, 0.5785] & 1.00, 0.00, [1.00, 1.00] & 662.47, 241.63, [572.26, 752.69] \\
Ours $\lambda = 0.9$ (200) & 0.4828, 0.1070, [0.4429, 0.5227] & 1.07, 0.37, [0.93, 1.20] & 713.16, 127.55, [665.54, 760.79] \\
Ours $\lambda = 1.0$ (200) & 0.4536, 0.1259, [0.4066, 0.5006] & 1.20, 0.61, [0.97, 1.43] & 729.71, 92.86, [695.04, 764.38] \\
\end{tabular}
}
\end{table}

\begin{table}[H]
\centering
\caption{\textcolor{black}{Performance comparison (Case II, Regime I)}}
\label{table:8}
\resizebox{\textwidth}{!}{
\begin{tabular}{cccc}
\thicktoprule
 Model & \textcolor{black}{Mean optimal-solution probability} & Sampling rank & Expected energy gap   \\
\midrule
Standard QAOA (500)& 0.0645, 0.0133, [0.0595, 0.0694]   & 2.90,1.47,[2.35,3.45] & 1340.77,98.35, [1304.05, 1377.49]  \\
Block-wise XY-QAOA (1700)& 0.1041, 0.0738, [0.0765, 0.1317]   & 8.30,5.48,[6.25,10.35] & 720.62,134.75, [670.30, 770.93]  \\
Ours $\lambda = 0.5$ (800) & 0.1139, 0.0138, [0.1087, 0.1191] & 3.97,1.45,[3.42,4.51] & 888.80, 41.20, [873.42, 904.18] \\
Ours $\lambda = 0.6$ (500) & 0.1190, 0.0247, [0.1098, 0.1283] & 3.90,2.28,[3.05,4.75] & 906.66, 99.68, [869.44, 943.88] \\
Ours $\lambda = 0.7$ (400) & 0.1372, 0.0301, [0.1260, 0.1484] & 3.13,2.05,[2.37,3.90] & 913.76, 103.65, [875.06, 952.46] \\
Ours $\lambda = 0.8$ (400) & 0.1528, 0.0326, [0.1406, 0.1649] & \textbf{2.77,1.68,[2.14,3.39]} & 890.85, 119.41, [846.27, 935.43] \\
Ours $\lambda = 0.9$ (1000) & 0.1500, 0.0323, [0.1380, 0.1621] & 3.33,1.97,[2.60,4.07] & 876.05, 121.29, [830.77, 921.34] \\
\end{tabular}
}
\end{table}

\begin{table}[H]
\centering
\caption{\textcolor{black}{Performance comparison (Case II, Regime II)}}
\label{table:9}
\resizebox{\textwidth}{!}{
\begin{tabular}{cccc}
\thicktoprule
Model & \textcolor{black}{Mean optimal-solution probability} & Sampling rank & Expected energy gap \\
\midrule
Standard QAOA (200)& 0.0494,0.0219,[0.0412,0.0576] & 3.33,2.09,[2.55,4.11] & 1560.64,294.05,[1450.85,1670.43]  \\
Block-wise XY-QAOA (200)& 0.0788,0.0681,[0.0533,0.1042] & 11.63,9.19,[8.20,15.07] & 878.08,115.26,[835.04,921.11] \\
Ours $\lambda = 0.5$ (200) & 0.1014, 0.0263, [0.0916, 0.1112] & 4.73, 5.46, [2.70, 6.77] & 1028.45, 119.48, [983.84, 1073.06] \\
Ours $\lambda = 0.6$ (200) & 0.1082, 0.0337, [0.0957, 0.1208] & 3.93, 4.23, [2.35, 5.51] & 1039.07, 124.61, [992.54, 1085.59] \\
Ours $\lambda = 0.7$ (200) & 0.1146, 0.0411, [0.0993, 0.1300] & 3.37, 2.37, [2.48, 4.25] & 1040.10, 143.37, [986.57, 1093.63] \\
Ours $\lambda = 0.8$ (200) & 0.1315, 0.0381, [0.1172, 0.1457] & 3.10, 2.12, [2.31, 3.89] & 1008.32, 151.21, [951.86, 1064.77] \\
Ours $\lambda = 0.9$ (200) & 0.1266, 0.0282, [0.1161, 0.1371] & 2.67, 1.79, [2.00, 3.33] & 1030.03, 175.07, [964.67, 1095.40] \\
\end{tabular}
}
\end{table}

\begin{table}[H]
\centering
\caption{\textcolor{black}{Performance comparison (Case II, Regime III)}}
\label{table:10}
\resizebox{\textwidth}{!}{
\begin{tabular}{cccc}
\thicktoprule
Model & \textcolor{black}{Mean optimal-solution probability} & Sampling rank & Expected energy gap \\
\midrule
Standard QAOA (200)& 0.0480,0.0170,[0.0418,0.0542] & 3.60,2.99,[2.48,4.71] & 1547.79,258.02,[1451.46,1644.13]  \\
Block-wise XY-QAOA (200)& 0.0483,0.0522,[0.0288,0.0678] & 14.27,10.90,[10.20,158.33] & 1106.98,163.50,[1045.94,1168.03] \\
Ours $\lambda = 0.5$ (200) & 0.0944, 0.0361, [0.0809, 0.1079] & 4.77, 5.69, [2.64, 6.89] & 1186.35, 181.67, [1118.52, 1254.18] \\
Ours $\lambda = 1.0$ (200) & \textbf{0.1141, 0.0249, [0.1048, 0.1235]} & 2.53, 1.74, [1.88, 3.18] & 1197.25, 233.46, [1110.08, 1284.41] \\
Ours $\lambda = 1.5$ (200) & 0.1000, 0.0365, [0.0864, 0.1137] & 4.33, 2.82, [3.28, 5.39] & 1128.50, 270.96, [1027.33, 1229.67] \\
Ours $\lambda = 2.0$ (200) & 0.1040, 0.0487, [0.0858, 0.1222] & \textbf{6.13, 6.14, [3.84, 8.43]} & \textbf{941.03, 139.97, [888.77, 993.29]} \\
Ours $\lambda = 2.5$ (200) & 0.1049, 0.0265, [0.0950, 0.1148] & 3.23, 2.13, [2.44, 4.03] & 1088.14, 165.65, [1026.30, 1149.99] \\
\thickbottomrule
\end{tabular}
}
\end{table}

\begin{table}[H]
\centering
\caption{\textcolor{black}{Performance comparison (Case III, Regime I)}}
\label{table:11}
\resizebox{\textwidth}{!}{
\begin{tabular}{cccc}
\thicktoprule
Model & \textcolor{black}{Mean optimal-solution probability} & Sampling rank & Expected energy gap \\
\midrule
Standard QAOA (1100)& 0.0660, 0.0121, [0.0615, 0.0705] & 2.37,1.16,[1.93,2.80] & 1509.47,123.01, [1463.55, 1555.40] \\
Block-wise XY-QAOA (1700)& 0.2057, 0.0473, [0.1881, 0.2233]   & 1.83,0.87,[1.51,2.16] & 783.78,163.84, [722.60, 844.95]  \\
Ours $\lambda = 0.5$ (800) & 0.1440, 0.0419, [0.1284, 0.1597] & 3.13, 4.36, [1.51, 4.76] & 1017.85, 62.20, [994.63, 1041.07] \\
Ours $\lambda = 0.6$ (500) & 0.1339, 0.0320, [0.1219, 0.1458] & 2.87, 3.49, [1.56, 4.17] & 1039.14, 114.55, [996.37, 1081.91] \\
Ours $\lambda = 0.7$ (1500) & 0.1224, 0.0298, [0.1113, 0.1336] & 4.87, 5.86, [2.68, 7.05] & 1019.47, 109.07, [978.75, 1060.19] \\
Ours $\lambda = 0.8$ (500) & 0.1152, 0.0405, [0.1001, 0.1304] & 5.60, 7.09, [2.95, 8.25] & 1012.08, 139.63, [959.94, 1064.21] \\
Ours $\lambda = 0.9$ (800) & 0.1239, 0.0513, [0.1048, 0.1431] & 4.67, 7.28, [1.95, 7.39] & 983.16, 142.10, [930.11, 1036.21] \\
\end{tabular}
}
\end{table}

\begin{table}[H]
\centering
\caption{\textcolor{black}{Performance comparison (Case III, Regime II)}}
\label{table:12}
\resizebox{\textwidth}{!}{
\begin{tabular}{cccc}
\thicktoprule
Model & \textcolor{black}{Mean optimal-solution probability} & Sampling rank & Expected energy gap \\
\midrule
Standard QAOA (200) & 0.0479,0.0210,[0.0400,0.0558]& 2.87,1.57,[2.28,3.45] & 1801.32,322.82,[1680.79,1921.85]\\
Block-wise XY-QAOA (200) & 0.1509,0.0466,[0.1335,0.1682] & 2.1,0.92,[1.76,2.44] & 1002
.56,120.47,[957.58,1047.53] \\
Ours $\lambda = 0.5$ (200) & 0.1277, 0.0438, [0.1113, 0.1440] & 3.10, 3.92, [1.64, 4.56] & 1129.94, 110.80, [1088.57, 1171.31] \\
Ours $\lambda = 0.6$ (200) & 0.0997, 0.0452, [0.0828, 0.1166] & 5.60, 5.95, [3.38, 7.82] & 1183.92, 141.60, [1131.05, 1236.79] \\
Ours $\lambda = 0.7$ (200) & 0.1056, 0.0352, [0.0924, 0.1187] & 4.27, 3.10, [3.11, 5.42] & 1156.33, 171.95, [1092.13, 1220.53] \\
Ours $\lambda = 0.8$ (200) & 0.1007, 0.0470, [0.0832, 0.1183] & 7.50, 12.59, [2.80, 12.20] & 1140.95, 172.69, [1076.47, 1205.43] \\
Ours $\lambda = 0.9$ (200) & 0.0987, 0.0403, [0.0837, 0.1138] & 4.53, 2.32, [3.67, 5.40] & 1148.41, 193.62, [1076.12, 1220.70] \\
\end{tabular}
}
\end{table}

\begin{table}[H]
\centering
\caption{\textcolor{black}{Performance comparison (Case III, Regime III)}}
\label{table:13}
\resizebox{\textwidth}{!}{
\begin{tabular}{cccc}
\thicktoprule
Model & \textcolor{black}{Mean optimal-solution probability} & Sampling rank & Expected energy gap \\
\midrule
Standard QAOA (200)& 0.0531,0.0211,[0.0453,0.0610] & 2.56,1.50,[2.00,3.13] & 1762.65,300.37,[1650.50,1874.79]  \\
Block-wise XY-QAOA (200)& 0.1014,0.0473,[0.0838,0.1191] & 4.63,10.03,[0.89,8.38] & 1264.00,224.79,[1180.07,1347.93] \\
Ours $\lambda = 0.5$ (200) & 0.0856, 0.0477, [0.0678, 0.1034] & 7.20, 9.26, [3.74, 10.66] & 1321.33, 184.41, [1252.47, 1390.18] \\
Ours $\lambda = 1.0$ (200) & 0.0634, 0.0480, [0.0454, 0.0813] & 29.10, 67.02, [4.08, 54.12] & 1390.81, 258.10, [1294.45, 1487.18] \\
Ours $\lambda = 1.5$ (200) & 0.1079, 0.0427, [0.0920, 0.1239] & 5.37, 11.64, [1.02, 9.71] & 1283.46, 297.30, [1172.46, 1394.46] \\
Ours $\lambda = 2.0$ (200) & \textbf{0.1683, 0.0610, [0.1455, 0.1911]} & \textbf{2.03, 1.56, [1.45, 2.62]} & \textbf{1079.60, 160.61, [1019.63, 1139.56]} \\
Ours $\lambda = 2.5$ (200) & 0.1103, 0.0418, [0.0947, 0.1259] & 3.63, 2.59, [2.67, 4.60] & 1253.76, 177.39, [1187.53, 1319.99] \\
\thickbottomrule
\end{tabular}
}
\end{table}
\FloatBarrier

\subsection{\textcolor{black}{Component Ablation Study Tables}}
\label{sec:7}\label{line:7}
\textcolor{black}{This subsection gives the full ablation statistics for the complete proposed ansatz, Init-only, and Mixer-only variants across all benchmark cases and evaluation regimes.}

\begin{table}[H]
  \centering
\caption{\textcolor{black}{Case~I ablation under the statevector regime ($\lambda=1.3$, maxiter $=1300$, $N=30$).}}
  \label{table:14}
  \scriptsize
  \setlength{\tabcolsep}{3pt}
  \begin{tabular}{l ccc ccc ccc}
    \thicktoprule
    & \multicolumn{3}{c}{$p^{*}$}
    & \multicolumn{3}{c}{$E[C]-C^{*}_{\mathrm{feas}}$}
    & \multicolumn{3}{c}{Sampling rank} \\
    \cmidrule(lr){2-4}\cmidrule(lr){5-7}\cmidrule(lr){8-10}
    Variant & Mean & Std & [95\% CI]
            & Mean & Std & [95\% CI]
            & Mean & Std & [95\% CI] \\
    \midrule
    Full proposed
      & \textbf{0.7201} & 0.1319 & [0.6709,\,0.7694]
      & \textbf{375.53} & 190.45 & [304.42,\,446.63]
      & \textbf{1.00} & 0.00 & [1.00,\,1.00] \\[0.3em]
    Init-only
      & 0.2439 & 0.0103 & [0.2400,\,0.2478]
      & 1396.10 & 25.05 & [1386.75,\,1405.45]
      & 2.50 & 1.25 & [2.03,\,2.97] \\[0.3em]
    Mixer-only
      & 0.1381 & 0.0260 & [0.1284,\,0.1478]
      & 1290.87 & 57.48 & [1269.41,\,1312.33]
      & 1.07 & 0.37 & [0.93,\,1.20] \\
    \thickbottomrule
  \end{tabular}
\end{table}

\begin{table}[H]
  \centering
\caption{\textcolor{black}{Case~I ablation under the ideal finite-shot regime ($\lambda=1.3$, maxiter $=200$, $N=30$).}}
  \label{table:15}
  \scriptsize
  \setlength{\tabcolsep}{3pt}
  \begin{tabular}{l ccc ccc ccc}
    \thicktoprule
    & \multicolumn{3}{c}{$p^{*}$}
    & \multicolumn{3}{c}{$E[C]-C^{*}_{\mathrm{feas}}$}
    & \multicolumn{3}{c}{Sampling rank} \\
    \cmidrule(lr){2-4}\cmidrule(lr){5-7}\cmidrule(lr){8-10}
    Variant & Mean & Std & [95\% CI]
            & Mean & Std & [95\% CI]
            & Mean & Std & [95\% CI] \\
    \midrule
    Full proposed
      & \textbf{0.7036} & 0.1419 & [0.6507,\,0.7566]
      & \textbf{401.76} & 196.33 & [328.45,\,475.06]
      & \textbf{1.07} & 0.37 & [0.93,\,1.20] \\[0.3em]
    Init-only
      & 0.2462 & 0.0049 & [0.2443,\,0.2480]
      & 1402.15 & 14.07 & [1396.90,\,1407.41]
      & 2.77 & 1.28 & [2.29,\,3.24] \\[0.3em]
    Mixer-only
      & 0.1236 & 0.0253 & [0.1141,\,0.1330]
      & 1341.03 & 56.32 & [1320.00,\,1362.06]
      & 1.30 & 0.84 & [0.99,\,1.61] \\
    \thickbottomrule
  \end{tabular}
\end{table}

\begin{table}[H]
  \centering
\caption{\textcolor{black}{Case~I ablation under the noisy finite-shot regime ($\lambda=1.3$, maxiter $=200$, $N=30$).}}
  \label{table:16}
  \scriptsize
  \setlength{\tabcolsep}{3pt}
  \begin{tabular}{l ccc ccc ccc}
    \thicktoprule
    & \multicolumn{3}{c}{$p^{*}$}
    & \multicolumn{3}{c}{$E[C]-C^{*}_{\mathrm{feas}}$}
    & \multicolumn{3}{c}{Sampling rank} \\
    \cmidrule(lr){2-4}\cmidrule(lr){5-7}\cmidrule(lr){8-10}
    Variant & Mean & Std & [95\% CI]
            & Mean & Std & [95\% CI]
            & Mean & Std & [95\% CI] \\
    \midrule
    Full proposed
      & \textbf{0.6834} & 0.1459 & [0.6290,\,0.7379]
      & \textbf{434.45} & 201.61 & [359.18,\,509.73]
      & \textbf{1.07} & 0.37 & [0.93,\,1.20] \\[0.3em]
    Init-only
      & 0.2420 & 0.0052 & [0.2400,\,0.2439]
      & 1412.16 & 15.21 & [1406.49,\,1417.84]
      & 2.93 & 1.17 & [2.50,\,3.37] \\[0.3em]
    Mixer-only
      & 0.1173 & 0.0294 & [0.1064,\,0.1283]
      & 1359.31 & 57.88 & [1337.70,\,1380.92]
      & 1.60 & 2.11 & [0.81,\,2.39] \\
    \thickbottomrule
  \end{tabular}
\end{table}

\begin{table}[H]
  \centering
\caption{\textcolor{black}{Case~II ablation under the statevector regime ($\lambda=2$, maxiter $=2000$, $N=30$).}}
  \label{table:17}
  \scriptsize
  \setlength{\tabcolsep}{3pt}
  \begin{tabular}{l ccc ccc ccc}
    \thicktoprule
    & \multicolumn{3}{c}{$p^{*}$}
    & \multicolumn{3}{c}{$E[C]-C^{*}_{\mathrm{feas}}$}
    & \multicolumn{3}{c}{Sampling rank} \\
    \cmidrule(lr){2-4}\cmidrule(lr){5-7}\cmidrule(lr){8-10}
    Variant & Mean & Std & [95\% CI]
            & Mean & Std & [95\% CI]
            & Mean & Std & [95\% CI] \\
    \midrule
    Full proposed
      & \textbf{0.1553} & 0.0086 & [0.1521,\,0.1585]
      & \textbf{686.24} & 33.09 & [673.89,\,698.60]
      & \textbf{4.17} & 1.09 & [3.76,\,4.57] \\[0.3em]
    Init-only
      & 0.0283 & 0.0025 & [0.0274,\,0.0293]
      & 2177.12 & 23.19 & [2168.46,\,2185.77]
      & 7.13 & 4.06 & [5.62,\,8.65] \\[0.3em]
    Mixer-only
      & 0.0005 & 0.0010 & [0.0001,\,0.0009]
      & 3338.04 & 30.61 & [3326.61,\,3349.47]
      & 701.60 & 266.85 & [601.97,\,801.23] \\
    \thickbottomrule
  \end{tabular}
\end{table}

\begin{table}[H]
  \centering
\caption{\textcolor{black}{Case~II ablation under the ideal finite-shot regime ($\lambda=2$, maxiter $=200$, $N=30$).}}
  \label{table:18}
  \scriptsize
  \setlength{\tabcolsep}{3pt}
  \begin{tabular}{l ccc ccc ccc}
    \thicktoprule
    & \multicolumn{3}{c}{$p^{*}$}
    & \multicolumn{3}{c}{$E[C]-C^{*}_{\mathrm{feas}}$}
    & \multicolumn{3}{c}{Sampling rank} \\
    \cmidrule(lr){2-4}\cmidrule(lr){5-7}\cmidrule(lr){8-10}
    Variant & Mean & Std & [95\% CI]
            & Mean & Std & [95\% CI]
            & Mean & Std & [95\% CI] \\
    \midrule
    Full proposed
      & \textbf{0.1327} & 0.0516 & [0.1135,\,0.1520]
      & \textbf{786.93} & 91.99 & [752.58,\,821.27]
      & \textbf{4.57} & 2.90 & [3.49,\,5.65] \\[0.3em]
    Init-only
      & 0.0264 & 0.0021 & [0.0256,\,0.0272]
      & 2235.98 & 37.85 & [2221.85,\,2250.11]
      & 7.43 & 4.27 & [5.84,\,9.03] \\[0.3em]
    Mixer-only
      & 0.0020 & 0.0014 & [0.0015,\,0.0026]
      & 3430.99 & 88.90 & [3397.80,\,3464.18]
      & 338.10 & 340.87 & [210.83,\,465.37] \\
    \thickbottomrule
  \end{tabular}
\end{table}

\begin{table}[H]
  \centering
\caption{\textcolor{black}{Case~II ablation under the noisy finite-shot regime ($\lambda=2$, maxiter $=200$, $p=2$, $N=30$).}}
  \label{table:19}
  \scriptsize
  \setlength{\tabcolsep}{3pt}
  \begin{tabular}{l ccc ccc ccc}
    \thicktoprule
    & \multicolumn{3}{c}{$p^{*}$}
    & \multicolumn{3}{c}{$E[C]-C^{*}_{\mathrm{feas}}$}
    & \multicolumn{3}{c}{Sampling rank} \\
    \cmidrule(lr){2-4}\cmidrule(lr){5-7}\cmidrule(lr){8-10}
    Variant & Mean & Std & [95\% CI]
            & Mean & Std & [95\% CI]
            & Mean & Std & [95\% CI] \\
    \midrule
    Full proposed
      & \textbf{0.1040} & 0.0487 & [0.0858,\,0.1222]
      & \textbf{941.03} & 139.97 & [888.77,\,993.29]
      & \textbf{6.13} & 6.14 & [3.84,\,8.43] \\[0.3em]
    Init-only
      & 0.0257 & 0.0018 & [0.0250,\,0.0263]
      & 2263.95 & 36.31 & [2250.40,\,2277.51]
      & 7.50 & 4.06 & [5.98,\,9.02] \\[0.3em]
    Mixer-only
      & 0.0018 & 0.0014 & [0.0013,\,0.0023]
      & 3411.05 & 73.88 & [3383.47,\,3438.64]
      & 441.23 & 420.74 & [284.14,\,598.32] \\
    \thickbottomrule
  \end{tabular}
\end{table}

\begin{table}[H]
  \centering
\caption{\textcolor{black}{Case~III ablation under the statevector regime ($\lambda=2$, maxiter $=600$, $N=30$).}}
  \label{table:20}
  \scriptsize
  \setlength{\tabcolsep}{3pt}
  \begin{tabular}{l ccc ccc ccc}
    \thicktoprule
    & \multicolumn{3}{c}{$p^{*}$}
    & \multicolumn{3}{c}{$E[C]-C^{*}_{\mathrm{feas}}$}
    & \multicolumn{3}{c}{Sampling rank} \\
    \cmidrule(lr){2-4}\cmidrule(lr){5-7}\cmidrule(lr){8-10}
    Variant & Mean & Std & [95\% CI]
            & Mean & Std & [95\% CI]
            & Mean & Std & [95\% CI] \\
    \midrule
    Full proposed
      & \textbf{0.2080} & 0.0262 & [0.1982,\,0.2178]
      & \textbf{772.43} & 34.94 & [759.39,\,785.48]
      & \textbf{1.37} & 1.13 & [0.95,\,1.79] \\[0.3em]
    Init-only
      & 0.0274 & 0.0022 & [0.0266,\,0.0282]
      & 2468.17 & 28.88 & [2457.39,\,2478.96]
      & 8.37 & 4.14 & [6.82,\,9.91] \\[0.3em]
    Mixer-only
      & 0.0075 & 0.0016 & [0.0069,\,0.0081]
      & 3783.50 & 35.55 & [3770.22,\,3796.77]
      & 41.90 & 21.19 & [33.99,\,49.81] \\
    \thickbottomrule
  \end{tabular}
\end{table}

\begin{table}[H]
  \centering
\caption{\textcolor{black}{Case~III ablation under the ideal finite-shot regime ($\lambda=2$, maxiter $=200$, $N=30$).}}
  \label{table:21}
  \scriptsize
  \setlength{\tabcolsep}{3pt}
  \begin{tabular}{l ccc ccc ccc}
    \thicktoprule
    & \multicolumn{3}{c}{$p^{*}$}
    & \multicolumn{3}{c}{$E[C]-C^{*}_{\mathrm{feas}}$}
    & \multicolumn{3}{c}{Sampling rank} \\
    \cmidrule(lr){2-4}\cmidrule(lr){5-7}\cmidrule(lr){8-10}
    Variant & Mean & Std & [95\% CI]
            & Mean & Std & [95\% CI]
            & Mean & Std & [95\% CI] \\
    \midrule
    Full proposed
      & \textbf{0.2037} & 0.0433 & [0.1875,\,0.2199]
      & \textbf{887.14} & 93.66 & [852.17,\,922.11]
      & \textbf{2.13} & 1.53 & [1.56,\,2.70] \\[0.3em]
    Init-only
      & 0.0260 & 0.0023 & [0.0252,\,0.0269]
      & 2538.54 & 44.69 & [2521.85,\,2555.22]
      & 7.17 & 3.74 & [5.77,\,8.56] \\[0.3em]
    Mixer-only
      & 0.0060 & 0.0015 & [0.0054,\,0.0065]
      & 3904.48 & 104.80 & [3865.35,\,3943.61]
      & 28.67 & 24.89 & [19.37,\,37.96] \\
    \thickbottomrule
  \end{tabular}
\end{table}

\begin{table}[H]
  \centering
\caption{\textcolor{black}{Case~III ablation under the noisy finite-shot regime ($\lambda=2$, maxiter $=200$, $N=30$).}}
  \label{table:22}
  \scriptsize
  \setlength{\tabcolsep}{3pt}
  \begin{tabular}{l ccc ccc ccc}
    \thicktoprule
    & \multicolumn{3}{c}{$p^{*}$}
    & \multicolumn{3}{c}{$E[C]-C^{*}_{\mathrm{feas}}$}
    & \multicolumn{3}{c}{Sampling rank} \\
    \cmidrule(lr){2-4}\cmidrule(lr){5-7}\cmidrule(lr){8-10}
    Variant & Mean & Std & [95\% CI]
            & Mean & Std & [95\% CI]
            & Mean & Std & [95\% CI] \\
    \midrule
    Full proposed
      & \textbf{0.1683} & 0.0610 & [0.1455,\,0.1911]
      & \textbf{1079.60} & 160.61 & [1019.63,\,1139.56]
      & \textbf{2.03} & 1.56 & [1.45,\,2.62] \\[0.3em]
    Init-only
      & 0.0250 & 0.0017 & [0.0244,\,0.0257]
      & 2570.70 & 46.77 & [2553.23,\,2588.16]
      & 8.23 & 3.59 & [6.89,\,9.57] \\[0.3em]
    Mixer-only
      & 0.0058 & 0.0013 & [0.0053,\,0.0063]
      & 3883.10 & 83.41 & [3851.96,\,3914.24]
      & 30.03 & 22.62 & [21.59,\,38.48] \\
    \thickbottomrule
  \end{tabular}
\end{table}
\FloatBarrier

\subsection{\textcolor{black}{QAOA Depth Sensitivity Tables}}
\label{sec:10}\label{line:10}\label{sec:24}\label{line:24}
\textcolor{black}{This subsection contains the \(p\)-depth sensitivity results for the standard, proposed, and block-wise XY-QAOA ansatz families.}

\subsubsection{\textcolor{black}{Standard QAOA Depth Study}}

\textcolor{black}{The following tables summarize how standard QAOA changes with \(p\in\{1,2,3\}\) for the three benchmark cases under the stated evaluation regimes.}

\begin{table}[H]
  \centering
\caption{\textcolor{black}{Standard QAOA $p$-depth study — Case I , Regime I (statevector)}}
  \label{table:23}
  \scriptsize
  \setlength{\tabcolsep}{3pt}
\begin{tabular}{c ccc ccc ccc}
    \thicktoprule
    & \multicolumn{3}{c}{$p^{*}$}
    & \multicolumn{3}{c}{$E[C]-C^{*}_{\mathrm{feas}}$}
    & \multicolumn{3}{c}{Sampling rank} \\
    \cmidrule(lr){2-4}\cmidrule(lr){5-7}\cmidrule(lr){8-10}
    $p$ & Mean & Std & [95\% CI]
        & Mean & Std & [95\% CI]
        & Mean & Std & [95\% CI] \\
    \midrule
    1 & 0.1641 & 0.0000 & [0.1641, 0.1641] & 1117.4 & 0.0 & [1117.4, 1117.4] & 1.00 & 0.00 & [1.00, 1.00] \\
    2 & 0.4994 & 0.0925 & [0.4649, 0.5339] & 638.8 & 118.3 & [594.6, 682.9] & 1.00 & 0.00 & [1.00, 1.00] \\
    3 & 0.5867 & 0.1025 & [0.5484, 0.6249] & 489.1 & 132.9 & [439.5, 538.7] & 1.00 & 0.00 & [1.00, 1.00] \\
    \thickbottomrule
  \end{tabular}
\end{table}

\begin{table}[H]
  \centering
\caption{\textcolor{black}{Standard QAOA $p$-depth study — Case I, Regime II (ideal finite-shot)}}
  \label{table:24}
  \scriptsize
  \setlength{\tabcolsep}{3pt}
  \begin{tabular}{c ccc ccc ccc}
    \thicktoprule
    & \multicolumn{3}{c}{$p^{*}$}
    & \multicolumn{3}{c}{$E[C]-C^{*}_{\mathrm{feas}}$}
    & \multicolumn{3}{c}{Sampling rank} \\
    \cmidrule(lr){2-4}\cmidrule(lr){5-7}\cmidrule(lr){8-10}
    $p$ & Mean & Std & [95\% CI]
        & Mean & Std & [95\% CI]
        & Mean & Std & [95\% CI] \\
    \midrule
    1 & 0.1523 & 0.0074 & [0.1496, 0.1551] & 1107.6 & 11.5 & [1103.3, 1111.9] & 1.00 & 0.00 & [1.00, 1.00] \\
    2 & 0.4341 & 0.1311 & [0.3852, 0.4831] & 731.7 & 169.3 & [668.5, 794.9] & 1.00 & 0.00 & [1.00, 1.00] \\
    3 & 0.4760 & 0.1156 & [0.4328, 0.5191] & 672.1 & 169.2 & [608.9, 735.3] & 1.00 & 0.00 & [1.00, 1.00] \\
    \thickbottomrule
  \end{tabular}
\end{table}

\begin{table}[H]
  \centering
\caption{\textcolor{black}{Standard QAOA $p$-depth study — Case I, Regime III (noisy finite-shot)}}
  \label{table:25}
  \scriptsize
  \setlength{\tabcolsep}{3pt}
  \begin{tabular}{c ccc ccc ccc}
    \thicktoprule
    & \multicolumn{3}{c}{$p^{*}$}
    & \multicolumn{3}{c}{$E[C]-C^{*}_{\mathrm{feas}}$}
    & \multicolumn{3}{c}{Sampling rank} \\
    \cmidrule(lr){2-4}\cmidrule(lr){5-7}\cmidrule(lr){8-10}
    $p$ & Mean & Std & [95\% CI]
        & Mean & Std & [95\% CI]
        & Mean & Std & [95\% CI] \\
    \midrule
    1 & 0.1498 & 0.0086 & [0.1466, 0.1531] & 1115.0 & 9.6 & [1111.4, 1118.6] & 1.00 & 0.00 & [1.00, 1.00] \\
    2 & 0.4170 & 0.1246 & [0.3705, 0.4635] & 767.9 & 172.5 & [703.5, 832.3] & 1.00 & 0.00 & [1.00, 1.00] \\
    3 & 0.4770 & 0.1033 & [0.4385, 0.5156] & 660.7 & 145.8 & [606.2, 715.1] & 1.00 & 0.00 & [1.00, 1.00] \\
    \thickbottomrule
  \end{tabular}
\end{table}

\begin{table}[H]
  \centering
\caption{\textcolor{black}{Standard QAOA $p$-depth study — Case II, Regime I (statevector)}}
  \label{table:26}
  \scriptsize
  \setlength{\tabcolsep}{3pt}
  \begin{tabular}{c ccc ccc ccc}
    \thicktoprule
    & \multicolumn{3}{c}{$p^{*}$}
    & \multicolumn{3}{c}{$E[C]-C^{*}_{\mathrm{feas}}$}
    & \multicolumn{3}{c}{Sampling rank} \\
    \cmidrule(lr){2-4}\cmidrule(lr){5-7}\cmidrule(lr){8-10}
    $p$ & Mean & Std & [95\% CI]
        & Mean & Std & [95\% CI]
        & Mean & Std & [95\% CI] \\
    \midrule
    1 & 0.0171 & 0.0017 & [0.0165, 0.0178] & 2097.5 & 18.7 & [2090.5, 2104.5] & 2.47 & 1.17 & [2.03, 2.90] \\
    2 & 0.0645 & 0.0133 & [0.0595, 0.0694] & 1340.77 & 98.35 & [1304.05, 1377.49] & 2.9 & 1.47 & [2.35, 3.45] \\
    3 & 0.1085 & 0.0175 & [0.1020, 0.1150] & 983.7 & 124.3 & [937.3, 1030.1] & 2.90 & 1.47 & [2.35, 3.45] \\
    \thickbottomrule
  \end{tabular}
\end{table}

\begin{table}[H]
  \centering
\caption{\textcolor{black}{Standard QAOA $p$-depth study — Case II, Regime II (ideal finite-shot)}}
  \label{table:27}
  \scriptsize
  \setlength{\tabcolsep}{3pt}
  \begin{tabular}{c ccc ccc ccc}
    \thicktoprule
    & \multicolumn{3}{c}{$p^{*}$}
    & \multicolumn{3}{c}{$E[C]-C^{*}_{\mathrm{feas}}$}
    & \multicolumn{3}{c}{Sampling rank} \\
    \cmidrule(lr){2-4}\cmidrule(lr){5-7}\cmidrule(lr){8-10}
    $p$ & Mean & Std & [95\% CI]
        & Mean & Std & [95\% CI]
        & Mean & Std & [95\% CI] \\
    \midrule
    1 & 0.0172 & 0.0020 & [0.0164, 0.0179] & 2114.0 & 23.0 & [2105.4, 2122.6] & 2.53 & 1.41 & [2.01, 3.06] \\
    2 & 0.0494 & 0.0219 & [0.0412, 0.0576] & 1560.6 & 294.1 & [1450.9, 1670.4] & 3.33 & 2.09 & [2.55, 4.11] \\
    3 & 0.0839 & 0.0289 & [0.0731, 0.0947] & 1230.9 & 223.5 & [1147.4, 1314.3] & 3.77 & 3.78 & [2.35, 5.18] \\
    \thickbottomrule
  \end{tabular}
\end{table}

\begin{table}[H]
  \centering
\caption{\textcolor{black}{Standard QAOA $p$-depth study — Case II, Regime III (noisy finite-shot)}}
  \label{table:28}
  \scriptsize
  \setlength{\tabcolsep}{3pt}
  \begin{tabular}{c ccc ccc ccc}
    \thicktoprule
    & \multicolumn{3}{c}{$p^{*}$}
    & \multicolumn{3}{c}{$E[C]-C^{*}_{\mathrm{feas}}$}
    & \multicolumn{3}{c}{Sampling rank} \\
    \cmidrule(lr){2-4}\cmidrule(lr){5-7}\cmidrule(lr){8-10}
    $p$ & Mean & Std & [95\% CI]
        & Mean & Std & [95\% CI]
        & Mean & Std & [95\% CI] \\
    \midrule
    1 & 0.0168 & 0.0016 & [0.0163, 0.0174] & 2127.2 & 22.0 & [2119.0, 2135.4] & 2.23 & 1.10 & [1.82, 2.65] \\
    2 & 0.0480 & 0.0167 & [0.0418, 0.0543] & 1547.8 & 258.0 & [1451.5, 1644.1] & 3.60 & 2.99 & [2.48, 4.72] \\
    3 & 0.0771 & 0.0258 & [0.0674, 0.0867] & 1277.7 & 200.7 & [1202.8, 1352.7] & 3.53 & 3.88 & [2.08, 4.98] \\
    \thickbottomrule
  \end{tabular}
\end{table}

\begin{table}[H]
  \centering
\caption{\textcolor{black}{Standard QAOA $p$-depth study — Case III, Regime I (statevector)}}
  \label{table:29}
  \scriptsize
  \setlength{\tabcolsep}{3pt}
  \begin{tabular}{c ccc ccc ccc}
    \thicktoprule
    & \multicolumn{3}{c}{$p^{*}$}
    & \multicolumn{3}{c}{$E[C]-C^{*}_{\mathrm{feas}}$}
    & \multicolumn{3}{c}{Sampling rank} \\
    \cmidrule(lr){2-4}\cmidrule(lr){5-7}\cmidrule(lr){8-10}
    $p$ & Mean & Std & [95\% CI]
        & Mean & Std & [95\% CI]
        & Mean & Std & [95\% CI] \\
    \midrule
    1 & 0.0171 & 0.0014 & [0.0166, 0.0176] & 2359.2 & 26.5 & [2349.3, 2369.1] & 2.33 & 1.37 & [1.82, 2.85] \\
    2 & 0.0660 & 0.0121 & [0.0615, 0.0705] & 1509.47 & 123.01 & [1463.55, 1555.40] & 2.37 & 1.16 & [2.37, 2.80] \\
    3 & 0.1075 & 0.0193 & [0.1003, 0.1147] & 1133.9 & 161.3 & [1073.7, 1194.2] & 2.77 & 1.52 & [2.20, 3.34] \\
    \thickbottomrule
  \end{tabular}
\end{table}

\begin{table}[H]
  \centering
\caption{\textcolor{black}{Standard QAOA $p$-depth study — Case III (4-node, case2), Regime II (ideal finite-shot)}}
  \label{table:30}
  \scriptsize
  \setlength{\tabcolsep}{3pt}
  \begin{tabular}{c ccc ccc ccc}
    \thicktoprule
    & \multicolumn{3}{c}{$p^{*}$}
    & \multicolumn{3}{c}{$E[C]-C^{*}_{\mathrm{feas}}$}
    & \multicolumn{3}{c}{Sampling rank} \\
    \cmidrule(lr){2-4}\cmidrule(lr){5-7}\cmidrule(lr){8-10}
    $p$ & Mean & Std & [95\% CI]
        & Mean & Std & [95\% CI]
        & Mean & Std & [95\% CI] \\
    \midrule
    1 & 0.0169 & 0.0016 & [0.0163, 0.0175] & 2375.2 & 36.0 & [2361.8, 2388.6] & 2.67 & 1.32 & [2.17, 3.16] \\
    2 & 0.0479 & 0.0210 & [0.0400, 0.0558] & 1801.3 & 322.8 & [1680.8, 1921.8] & 2.87 & 1.57 & [2.28, 3.45] \\
    3 & 0.0740 & 0.0343 & [0.0612, 0.0868] & 1462.0 & 267.3 & [1362.2, 1561.8] & 5.03 & 8.44 & [1.88, 8.19] \\
    \thickbottomrule
  \end{tabular}
\end{table}

\begin{table}[H]
  \centering
\caption{\textcolor{black}{Standard QAOA $p$-depth study — Case III (4-node, case2), Regime III (noisy finite-shot).}}
  \label{table:31}
  \scriptsize
  \setlength{\tabcolsep}{3pt}
  \begin{tabular}{c ccc ccc ccc}
    \thicktoprule
    & \multicolumn{3}{c}{$p^{*}$}
    & \multicolumn{3}{c}{$E[C]-C^{*}_{\mathrm{feas}}$}
    & \multicolumn{3}{c}{Sampling rank} \\
    \cmidrule(lr){2-4}\cmidrule(lr){5-7}\cmidrule(lr){8-10}
    $p$ & Mean & Std & [95\% CI]
        & Mean & Std & [95\% CI]
        & Mean & Std & [95\% CI] \\
    \midrule
    1 & 0.0168 & 0.0019 & [0.0161, 0.0175] & 2390.8 & 28.9 & [2380.0, 2401.6] & 2.50 & 0.97 & [2.14, 2.86] \\
    2 & 0.0531 & 0.0211 & [0.0453, 0.0610] & 1762.6 & 300.4 & [1650.5, 1874.8] & 2.57 & 1.50 & [2.01, 3.13] \\
    3 & 0.0772 & 0.0263 & [0.0674, 0.0870] & 1470.69 & 272.96 & [1368.77, 1572.60] & 2.90 & 3.17 & [1.61, 4.19] \\
    \thickbottomrule
  \end{tabular}
\end{table}
\FloatBarrier

\subsubsection{\textcolor{black}{Proposed QAOA Depth Study}}

\textcolor{black}{The following tables summarize the depth response of the proposed constraint-aware initialization and hybrid \(XY\)--\(X\) mixer under the same benchmark settings.}

\begin{table}[H]
  \centering
\caption{\textcolor{black}{Proposed QAOA $p$-depth study — Case I, Regime I (statevector)}}
  \label{table:32}
  \scriptsize
  \setlength{\tabcolsep}{3pt}
  \begin{tabular}{c ccc ccc ccc}
    \thicktoprule
    & \multicolumn{3}{c}{$p^{*}$}
    & \multicolumn{3}{c}{$E[C]-C^{*}_{\mathrm{feas}}$}
    & \multicolumn{3}{c}{Sampling rank} \\
    \cmidrule(lr){2-4}\cmidrule(lr){5-7}\cmidrule(lr){8-10}
    $p$ & Mean & Std & [95\% CI]
        & Mean & Std & [95\% CI]
        & Mean & Std & [95\% CI] \\
    \midrule
    1 & 0.2695 & 0.0000 & [0.2695, 0.2695] & 1376.0 & 0.0 & [1376.0, 1376.0] & 1.00 & 0.00 & [1.00, 1.00] \\
    2 & 0.7201 & 0.1319 & [0.6709, 0.7694] & 375.5 & 190.4 & [304.4, 446.6] & 1.00 & 0.00 & [1.00, 1.00] \\
    3 & 0.8578 & 0.0472 & [0.8402, 0.8754] & 184.8 & 63.1 & [161.2, 208.4] & 1.00 & 0.00 & [1.00, 1.00] \\
    \thickbottomrule
  \end{tabular}
\end{table}

\begin{table}[H]
  \centering
\caption{\textcolor{black}{Proposed QAOA $p$-depth study — Case I, Regime II (ideal finite-shot).}}
  \label{table:33}
  \scriptsize
  \setlength{\tabcolsep}{3pt}
  \begin{tabular}{c ccc ccc ccc}
    \thicktoprule
    & \multicolumn{3}{c}{$p^{*}$}
    & \multicolumn{3}{c}{$E[C]-C^{*}_{\mathrm{feas}}$}
    & \multicolumn{3}{c}{Sampling rank} \\
    \cmidrule(lr){2-4}\cmidrule(lr){5-7}\cmidrule(lr){8-10}
    $p$ & Mean & Std & [95\% CI]
        & Mean & Std & [95\% CI]
        & Mean & Std & [95\% CI] \\
    \midrule
    1 & 0.2487 & 0.0058 & [0.2466, 0.2509] & 1400.7 & 23.9 & [1391.8, 1409.7] & 2.60 & 1.07 & [2.20, 3.00] \\
    2 & 0.7036 & 0.1419 & [0.6507, 0.7566] & 401.8 & 196.3 & [328.5, 475.1] & 1.07 & 0.37 & [0.93, 1.20] \\
    3 & 0.7846 & 0.1068 & [0.7447, 0.8244] & 282.0 & 137.4 & [230.7, 333.3] & 1.00 & 0.00 & [1.00, 1.00] \\
    \thickbottomrule
  \end{tabular}
\end{table}

\begin{table}[H]
  \centering
\caption{\textcolor{black}{Proposed QAOA $p$-depth study — Case I, Regime III (noisy finite-shot)}}
  \label{table:34}
  \scriptsize
  \setlength{\tabcolsep}{3pt}
  \begin{tabular}{c ccc ccc ccc}
    \thicktoprule
    & \multicolumn{3}{c}{$p^{*}$}
    & \multicolumn{3}{c}{$E[C]-C^{*}_{\mathrm{feas}}$}
    & \multicolumn{3}{c}{Sampling rank} \\
    \cmidrule(lr){2-4}\cmidrule(lr){5-7}\cmidrule(lr){8-10}
    $p$ & Mean & Std & [95\% CI]
        & Mean & Std & [95\% CI]
        & Mean & Std & [95\% CI] \\
    \midrule
    1 & 0.2450 & 0.0055 & [0.2429, 0.2470] & 1409.1 & 23.4 & [1400.4, 1417.9] & 2.53 & 1.04 & [2.14, 2.92] \\
    2 & 0.6834 & 0.1459 & [0.6290, 0.7379] & 434.5 & 201.6 & [359.2, 509.7] & 1.07 & 0.37 & [0.93, 1.20] \\
    3 & 0.7241 & 0.1394 & [0.6720, 0.7762] & 351.5 & 166.1 & [289.5, 413.5] & 1.00 & 0.00 & [1.00, 1.00] \\
    \thickbottomrule
  \end{tabular}
\end{table}

\begin{table}[H]
  \centering
\caption{\textcolor{black}{Proposed QAOA $p$-depth study — Case II, Regime I (statevector)}}
  \label{table:35}
  \scriptsize
  \setlength{\tabcolsep}{3pt}
  \begin{tabular}{c ccc ccc ccc}
    \thicktoprule
    & \multicolumn{3}{c}{$p^{*}$}
    & \multicolumn{3}{c}{$E[C]-C^{*}_{\mathrm{feas}}$}
    & \multicolumn{3}{c}{Sampling rank} \\
    \cmidrule(lr){2-4}\cmidrule(lr){5-7}\cmidrule(lr){8-10}
    $p$ & Mean & Std & [95\% CI]
        & Mean & Std & [95\% CI]
        & Mean & Std & [95\% CI] \\
    \midrule
    1 & 0.0234 & 0.0000 & [0.0234, 0.0234] & 2403.7 & 0.0 & [2403.7, 2403.7] & 20.00 & 0.00 & [20.00, 20.00] \\
    2 & 0.1553 & 0.0086 & [0.1521, 0.1585] & 686.2 & 33.1 & [673.9, 698.6] & 4.17 & 1.09 & [3.76, 4.57] \\
    3 & 0.1944 & 0.0190 & [0.1873, 0.2015] & 521.9 & 66.5 & [497.0, 546.7] & 4.03 & 1.19 & [3.59, 4.48] \\
    \thickbottomrule
  \end{tabular}
\end{table}

\begin{table}[H]
  \centering
\caption{\textcolor{black}{Proposed QAOA $p$-depth study — Case II, Regime II (ideal finite-shot)}}
  \label{table:36}
  \scriptsize
  \setlength{\tabcolsep}{3pt}
  \begin{tabular}{c ccc ccc ccc}
    \thicktoprule
    & \multicolumn{3}{c}{$p^{*}$}
    & \multicolumn{3}{c}{$E[C]-C^{*}_{\mathrm{feas}}$}
    & \multicolumn{3}{c}{Sampling rank} \\
    \cmidrule(lr){2-4}\cmidrule(lr){5-7}\cmidrule(lr){8-10}
    $p$ & Mean & Std & [95\% CI]
        & Mean & Std & [95\% CI]
        & Mean & Std & [95\% CI] \\
    \midrule
    1 & 0.0953 & 0.0048 & [0.0935, 0.0971] & 1040.8 & 15.6 & [1035.0, 1046.6] & 3.17 & 1.23 & [2.71, 3.63] \\
    2 & 0.1327 & 0.0516 & [0.1135, 0.1520] & 786.9 & 92.0 & [752.6, 821.3] & 4.57 & 2.90 & [3.49, 5.65] \\
    3 & 0.1551 & 0.0412 & [0.1397, 0.1705] & 715.8 & 105.4 & [676.4, 755.1] & 3.97 & 2.01 & [3.22, 4.72] \\
    \thickbottomrule
  \end{tabular}
\end{table}

\begin{table}[H]
  \centering
\caption{\textcolor{black}{Proposed QAOA $p$-depth study — Case II, Regime III (noisy finite-shot)}}
  \label{table:37}
  \scriptsize
  \setlength{\tabcolsep}{3pt}
  \begin{tabular}{c ccc ccc ccc}
    \thicktoprule
    & \multicolumn{3}{c}{$p^{*}$}
    & \multicolumn{3}{c}{$E[C]-C^{*}_{\mathrm{feas}}$}
    & \multicolumn{3}{c}{Sampling rank} \\
    \cmidrule(lr){2-4}\cmidrule(lr){5-7}\cmidrule(lr){8-10}
    $p$ & Mean & Std & [95\% CI]
        & Mean & Std & [95\% CI]
        & Mean & Std & [95\% CI] \\
    \midrule
    1 & 0.0914 & 0.0049 & [0.0896, 0.0932] & 1083.9 & 14.5 & [1078.5, 1089.3] & 3.43 & 1.19 & [2.99, 3.88] \\
    2 & 0.1040 & 0.0487 & [0.0858, 0.1222] & 941.0 & 140.0 & [888.8, 993.3] & 6.13 & 6.14 & [3.84, 8.43] \\
    3 & 0.1365 & 0.0395 & [0.1217, 0.1512] & 844.0 & 105.8 & [804.5, 883.5] & 4.27 & 2.18 & [3.45, 5.08] \\
    \thickbottomrule
  \end{tabular}
\end{table}

\begin{table}[H]
  \centering
\caption{\textcolor{black}{Proposed QAOA $p$-depth study — Case III, Regime I (statevector)}}
  \label{table:38}
  \scriptsize
  \setlength{\tabcolsep}{3pt}
  \begin{tabular}{c ccc ccc ccc}
    \thicktoprule
    & \multicolumn{3}{c}{$p^{*}$}
    & \multicolumn{3}{c}{$E[C]-C^{*}_{\mathrm{feas}}$}
    & \multicolumn{3}{c}{Sampling rank} \\
    \cmidrule(lr){2-4}\cmidrule(lr){5-7}\cmidrule(lr){8-10}
    $p$ & Mean & Std & [95\% CI]
        & Mean & Std & [95\% CI]
        & Mean & Std & [95\% CI] \\
    \midrule
    1 & 0.0989 & 0.0053 & [0.0969, 0.1009] & 1160.6 & 16.2 & [1154.5, 1166.6] & 2.03 & 1.33 & [1.54, 2.53] \\
    2 & 0.2080 & 0.0262 & [0.1982, 0.2178] & 772.4 & 34.9 & [759.4, 785.5] & 1.37 & 1.13 & [0.95, 1.79] \\
    3 & 0.2367 & 0.0231 & [0.2281, 0.2454] & 581.4 & 63.7 & [557.6, 605.1] & 1.80 & 1.45 & [1.26, 2.34] \\
    \thickbottomrule
  \end{tabular}
\end{table}

\begin{table}[H]
  \centering
\caption{\textcolor{black}{Proposed QAOA $p$-depth study — Case III, Regime II (ideal finite-shot)}}
  \label{table:39}
  \scriptsize
  \setlength{\tabcolsep}{3pt}
  \begin{tabular}{c ccc ccc ccc}
    \thicktoprule
    & \multicolumn{3}{c}{$p^{*}$}
    & \multicolumn{3}{c}{$E[C]-C^{*}_{\mathrm{feas}}$}
    & \multicolumn{3}{c}{Sampling rank} \\
    \cmidrule(lr){2-4}\cmidrule(lr){5-7}\cmidrule(lr){8-10}
    $p$ & Mean & Std & [95\% CI]
        & Mean & Std & [95\% CI]
        & Mean & Std & [95\% CI] \\
    \midrule
    1 & 0.0985 & 0.0075 & [0.0957, 0.1013] & 1170.5 & 15.8 & [1164.6, 1176.3] & 1.93 & 1.31 & [1.44, 2.42] \\
    2 & 0.2037 & 0.0433 & [0.1875, 0.2199] & 887.1 & 93.7 & [852.2, 922.1] & 2.13 & 1.53 & [1.56, 2.70] \\
    3 & 0.2068 & 0.0442 & [0.1903, 0.2232] & 800.9 & 123.0 & [754.9, 846.8] & 1.97 & 1.54 & [1.39, 2.54] \\
    \thickbottomrule
  \end{tabular}
\end{table}

\begin{table}[H]
  \centering
\caption{\textcolor{black}{Proposed QAOA $p$-depth study — Case III, Regime III (noisy finite-shot)}}
  \label{table:40}
  \scriptsize
  \setlength{\tabcolsep}{3pt}
  \begin{tabular}{c ccc ccc ccc}
    \thicktoprule
    & \multicolumn{3}{c}{$p^{*}$}
    & \multicolumn{3}{c}{$E[C]-C^{*}_{\mathrm{feas}}$}
    & \multicolumn{3}{c}{Sampling rank} \\
    \cmidrule(lr){2-4}\cmidrule(lr){5-7}\cmidrule(lr){8-10}
    $p$ & Mean & Std & [95\% CI]
        & Mean & Std & [95\% CI]
        & Mean & Std & [95\% CI] \\
    \midrule
    1 & 0.0964 & 0.0068 & [0.0939, 0.0990] & 1216.9 & 17.6 & [1210.3, 1223.5] & 1.80 & 1.35 & [1.30, 2.30] \\
    2 & 0.1683 & 0.0610 & [0.1455, 0.1911] & 1079.6 & 160.6 & [1019.6, 1139.6] & 2.03 & 1.56 & [1.45, 2.62] \\
    3 & 0.1923 & 0.0325 & [0.1801, 0.2044] & 938.3 & 107.2 & [898.3, 978.4] & 1.80 & 1.52 & [1.23, 2.37] \\
    \thickbottomrule
  \end{tabular}
\end{table}
\FloatBarrier
%\clearpage

\subsubsection{\textcolor{black}{Block-wise XY-QAOA Baseline Depth Study}}

\textcolor{black}{The following tables report the \(p\)-depth sensitivity of the block-wise XY-QAOA baseline for the two four-node benchmark cases.}

\begin{table}[H]
  \centering
\caption{\textcolor{black}{Block-wise XY-QAOA $p$-depth study — Case II, Regime I (statevector)}}
  \label{table:41}
  \scriptsize
  \setlength{\tabcolsep}{3pt}
  \begin{tabular}{c ccc ccc ccc}
    \thicktoprule
    & \multicolumn{3}{c}{$p^{*}$}
    & \multicolumn{3}{c}{$E[C]-C^{*}_{\mathrm{feas}}$}
    & \multicolumn{3}{c}{Sampling rank} \\
    \cmidrule(lr){2-4}\cmidrule(lr){5-7}\cmidrule(lr){8-10}
    $p$ & Mean & Std & [95\% CI]
        & Mean & Std & [95\% CI]
        & Mean & Std & [95\% CI] \\
    \midrule
    1 & 0.0952 & 0.0450 & [0.0784, 0.1120] & 905.1 & 63.6 & [881.3, 928.8] & 7.67 & 8.31 & [4.56, 10.77] \\
    2 & 0.1041 & 0.0738 & [0.0765, 0.1317] & 720.6 & 134.8 & [670.3, 770.9] & 8.30 & 5.48 & [6.25, 10.35] \\
    3 & 0.1379 & 0.0810 & [0.1077, 0.1681] & 620.6 & 137.8 & [569.1, 672.0] & 6.03 & 4.44 & [4.38, 7.69] \\
    \thickbottomrule
  \end{tabular}
\end{table}

\begin{table}[H]
  \centering
\caption{\textcolor{black}{Block-wise XY-QAOA $p$-depth study — Case II, Regime II (ideal finite-shot)}}
  \label{table:42}
  \scriptsize
  \setlength{\tabcolsep}{3pt}
  \begin{tabular}{c ccc ccc ccc}
    \thicktoprule
    & \multicolumn{3}{c}{$p^{*}$}
    & \multicolumn{3}{c}{$E[C]-C^{*}_{\mathrm{feas}}$}
    & \multicolumn{3}{c}{Sampling rank} \\
    \cmidrule(lr){2-4}\cmidrule(lr){5-7}\cmidrule(lr){8-10}
    $p$ & Mean & Std & [95\% CI]
        & Mean & Std & [95\% CI]
        & Mean & Std & [95\% CI] \\
    \midrule
    1 & 0.1089 & 0.0290 & [0.0981, 0.1197] & 902.7 & 56.3 & [881.7, 923.8] & 4.90 & 2.92 & [3.81, 5.99] \\
    2 & 0.0788 & 0.0681 & [0.0533, 0.1042] & 878.1 & 115.3 & [835.0, 921.1] & 11.63 & 9.19 & [8.20, 15.07] \\
    3 & 0.0536 & 0.0642 & [0.0296, 0.0776] & 870.3 & 120.5 & [825.3, 915.3] & 14.00 & 10.84 & [9.95, 18.05] \\
    \thickbottomrule
  \end{tabular}
\end{table}

\begin{table}[H]
  \centering
\caption{\textcolor{black}{Block-wise XY-QAOA $p$-depth study — Case II, Regime III (noisy finite-shot)}}
  \label{table:43}
  \scriptsize
  \setlength{\tabcolsep}{3pt}
  \begin{tabular}{c ccc ccc ccc}
    \thicktoprule
    & \multicolumn{3}{c}{$p^{*}$}
    & \multicolumn{3}{c}{$E[C]-C^{*}_{\mathrm{feas}}$}
    & \multicolumn{3}{c}{Sampling rank} \\
    \cmidrule(lr){2-4}\cmidrule(lr){5-7}\cmidrule(lr){8-10}
    $p$ & Mean & Std & [95\% CI]
        & Mean & Std & [95\% CI]
        & Mean & Std & [95\% CI] \\
    \midrule
    1 & 0.0885 & 0.0445 & [0.0719, 0.1052] & 979.0 & 65.8 & [954.5, 1003.6] & 10.57 & 14.36 & [5.21, 15.93] \\
    2 & 0.0483 & 0.0522 & [0.0288, 0.0677] & 1107.0 & 163.5 & [1045.9, 1168.0] & 14.27 & 10.90 & [10.20, 18.34] \\
    3 & 0.0667 & 0.0621 & [0.0435, 0.0899] & 1009.0 & 132.1 & [959.7, 1058.3] & 10.90 & 8.72 & [7.64, 14.16] \\
    \thickbottomrule
  \end{tabular}
\end{table}

\begin{table}[H]
  \centering
\caption{\textcolor{black}{Block-wise XY-QAOA $p$-depth study — Case III, Regime I (statevector)}}
  \label{table:44}
  \scriptsize
  \setlength{\tabcolsep}{3pt}
  \begin{tabular}{c ccc ccc ccc}
    \thicktoprule
    & \multicolumn{3}{c}{$p^{*}$}
    & \multicolumn{3}{c}{$E[C]-C^{*}_{\mathrm{feas}}$}
    & \multicolumn{3}{c}{Sampling rank} \\
    \cmidrule(lr){2-4}\cmidrule(lr){5-7}\cmidrule(lr){8-10}
    $p$ & Mean & Std & [95\% CI]
        & Mean & Std & [95\% CI]
        & Mean & Std & [95\% CI] \\
    \midrule
    1 & 0.1288 & 0.0298 & [0.1177, 0.1399] & 1021.7 & 75.2 & [993.6, 1049.8] & 1.77 & 1.25 & [1.30, 2.23] \\
    2 & 0.2057 & 0.0473 & [0.1881, 0.2233] & 783.8 & 163.8 & [722.6, 845.0] & 1.83 & 0.87 & [1.51, 2.16] \\
    3 & 0.2371 & 0.0474 & [0.2194, 0.2548] & 677.8 & 149.4 & [622.0, 733.6] & 1.60 & 0.67 & [1.35, 1.85] \\
    \thickbottomrule
  \end{tabular}
\end{table}

\begin{table}[H]
  \centering
\caption{\textcolor{black}{Block-wise XY-QAOA $p$-depth study — Case III, Regime II (ideal finite-shot)}}
  \label{table:45}
  \scriptsize
  \setlength{\tabcolsep}{3pt}
  \begin{tabular}{c ccc ccc ccc}
    \thicktoprule
    & \multicolumn{3}{c}{$p^{*}$}
    & \multicolumn{3}{c}{$E[C]-C^{*}_{\mathrm{feas}}$}
    & \multicolumn{3}{c}{Sampling rank} \\
    \cmidrule(lr){2-4}\cmidrule(lr){5-7}\cmidrule(lr){8-10}
    $p$ & Mean & Std & [95\% CI]
        & Mean & Std & [95\% CI]
        & Mean & Std & [95\% CI] \\
    \midrule
    1 & 0.1295 & 0.0322 & [0.1175, 0.1415] & 1041.7 & 78.4 & [1012.4, 1071.0] & 1.80 & 1.37 & [1.29, 2.31] \\
    2 & 0.1509 & 0.0466 & [0.1335, 0.1682] & 1002.6 & 120.5 & [957.6, 1047.5] & 2.10 & 0.92 & [1.76, 2.44] \\
    3 & 0.1682 & 0.0424 & [0.1524, 0.1840] & 984.0 & 111.3 & [942.5, 1025.6] & 2.33 & 1.27 & [1.86, 2.81] \\
    \thickbottomrule
  \end{tabular}
\end{table}

\begin{table}[H]
  \centering
\caption{\textcolor{black}{Block-wise XY-QAOA $p$-depth study — Case III, Regime III (noisy finite-shot)}}
  \label{table:46}
  \scriptsize
  \setlength{\tabcolsep}{3pt}
  \begin{tabular}{c ccc ccc ccc}
    \thicktoprule
    & \multicolumn{3}{c}{$p^{*}$}
    & \multicolumn{3}{c}{$E[C]-C^{*}_{\mathrm{feas}}$}
    & \multicolumn{3}{c}{Sampling rank} \\
    \cmidrule(lr){2-4}\cmidrule(lr){5-7}\cmidrule(lr){8-10}
    $p$ & Mean & Std & [95\% CI]
        & Mean & Std & [95\% CI]
        & Mean & Std & [95\% CI] \\
    \midrule
    1 & 0.1249 & 0.0269 & [0.1149, 0.1350] & 1097.9 & 63.1 & [1074.4, 1121.5] & 1.60 & 1.16 & [1.17, 2.03] \\
    2 & 0.1014 & 0.0473 & [0.0838, 0.1191] & 1264.0 & 224.8 & [1180.1, 1347.9] & 4.63 & 10.03 & [0.89, 8.38] \\
    3 & 0.1388 & 0.0439 & [0.1225, 0.1552] & 1146.9 & 128.2 & [1099.0, 1194.8] & 2.40 & 1.25 & [1.93, 2.87] \\
    \thickbottomrule
  \end{tabular}
\end{table}
\FloatBarrier

\clearpage
\subsection{\textcolor{black}{QAOA Circuit Realizations for the Benchmark Cases}}
\label{app:circuit_realizations}

\textcolor{black}{This subsection presents the explicit circuit realizations used for the three benchmark cases. Each case is shown as a matched pair, allowing direct inspection of the conventional standard-QAOA circuit and the proposed circuit with constraint-aware initialization and hybrid \(XY\)--\(X\) mixing.}

\subsubsection{\textcolor{black}{Case I circuit realizations}}

\textcolor{black}{The following circuits correspond to the three-node toy instance used for Case~I.}

\begin{figure}[H]
    \centering
    \includegraphics[width=1.0\linewidth]{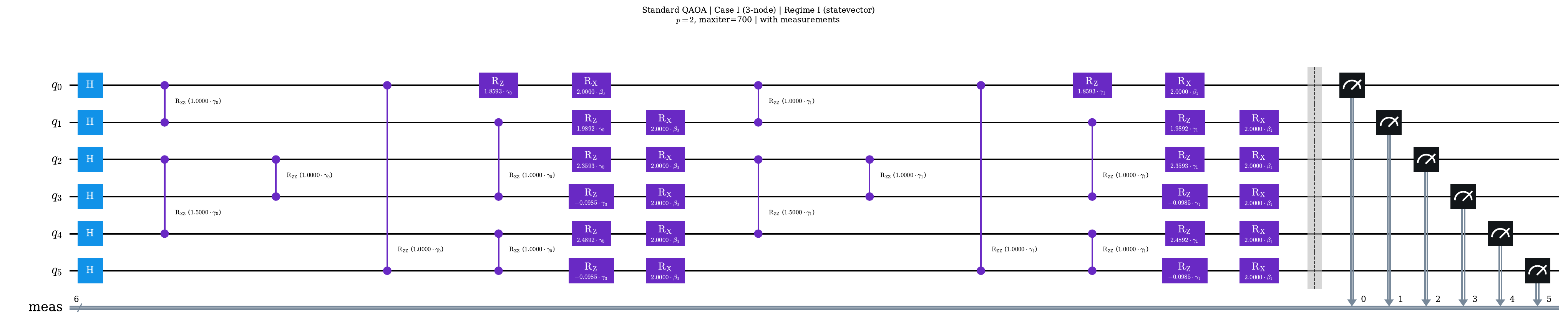}
    \caption{\textcolor{black}{Standard QAOA circuit (Case I)}}
    \label{fig:16}
\end{figure}

\begin{figure}[H]
    \centering
    \includegraphics[width=1.0\linewidth]{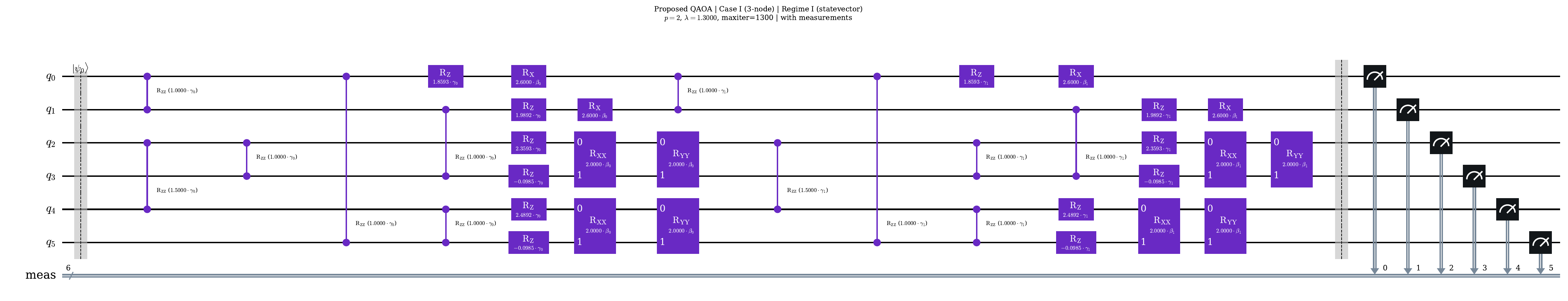}
    \caption{\textcolor{black}{Proposed QAOA circuit (Case I)}}
    \label{fig:17}
\end{figure}
\FloatBarrier

\subsubsection{\textcolor{black}{Case II circuit realizations}}

\textcolor{black}{The following circuits correspond to the balanced symmetric four-node instance used for Case~II.}

\begin{figure}[H]
    \centering
    \includegraphics[width=1.0\linewidth]{case1_balanced_symmetric_statevector_p2_maxiter700.pdf}
    \caption{\textcolor{black}{Standard QAOA circuit (Case II)}}
    \label{fig:18}
\end{figure}

\begin{figure}[H]
    \centering
    \includegraphics[width=1.0\linewidth]{case1_balanced_symmetric_statevector_p2_lambda2p0.pdf}
    \caption{\textcolor{black}{Proposed QAOA circuit (Case II)}}
    \label{fig:19}
\end{figure}
\FloatBarrier

\subsubsection{\textcolor{black}{Case III circuit realizations}}

\textcolor{black}{The following circuits correspond to the customer-cluster four-node instance used for Case~III.}

\begin{figure}[H]
    \centering
    \includegraphics[width=1.0\linewidth]{case2_customer_cluster_statevector_p2_maxiter700.pdf}
    \caption{\textcolor{black}{Standard QAOA circuit (Case III)}}
    \label{fig:20}
\end{figure}

\begin{figure}[H]
    \centering
    \includegraphics[width=1.0\linewidth]{case2_customer_cluster_statevector_p2_lambda2p0.pdf}
    \caption{\textcolor{black}{Proposed QAOA circuit (Case III)}}
    \label{fig:21}
\end{figure}
\FloatBarrier

{\color{black}
\subsection{Algorithmic Optimization Procedures}\phantomsection\label{app:algorithmic_optimization_procedures}\label{sec:16}\label{line:16}

This subsection collects the four optimization procedures used to construct, optimize, and evaluate the QAOA circuits. The procedures distinguish exact statevector optimization, ideal finite-shot optimization, noisy finite-shot optimization, and the shared multi-restart COBYLA routine.

\subsubsection{Statevector QAOA procedure}

This procedure specifies the QUBO-to-Ising conversion, statevector objective evaluation, multi-restart parameter optimization, and final sampling used in the statevector regime.

\begin{algorithm}[htbp]
\caption{Standard QAOA for a QUBO}
\label{algorithm:1}
\DontPrintSemicolon
\small
\KwIn{
QUBO coefficients $(C_{\mathrm{Q}},\{a_i\},\{b_{ij}\})$ with
$C(\mathbf{x})=C_{\mathrm{Q}}+\sum_i a_i x_i+\sum_{i<j} b_{ij} x_i x_j$, \\
number of qubits $n$, QAOA depth $p$, energy scaling factor $s>0$, \\
number of restarts $R$, optimizer budget $T$, sampling shots $S$.
}
\KwOut{
Best sampled bitstring $\hat{\mathbf{x}}\in\{0,1\}^n$ and its QUBO value $C(\hat{\mathbf{x}})$.
}

\BlankLine
\textbf{(1) Convert QUBO to Ising in Pauli-$Z$ basis.}\\
Initialize $C_{\mathrm{I}} \leftarrow C_{\mathrm{Q}}$, $h_i\leftarrow 0$ for all $i$, and $J_{ij}\leftarrow 0$ for all $i<j$.\\
\ForEach{$(i,j)$ with coefficient $b_{ij}$}{
    $J_{ij} \leftarrow J_{ij} + \frac{b_{ij}}{4}$\;
    $h_i \leftarrow h_i - \frac{b_{ij}}{4}$,\quad
    $h_j \leftarrow h_j - \frac{b_{ij}}{4}$\;
    $C_{\mathrm{I}} \leftarrow C_{\mathrm{I}} + \frac{b_{ij}}{4}$\;
}
\ForEach{$i$ with coefficient $a_i$}{
    $h_i \leftarrow h_i - \frac{a_i}{2}$\;
    $C_{\mathrm{I}} \leftarrow C_{\mathrm{I}} + \frac{a_i}{2}$\;
}
\BlankLine
\textbf{(2) Build the (scaled) cost Hamiltonian (without constant).}\\
Define
$
\widetilde{H}_C
=
\sum_{i<j}\frac{J_{ij}}{s}Z_iZ_j
+
\sum_{i=0}^{n-1}\frac{h_i}{s}Z_i
$\;
\BlankLine
\textbf{(3) QAOA objective via statevector expectation.}\\
Define the mixer Hamiltonian $H_M=\sum_{i=0}^{n-1}X_i$.\\
For parameters $\bm{\gamma}=(\gamma_1,\dots,\gamma_p)$ and $\bm{\beta}=(\beta_1,\dots,\beta_p)$, define
\[
U_\ell
=
e^{-i\beta_\ell H_M}
e^{-i\gamma_\ell\widetilde{H}_C}
\qquad
\ell\in\{1,\ldots,p\}
\]
and
\[
\ket{\psi(\bm{\gamma},\bm{\beta})}
=
U_pU_{p-1}\cdots U_1\ket{+}^{\bigotimes n}
\]
Define objective (constant omitted):
$
f(\bm{\gamma},\bm{\beta})
=
\langle\psi(\bm{\gamma},\bm{\beta})|\widetilde{H}_C|\psi(\bm{\gamma},\bm{\beta})\rangle
$\;

\BlankLine
\textbf{(4) Classical optimization with multiple restarts.}\\
Set $f^\star \leftarrow +\infty$.\;
\For{$r=1$ \KwTo $R$}{
    Randomly initialize $\bm{\gamma}^{(0)}\in[-\pi,\pi]^p$ and $\bm{\beta}^{(0)}\in[0,\frac{\pi}{2}]^p$.\;
    Use a derivative-free optimizer (e.g., COBYLA) for at most $T$ iterations to obtain
    $(\bm{\gamma}^{(r)},\bm{\beta}^{(r)}) \approx \arg\min f(\bm{\gamma},\bm{\beta})$.\;
    \If{$f(\bm{\gamma}^{(r)},\bm{\beta}^{(r)}) < f^\star$}{
        $f^\star \leftarrow f(\bm{\gamma}^{(r)},\bm{\beta}^{(r)})$;\;
        $(\bm{\gamma}^\star,\bm{\beta}^\star)\leftarrow (\bm{\gamma}^{(r)},\bm{\beta}^{(r)})$;\;
    }
}

\BlankLine
\textbf{(5) Sampling and decoding.}\\
Prepare the measured QAOA circuit with $(\bm{\gamma}^\star,\bm{\beta}^\star)$ and sample $S$ shots to obtain counts over bitstrings.\;
Let $\hat{\mathbf{x}}$ be the most frequent bitstring after reordering bits to match the variable indexing.\;
Compute the original QUBO value $C(\hat{\mathbf{x}}) = C_{\mathrm{Q}} + \sum_i a_i \hat{x}_i + \sum_{i<j} b_{ij}\hat{x}_i\hat{x}_j$.\;

\Return $(\hat{\mathbf{x}}\quad C(\hat{\mathbf{x}}))$\;
\end{algorithm}
\FloatBarrier

\clearpage
\subsubsection{Ideal finite-shot QAOA procedure}

This procedure uses finite sampling to estimate the objective in independent shot batches while retaining a noiseless circuit model.

\begin{algorithm}[H]
\caption{Shot-based QAOA optimization for a QUBO using sampling}
\label{algorithm:2}
\DontPrintSemicolon
\footnotesize
\KwIn{
QUBO coefficients $(C_{\mathrm{Q}},\{a_i\},\{b_{ij}\})$ with
$C(\mathbf{x})=C_{\mathrm{Q}}+\sum_i a_i x_i+\sum_{i<j} b_{ij} x_i x_j$, 
number of qubits $n$, QAOA depth $p$, scaling factor $s>0$, 
objective shots $S_{\mathrm{obj}}$, final shots $S_{\mathrm{final}}$, batches $B$, 
number of restarts $R$, optimizer budget $T$.
}
\KwOut{
Optimized parameters $\bm{\theta}^{\star}=(\bm{\gamma}^\star,\bm{\beta}^\star)$, final histogram, and best sampled bitstring $\hat{\mathbf{x}}$.
}

\BlankLine
\textbf{(1) Convert QUBO to Ising coefficients in Pauli-$Z$ basis.}\\
Using $x_i=(I-Z_i)/2$, compute $J_{ij}$, $h_i$, and constant $C_{\mathrm{I}}$ such that
\[
\widehat{C}=C_{\mathrm{I}}I + \sum_{i<j} J_{ij} Z_i Z_j + \sum_i h_i Z_i
\]
Define the scaled constant-free Hamiltonian
\[
\widetilde{H}_C
=
\sum_{i<j}\frac{J_{ij}}{s}Z_iZ_j
+
\sum_{i=0}^{n-1}\frac{h_i}{s}Z_i
\]
Only $J_{ij}$ and $h_i$ are needed to build the cost unitary.\;

\BlankLine
\textbf{(2) Define the $p$-layer QAOA circuit.}\\
For parameters $\bm{\gamma}=(\gamma_1,\dots,\gamma_p)$ and $\bm{\beta}=(\beta_1,\dots,\beta_p)$, prepare
\[
U_\ell
=
e^{-i\beta_\ell\sum_{i=0}^{n-1}X_i}
e^{-i\gamma_\ell\widetilde{H}_C}
\qquad
\ell\in\{1,\ldots,p\}
\]
and
\[
|\psi(\bm{\gamma},\bm{\beta})\rangle
=
U_pU_{p-1}\cdots U_1|+\rangle^{\bigotimes n}
\]
Implement $e^{-i\gamma_\ell (J_{ij}/s)Z_iZ_j}$ with $R_{ZZ}$$(2\gamma_\ell J_{ij}/s)$ and
$e^{-i\gamma_\ell (h_i/s)Z_i}$ with $R_{Z}$$(2\gamma_\ell h_i/s)$; implement the mixer with $R_{X}$$(2\beta_\ell)$.\;

\BlankLine
\textbf{(3) Shot-based objective evaluation.}\\
Define the stochastic objective $\widehat{f}(\bm{\gamma},\bm{\beta})$:
\begin{enumerate}
\item For $b=1,\dots,B$:
  run the measured QAOA circuit $S_{\mathrm{obj}}$ shots to obtain samples $\{\mathbf{x}^{(s)}\}$.
  Compute the sample mean
  $\widehat{E}_b=\frac{1}{S_{\mathrm{obj}}}\sum_{s=1}^{S_{\mathrm{obj}}} C(\mathbf{x}^{(s)})$.
\item Return $\widehat{f}(\bm{\gamma},\bm{\beta})=\frac{1}{B}\sum_{b=1}^B \widehat{E}_b / s$.
\end{enumerate}
The sampled objective $\widehat f$ differs from the expectation of $\widetilde H_C$ only by the parameter-independent constant $C_{\mathrm I}/s$, so both objectives have the same minimizers.\;

\BlankLine
\textbf{(4) Shot-based black-box optimization with restarts.}\\
Set $\widehat{f}^\star \leftarrow +\infty$.\;
\For{$r=1$ \KwTo $R$}{
  Randomly initialize $\bm{\gamma}^{(0)}\in[-\pi,\pi]^p$ and $\bm{\beta}^{(0)}\in[0,\pi/2]^p$.\;
  Use a derivative-free optimizer (e.g., COBYLA) for at most $T$ iterations to obtain
  $(\bm{\gamma}^{(r)},\bm{\beta}^{(r)}) \approx \arg\min \widehat{f}(\bm{\gamma},\bm{\beta})$.\;
  \If{$\widehat{f}(\bm{\gamma}^{(r)},\bm{\beta}^{(r)}) < \widehat{f}^\star$}{
    $\widehat{f}^\star \leftarrow \widehat{f}(\bm{\gamma}^{(r)},\bm{\beta}^{(r)})$;\;
    $(\bm{\gamma}^\star,\bm{\beta}^\star)\leftarrow(\bm{\gamma}^{(r)},\bm{\beta}^{(r)})$;\;
  }
}

\BlankLine
\textbf{(5) Final sampling and solution extraction.}\\
Run the measured QAOA circuit with $(\bm{\gamma}^\star,\bm{\beta}^\star)$ for $S_{\mathrm{final}}$ shots to get a histogram over bitstrings.\;
Let $\hat{\mathbf{x}}$ be the most frequent bitstring after bit-order correction if necessary.\;
Compute and report $C(\hat{\mathbf{x}})$ and the full histogram.\;

\Return $(\bm{\theta}^{\star},\hat{\mathbf{x}})$\;
\end{algorithm}
\FloatBarrier

\clearpage
\subsubsection{Noisy finite-shot QAOA procedure}

This procedure extends the finite-shot optimization workflow by applying the prescribed gate and readout noise model during objective estimation and final sampling.

\begin{algorithm}[H]
\caption{Noisy shot-based QAOA optimization for a QUBO}
\label{algorithm:3}
\DontPrintSemicolon
\footnotesize
\KwIn{
QUBO coefficients $(C_{\mathrm{Q}},\{a_i\},\{b_{ij}\})$ with
$C(\mathbf{x})=C_{\mathrm{Q}}+\sum_i a_i x_i+\sum_{i<j} b_{ij} x_i x_j$, \\
number of qubits $n$, QAOA depth $p$, scaling factor $s>0$, \\
noise model $\mathcal{M}_{\mathrm{noise}}$ (gate noise + readout noise), \\
objective shots $S_{\mathrm{obj}}$, final shots $S_{\mathrm{final}}$, batches $B$, \\
number of restarts $R$, optimizer budget $T$.
}
\KwOut{
Optimized parameters $\bm{\theta}^{\star}=(\bm{\gamma}^\star,\bm{\beta}^\star)$, final histogram, and best sampled bitstring $\hat{\mathbf{x}}$.
}

\BlankLine
\textbf{(1) Map QUBO to an Ising Hamiltonian in the Pauli-$Z$ basis.}\\
Using $x_i=(I-Z_i)/2$, compute coefficients $J_{ij}$, $h_i$, and $C_{\mathrm{I}}$ such that
\[
\widehat{C}=C_{\mathrm{I}}I+\sum_{i<j}J_{ij}Z_iZ_j+\sum_i h_i Z_i
\]
Define the scaled cost Hamiltonian with the constant omitted
\[
\widetilde{H}_C
=
\sum_{i<j}\frac{J_{ij}}{s}Z_iZ_j
+
\sum_{i=0}^{n-1}\frac{h_i}{s}Z_i
\]

\BlankLine
\textbf{(2) Define the $p$-layer QAOA circuit family.}\\
For $\bm{\gamma}=(\gamma_1,\dots,\gamma_p)$ and $\bm{\beta}=(\beta_1,\dots,\beta_p)$, prepare
\[
U_\ell
=
e^{-i\beta_\ell\sum_{i=0}^{n-1}X_i}
e^{-i\gamma_\ell\widetilde{H}_C}
\qquad
\ell\in\{1,\ldots,p\}
\]
and
\[
|\psi(\bm{\gamma},\bm{\beta})\rangle
=
U_pU_{p-1}\cdots U_1|+\rangle^{\bigotimes n}
\]
Implement $e^{-i\gamma_\ell (J_{ij}/s)Z_iZ_j}$ via $R_{ZZ}$$(2\gamma_\ell J_{ij}/s)$,
$e^{-i\gamma_\ell (h_i/s)Z_i}$ via $R_{Z}$$(2\gamma_\ell h_i/s)$,
and the mixer via $R_{X}$$(2\beta_\ell)$.

\BlankLine
\textbf{(3) Noisy shot-based objective evaluation.}\\
Let $\widehat{f}(\bm{\gamma},\bm{\beta})$ be a stochastic objective computed using a noisy sampler under a fixed noise model:
\begin{enumerate}
\item For $b=1,\dots,B$:
  execute the measured QAOA circuit on the noisy sampler with fixed gate noise and readout noise
  for $S_{\mathrm{obj}}$ shots to obtain samples $\{\mathbf{x}^{(s)}\}$.
\item Compute the batch sample mean
  $\widehat{E}_b=\frac{1}{S_{\mathrm{obj}}}\sum_{s=1}^{S_{\mathrm{obj}}} C(\mathbf{x}^{(s)})$.
\item Return $\widehat{f}(\bm{\gamma},\bm{\beta})=\frac{1}{B}\sum_{b=1}^B \widehat{E}_b / s$.
\end{enumerate}
As in the noiseless shot-based procedure, this objective and the expectation of $\widetilde H_C$ differ only by $C_{\mathrm I}/s$, which is independent of the variational parameters.\;

\BlankLine
\textbf{(4) Noise-adapted parameter optimization with restarts.}\\
Set $\widehat{f}^\star\leftarrow +\infty$.\;
\For{$r=1$ \KwTo $R$}{
  Randomly initialize $\bm{\gamma}^{(0)}\in[-\pi,\pi]^p$ and $\bm{\beta}^{(0)}\in[0,\pi/2]^p$.\;
  Use a derivative-free optimizer (e.g., COBYLA) for at most $T$ iterations to obtain
  $(\bm{\gamma}^{(r)},\bm{\beta}^{(r)}) \approx \arg\min \widehat{f}(\bm{\gamma},\bm{\beta})$.\;
  \If{$\widehat{f}(\bm{\gamma}^{(r)},\bm{\beta}^{(r)}) < \widehat{f}^\star$}{
    $\widehat{f}^\star\leftarrow \widehat{f}(\bm{\gamma}^{(r)},\bm{\beta}^{(r)})$;\;
    $(\bm{\gamma}^\star,\bm{\beta}^\star)\leftarrow(\bm{\gamma}^{(r)},\bm{\beta}^{(r)})$;\;
  }
}

\BlankLine
\textbf{(5) Final noisy sampling and solution extraction.}\\
Execute the measured QAOA circuit with $(\bm{\gamma}^\star,\bm{\beta}^\star)$ under noise model $\mathcal{M}_{\mathrm{noise}}$
for $S_{\mathrm{final}}$ shots to obtain a histogram over bitstrings.\;
Let $\hat{\mathbf{x}}$ be the most frequent bitstring after correcting the bit order if needed.\;
Report $C(\hat{\mathbf{x}})$ and the full histogram.\;

\Return $(\bm{\theta}^{\star},\hat{\mathbf{x}})$\;
\end{algorithm}
\FloatBarrier

\subsubsection{Multi-restart COBYLA procedure}

This procedure gives the common restart-based COBYLA routine that selects the best variational parameters for the regime-specific objective.

\begin{algorithm}[H]
\caption{Multi-restart COBYLA for generalized QAOA}
\label{algorithm:generalized-qaoa-cobyla-inner}
\DontPrintSemicolon
\small
\KwIn{
Regime-specific objective
$
F(\bm{\gamma},\bm{\beta})
$,
evaluated either by an exact statevector expectation or by the prescribed noiseless or noisy sampling procedure, \\
where $\bm{\gamma}=(\gamma_1,\dots,\gamma_p)$ and $\bm{\beta}=(\beta_1,\dots,\beta_p)$; \\
number of restarts $R$, optimizer budget $T$, \\
trust-region size $\rho_{\mathrm{beg}}>0$, stopping tolerance $\tau>0$.
}
\KwOut{
Best parameters $(\bm{\gamma}^\star,\bm{\beta}^\star)$ and best objective value $F^\star$.
}

\BlankLine
Set $F^\star \leftarrow +\infty$.\;
\For{$r=1$ \KwTo $R$}{
    Randomly initialize
    $\bm{\gamma}^{(0,r)}\in[-\pi,\pi]^p$ and
    $\bm{\beta}^{(0,r)}\in\left[0,\frac{\pi}{2}\right]^p$.\;
    Run COBYLA on $F(\bm{\gamma},\bm{\beta})$ for at most $T$ iterations,
    with initial trust-region size $\rho_{\mathrm{beg}}$ and tolerance $\tau$,
    to obtain $(\bm{\gamma}^{(r)},\bm{\beta}^{(r)})$.\;
    \If{$F(\bm{\gamma}^{(r)},\bm{\beta}^{(r)}) < F^\star$}{
        $F^\star \leftarrow F(\bm{\gamma}^{(r)},\bm{\beta}^{(r)})$\;
        $(\bm{\gamma}^\star,\bm{\beta}^\star)\leftarrow (\bm{\gamma}^{(r)},\bm{\beta}^{(r)})$\;
    }
}

\Return $(\bm{\gamma}^\star\quad\bm{\beta}^\star\quad F^\star)$\;
\end{algorithm}
}
\FloatBarrier

% \section{CRediT} \label{6}
% \textbf{Yuan-Zheng Lei}: Conceptualization, Methodology, Writing - original draft. \textbf{Yaobang Gong}: Writing - original draft. \textbf{Xianfeng Terry Yang}: Conceptualization, Methodology and Supervision, Writing - original draft.

\section{\textcolor{black}{Model Source Code}} \label{sec:model_source_code}\label{sec:20}\label{line:20}
\textcolor{black}{The source code of the proposed model is available for readers to download via the following link: \url{https://github.com/EdisonYLei/Improving-Feasibility-in-QAOA-for-VRP}.}

\section{Acknowledgments} \label{8}
\textcolor{black}{The authors thank the two anonymous reviewers and the Area Editor for their constructive comments and suggestions, which substantially improved the quality of this manuscript. Any remaining errors are the authors' own.}
\newpage
\bibliography{ref}

@article{carmo2025warm,
  title={Warm-Starting QAOA with XY Mixers: A Novel Approach for Quantum-Enhanced Vehicle Routing Optimization},
  author={Carmo, Rafael S do and Santana, Marcos and Fanchini, Felipe F and de Albuquerque, Victor Hugo C and Papa, Jo{\~a}o Paulo},
  journal={arXiv preprint arXiv:2504.19934},
  year={2025}
}

@article{azfar2025quantum,
  title={Quantum-Assisted Vehicle Routing: Realizing QAOA-based Approach on Gate-Based Quantum Computer},
  author={Azfar, Talha and Raisuddin, Osama Muhammad and Ke, Ruimin and Holguin-Veras, Jose},
  journal={arXiv preprint arXiv:2505.01614},
  year={2025}
}

@article{somvanshi2026quantum,
  title={Quantum computing in transportation engineering: a survey},
  author={Somvanshi, Shriyank and Das, Subasish and Islam, Md Monzurul and Polock, Sazzad Bin Bashar and Chhetri, Gaurab and Anderson, Darrell},
  journal={IEEE Transactions on Intelligent Transportation Systems},
  year={2026},
  publisher={IEEE}
}

@article{azad2022solving,
  title={Solving vehicle routing problem using quantum approximate optimization algorithm},
  author={Azad, Utkarsh and Behera, Bikash K and Ahmed, Emad A and Panigrahi, Prasanta K and Farouk, Ahmed},
  journal={IEEE Transactions on Intelligent Transportation Systems},
  volume={24},
  number={7},
  pages={7564--7573},
  year={2022},
  publisher={IEEE}
}

@article{du2026overcoming,
  title={Overcoming computational challenges in air transportation: A quantum computing perspective of the status quo and future applicability},
  author={Du, Zhuoming and Wandelt, Sebastian and Sun, Xiaoqian},
  journal={Transportation Research Part C: Emerging Technologies},
  volume={184},
  pages={105505},
  year={2026},
  publisher={Elsevier}
}

@article{clarke1964scheduling,
  title={Scheduling of vehicles from a central depot to a number of delivery points},
  author={Clarke, Geoff and Wright, John W},
  journal={Operations research},
  volume={12},
  number={4},
  pages={568--581},
  year={1964},
  publisher={Informs}
}

@article{flood1956traveling,
  title={The traveling-salesman problem},
  author={Flood, Merrill M},
  journal={Operations research},
  volume={4},
  number={1},
  pages={61--75},
  year={1956},
  publisher={INFORMS}
}

@article{preskill2018quantum,
  title={Quantum computing in the NISQ era and beyond},
  author={Preskill, John},
  journal={Quantum},
  volume={2},
  pages={79},
  year={2018},
  publisher={Verein zur F{\"o}rderung des Open Access Publizierens in den Quantenwissenschaften}
}

@article{montanez2024unbalanced,
  title={Unbalanced penalization: A new approach to encode inequality constraints of combinatorial problems for quantum optimization algorithms},
  author={Monta{\~n}ez-Barrera, Jhon Alejandro and Willsch, Dennis and Maldonado-Romo, Alberto and Michielsen, Kristel},
  journal={Quantum Science and Technology},
  volume={9},
  number={2},
  pages={025022},
  year={2024},
  publisher={IOP Publishing}
}

@inproceedings{palackal2023quantum,
  title={Quantum-assisted solution paths for the capacitated vehicle routing problem},
  author={Palackal, Lilly and Poggel, Benedikt and Wulff, Matthias and Ehm, Hans and Lorenz, Jeanette Miriam and Mendl, Christian B},
  booktitle={2023 IEEE International Conference on Quantum Computing and Engineering (QCE)},
  volume={1},
  pages={648--658},
  year={2023},
  organization={IEEE}
}

@article{blekos2024review,
  title={A review on quantum approximate optimization algorithm and its variants},
  author={Blekos, Kostas and Brand, Dean and Ceschini, Andrea and Chou, Chiao-Hui and Li, Rui-Hao and Pandya, Komal and Summer, Alessandro},
  journal={Physics Reports},
  volume={1068},
  pages={1--66},
  year={2024},
  publisher={Elsevier}
}

@article{marxer2025above,
  title={Above 99.9\% fidelity single-qubit gates, two-qubit gates, and readout in a single superconducting quantum device},
  author={Marxer, Fabian and Mro{\.z}ek, Jakub and Andersson, Joona and Abdurakhimov, Leonid and Adam, Janos and Bergholm, Ville and Beriwal, Rohit and Chan, Chun Fai and Dahl, Saga and Das, Soumya Ranjan and others},
  journal={arXiv preprint arXiv:2508.16437},
  year={2025}
}

@article{li2023error,
  title={Error per single-qubit gate below 10- 4 in a superconducting qubit},
  author={Li, Zhiyuan and Liu, Pei and Zhao, Peng and Mi, Zhenyu and Xu, Huikai and Liang, Xuehui and Su, Tang and Sun, Weijie and Xue, Guangming and Zhang, Jing-Ning and others},
  journal={npj Quantum Information},
  volume={9},
  number={1},
  pages={111},
  year={2023},
  publisher={Nature Publishing Group UK London}
}

@article{chen2023transmon,
  title={Transmon qubit readout fidelity at the threshold for quantum error correction without a quantum-limited amplifier},
  author={Chen, Liangyu and Li, Hang-Xi and Lu, Yong and Warren, Christopher W and Kri{\v{z}}an, Christian J and Kosen, Sandoko and Rommel, Marcus and Ahmed, Shahnawaz and Osman, Amr and Bizn{\'a}rov{\'a}, Janka and others},
  journal={npj Quantum Information},
  volume={9},
  number={1},
  pages={26},
  year={2023},
  publisher={Nature Publishing Group UK London}
}

@article{wang202499,
  title={99.9\%-fidelity in measuring a superconducting qubit},
  author={Wang, Can and Liu, Feng-Ming and Chen, He and Du, Yi-Fei and Ying, Chong and Wang, Jian-Wen and Huo, Yong-Heng and Peng, Cheng-Zhi and Zhu, Xiaobo and Chen, Ming-Cheng and others},
  journal={arXiv preprint arXiv:2412.13849},
  year={2024}
}

@book{toth2002vehicle,
  title={The vehicle routing problem},
  author={Toth, Paolo and Vigo, Daniele},
  year={2002},
  publisher={SIAM}
}

@article{ralphs2003capacitated,
  title={On the capacitated vehicle routing problem},
  author={Ralphs, Ted K and Kopman, Leonid and Pulleyblank, William R and Trotter, Leslie E},
  journal={Mathematical programming},
  volume={94},
  number={2},
  pages={343--359},
  year={2003},
  publisher={Springer}
}

@inproceedings{choi2019tutorial,
  title={A tutorial on quantum approximate optimization algorithm (QAOA): Fundamentals and applications},
  author={Choi, Jaeho and Kim, Joongheon},
  booktitle={2019 international conference on information and communication technology convergence (ICTC)},
  pages={138--142},
  year={2019},
  organization={IEEE}
}

@article{zhou2020quantum,
  title={Quantum approximate optimization algorithm: Performance, mechanism, and implementation on near-term devices},
  author={Zhou, Leo and Wang, Sheng-Tao and Choi, Soonwon and Pichler, Hannes and Lukin, Mikhail D},
  journal={Physical Review X},
  volume={10},
  number={2},
  pages={021067},
  year={2020},
  publisher={APS}
}

@article{streif2021beating,
  title={Beating classical heuristics for the binary paint shop problem with the quantum approximate optimization algorithm},
  author={Streif, Michael and Yarkoni, Sheir and Skolik, Andrea and Neukart, Florian and Leib, Martin},
  journal={Physical Review A},
  volume={104},
  number={1},
  pages={012403},
  year={2021},
  publisher={APS}
}

@article{borle2021quantum,
  title={Quantum approximate optimization for hard problems in linear algebra},
  author={Borle, Ajinkya and Elfving, Vincent and Lomonaco, Samuel J},
  journal={SciPost Physics Core},
  volume={4},
  number={4},
  pages={031},
  year={2021}
}

@article{farhi2014quantum,
  title={A Quantum Approximate Optimization Algorithm Applied to a Bounded Occurrence Constraint Problem},
  author={Farhi, Edward and Goldstone, Jeffrey and Gutmann, Sam},
  journal={arXiv preprint arXiv:1412.6062},
  year={2014}
}

@inproceedings{awasthi2023quantum,
  title={Quantum computing techniques for multi-knapsack problems},
  author={Awasthi, Abhishek and B{\"a}r, Francesco and Doetsch, Joseph and Ehm, Hans and Erdmann, Marvin and Hess, Maximilian and Klepsch, Johannes and Limacher, Peter A and Luckow, Andre and Niedermeier, Christoph and others},
  booktitle={Science and information conference},
  pages={264--284},
  year={2023},
  organization={Springer}
}

@article{moussa2022unsupervised,
  title={Unsupervised strategies for identifying optimal parameters in quantum approximate optimization algorithm},
  author={Moussa, Charles and Wang, Hao and B{\"a}ck, Thomas and Dunjko, Vedran},
  journal={EPJ Quantum Technology},
  volume={9},
  number={1},
  pages={11},
  year={2022},
  publisher={Springer Berlin Heidelberg}
}

@article{ising1925beitrag,
  title={Beitrag zur theorie des ferromagnetismus},
  author={Ising, Ernst},
  journal={Zeitschrift f{\"u}r Physik},
  volume={31},
  number={1},
  pages={253--258},
  year={1925},
  publisher={Springer}
}

@article{barahona1989experiments,
  title={Experiments in quadratic 0--1 programming},
  author={Barahona, Francisco and J{\"u}nger, Michael and Reinelt, Gerhard},
  journal={Mathematical programming},
  volume={44},
  number={1},
  pages={127--137},
  year={1989},
  publisher={Springer}
}

@inproceedings{shor1994algorithms,
  title={Algorithms for quantum computation: discrete logarithms and factoring},
  author={Shor, Peter W},
  booktitle={Proceedings 35th annual symposium on foundations of computer science},
  pages={124--134},
  year={1994},
  organization={Ieee}
}

@article{steane1998quantum,
  title={Quantum computing},
  author={Steane, Andrew},
  journal={Reports on Progress in Physics},
  volume={61},
  number={2},
  pages={117--173},
  year={1998}
}

@incollection{preskill2023quantum,
  title={Quantum computing 40 years later},
  author={Preskill, John},
  booktitle={Feynman lectures on computation},
  pages={193--244},
  year={2023},
  publisher={CRC Press}
}

@article{rieffel2000introduction,
  title={An introduction to quantum computing for non-physicists},
  author={Rieffel, Eleanor and Polak, Wolfgang},
  journal={ACM Computing Surveys (CSUR)},
  volume={32},
  number={3},
  pages={300--335},
  year={2000},
  publisher={ACM New York, NY, USA}
}

@article{james2001measurement,
  title={Measurement of qubits},
  author={James, Daniel FV and Kwiat, Paul G and Munro, William J and White, Andrew G},
  journal={Physical Review A},
  volume={64},
  number={5},
  pages={052312},
  year={2001},
  publisher={APS}
}

@article{chatterjee2021semiconductor,
  title={Semiconductor qubits in practice},
  author={Chatterjee, Anasua and Stevenson, Paul and De Franceschi, Silvano and Morello, Andrea and de Leon, Nathalie P and Kuemmeth, Ferdinand},
  journal={Nature Reviews Physics},
  volume={3},
  number={3},
  pages={157--177},
  year={2021},
  publisher={Nature Publishing Group UK London}
}

@article{wang2019accelerated,
  title={Accelerated variational quantum eigensolver},
  author={Wang, Daochen and Higgott, Oscar and Brierley, Stephen},
  journal={Physical review letters},
  volume={122},
  number={14},
  pages={140504},
  year={2019},
  publisher={APS}
}

@article{peruzzo2014variational,
  title={A variational eigenvalue solver on a photonic quantum processor},
  author={Peruzzo, Alberto and McClean, Jarrod and Shadbolt, Peter and Yung, Man-Hong and Zhou, Xiao-Qi and Love, Peter J and Aspuru-Guzik, Al{\'a}n and O’brien, Jeremy L},
  journal={Nature communications},
  volume={5},
  number={1},
  pages={4213},
  year={2014},
  publisher={Nature Publishing Group UK London}
}

@article{tilly2022variational,
  title={The variational quantum eigensolver: a review of methods and best practices},
  author={Tilly, Jules and Chen, Hongxiang and Cao, Shuxiang and Picozzi, Dario and Setia, Kanav and Li, Ying and Grant, Edward and Wossnig, Leonard and Rungger, Ivan and Booth, George H and others},
  journal={Physics Reports},
  volume={986},
  pages={1--128},
  year={2022},
  publisher={Elsevier}
}

@article{harwood2021formulating,
  title={Formulating and solving routing problems on quantum computers},
  author={Harwood, Stuart and Gambella, Claudio and Trenev, Dimitar and Simonetto, Andrea and Bernal, David and Greenberg, Donny},
  journal={IEEE transactions on quantum engineering},
  volume={2},
  pages={1--17},
  year={2021},
  publisher={IEEE}
}

@article{mohanty2023analysis,
  title={Analysis of the vehicle routing problem solved via hybrid quantum algorithms in the presence of noisy channels},
  author={Mohanty, Nishikanta and Behera, Bikash K and Ferrie, Christopher},
  journal={IEEE Transactions on Quantum Engineering},
  volume={4},
  pages={1--14},
  year={2023},
  publisher={IEEE}
}

@article{baker2022wasserstein,
  title={Wasserstein solution quality and the quantum approximate optimization algorithm: a portfolio optimization case study},
  author={Baker, Jack S and Radha, Santosh Kumar},
  journal={arXiv preprint arXiv:2202.06782},
  year={2022}
}

@article{vikstaal2020applying,
  title={Applying the quantum approximate optimization algorithm to the tail-assignment problem},
  author={Vikst{\aa}l, Pontus and Gr{\"o}nkvist, Mattias and Svensson, Marika and Andersson, Martin and Johansson, G{\"o}ran and Ferrini, Giulia},
  journal={Physical Review Applied},
  volume={14},
  number={3},
  pages={034009},
  year={2020},
  publisher={APS}
}

@article{egger2021warm,
  title={Warm-starting quantum optimization},
  author={Egger, Daniel J and Mare{\v{c}}ek, Jakub and Woerner, Stefan},
  journal={Quantum},
  volume={5},
  pages={479},
  year={2021},
  publisher={Verein zur F{\"o}rderung des Open Access Publizierens in den Quantenwissenschaften}
}

@inproceedings{bartschi2020grover,
  title={Grover mixers for QAOA: Shifting complexity from mixer design to state preparation},
  author={B{\"a}rtschi, Andreas and Eidenbenz, Stephan},
  booktitle={2020 IEEE International Conference on Quantum Computing and Engineering (QCE)},
  pages={72--82},
  year={2020},
  organization={IEEE}
}

@article{cooper2021exploring,
  title={Exploring potential applications of quantum computing in transportation modelling},
  author={Cooper, Crispin HV},
  journal={IEEE Transactions on Intelligent Transportation Systems},
  volume={23},
  number={9},
  pages={14712--14720},
  year={2021},
  publisher={IEEE}
}

@inproceedings{onah2025quest,
  title={QUEST: QUantum-Enhanced Shared Transportation},
  author={Onah, Chinonso and Misciasci, Neel and Othmer, Carsten and Michielsen, Kristel},
  booktitle={2025 IEEE International Conference on Quantum Computing and Engineering (QCE)},
  volume={1},
  pages={2149--2160},
  year={2025},
  organization={IEEE}
}

@inproceedings{ak2025quantum,
  title={Quantum-enhanced Optimization for LNG Ship Routing: Integrating Digital Twins with QAOA},
  author={Ak, Elif and Van Huynh, Dang and Duong, Trung Q},
  booktitle={2025 IEEE International Conference on Communications Workshops (ICC Workshops)},
  pages={769--774},
  year={2025},
  organization={IEEE}
}

@article{picariello2025quantum,
  title={Quantum Approaches to Urban Logistics: From Core QAOA to Clustered Scalability},
  author={Picariello, Fabio and Turati, Gloria and Antonelli, Riccardo and Bailo, Igor and Bonura, Susanna and Ciarfaglia, Gianmarco and Cipolla, Salvatore and Cremonesi, Paolo and Dacrema, M Ferrari and Gabusi, M and others},
  journal={arXiv preprint arXiv:2512.10813},
  year={2025}
}

@inproceedings{harikrishnakumar2021smart,
  title={Smart rebalancing for bike sharing systems using quantum approximate optimization algorithm},
  author={Harikrishnakumar, Ramkumar and Nannapaneni, Saideep},
  booktitle={2021 IEEE International Intelligent Transportation Systems Conference (ITSC)},
  pages={2257--2263},
  year={2021},
  organization={IEEE}
}

@book{nielsen2010quantum,
  title={Quantum computation and quantum information},
  author={Nielsen, Michael A and Chuang, Isaac L},
  year={2010},
  publisher={Cambridge university press}
}

@article{zhuang2024quantum,
  title={Quantum computing in intelligent transportation systems: A survey},
  author={Zhuang, Yifan and Azfar, Talha and Wang, Yinhai and Sun, Wei and Wang, Xiaokun and Guo, Qianwen and Ke, Ruimin},
  journal={Chain},
  volume={1},
  number={2},
  pages={138--149},
  year={2024},
  publisher={Youke Publishing}
}

@article{massimiliano2026quantum,
  title={Quantum Computation in Air Transport: A Short Overview of the Fundamentals, Challenges and Opportunities},
  author={Massimiliano, Zanin and Zhuoming, Du and Giorgi, Gian Luca and Sun, Xiaoqian and Wandelt, Sebastian},
  journal={Technologies},
  volume={14},
  number={2},
  pages={103},
  year={2026},
  publisher={MDPI AG}
}

@article{udekwe2025q,
  title={Q-restore: quantum-driven framework for resilient and equitable transportation network restoration},
  author={Udekwe, Daniel and Ke, Ruimin and Lu, Jiaqing and Guo, Qian-wen},
  journal={arXiv preprint arXiv:2501.11197},
  year={2025}
}

@article{ke6431107quantum,
  title={Quantum Optimization for Public Transit Systems: A Quantum Hardware-Aware Analysis of Higher-Order Formulations},
  author={Ke, Ruimin and Guo, Qian-wen},
  journal={Available at SSRN 6431107}
}

@article{ke2026impact,
  title={Impact-Driven Quantum Decomposition for Traffic Zone Partitioning: A Hybrid Gate-Model Framework},
  author={Ke, Ruimin and Azfar, Talha and Huang, Kaicong and Li, Shuyang},
  journal={arXiv preprint arXiv:2605.01127},
  year={2026}
}

@article{azfar2026hardware,
  title={Hardware-Efficient Quantum Optimization for Transportation Networks via Compressed Adiabatic Evolution},
  author={Azfar, Talha and Ke, Ruimin and He, Sean and Wang, Cara and Holgu{\'\i}n-Veras, Jos{\'e}},
  journal={arXiv preprint arXiv:2604.26175},
  year={2026}
}

@book{groenland2025introduction,
  title={Introduction to Quantum Computing for Business},
  author={Groenland, Koen},
  year={2025},
  publisher={Amsterdam University Press}
}

@misc{ibmhardware2026, author = {{IBM Quantum}}, title = {IBM Quantum Computing: Hardware and Roadmap}, year = {2026}, url = {https://www.ibm.com/quantum/hardware}, note = {Accessed 17 June 2026} }

@misc{ibm100k2023, author = {{IBM Quantum}}, title = {Charting the Course to 100,000 Qubits}, year = {2023}, month = may,
url = {https://www.ibm.com/quantum/blog/100k-qubit-supercomputer}, note = {Accessed 17 June 2026} }

@misc{ibm2022,
  author       = {{IBM Quantum}},
  title        = {{IBM Unveils 400 Qubit-Plus Quantum Processor and
                  Next-Generation IBM Quantum System Two}},
  year         = {2022},
  month        = nov,
  day          = {9},
  url = {https://newsroom.ibm.com/2022-11-09-IBM-Unveils-400-Qubit-Plus-Quantum-Processor-and-Next-Generation-IBM-Quantum-System-Two},
  note         = {Accessed 17 June 2026}
}

@misc{quantinuumH12022,
  author       = {{Quantinuum}},
  title        = {{Quantinuum Completes Hardware Upgrade, Achieves 20 Fully Connected Qubits}},
  year         = {2022},
  month        = jun,
  day          = {14},
  url = {https://www.quantinuum.com/press-releases/quantinuum-completes-hardware-upgrade-achieves-20-fully-connected-qubits},
  note         = {Accessed: 17 June 2026}
}

@misc{quantinuumH22023,
  author       = {{Quantinuum}},
  title        = {{For the First Time Ever, Quantinuum's New H2 Quantum Computer Has Created Non-Abelian Topological Quantum Matter and Braided Its Anyons}},
  year         = {2023},
  month        = may,
  day          = {9},
  url = {https://www.quantinuum.com/press-releases/for-the-first-time-ever-quantinuums-new-h2-quantum-computer-has-created-non-abelian-topological-quantum-matter-and-braided-its-anyons},
  note         = {Accessed: 17 June 2026}
}

@misc{quantinuumH22024,
  author       = {{Quantinuum}},
  title        = {{Quantinuum's H-Series Hits 56 Physical Qubits That Are All-to-All Connected, and Departs the Era of Classical Simulation}},
  year         = {2024},
  month        = jun,
  day          = {5},
  url = {https://www.quantinuum.com/blog/quantinuums-h-series-hits-56-physical-qubits-that-are-all-to-all-connected-and-departs-the-era-of-classical-simulation},
  note         = {Accessed: 17 June 2026}
}

@misc{quantinuumHelios2025,
  author       = {{Quantinuum}},
  title        = {{Quantinuum Announces Commercial Launch of New Helios Quantum Computer That Offers Unprecedented Accuracy to Enable Generative Quantum AI}},
  year         = {2025},
  month        = nov,
  day          = {5},
  url = {https://www.quantinuum.com/press-releases/quantinuum-announces-commercial-launch-of-new-helios-quantum-computer-that-offers-unprecedented-accuracy-to-enable-generative-quantum-ai-genqai},
  note         = {Accessed: 17 June 2026}
}

@misc{quantinuum2024roadmap,
  author       = {{Quantinuum}},
  title        = {{Quantinuum Unveils Accelerated Roadmap to Achieve Universal, Fully Fault-Tolerant Quantum Computing by 2030}},
  year         = {2024},
  month        = sep,
  day          = {10},
  url = {https://www.quantinuum.com/press-releases/quantinuum-unveils-accelerated-roadmap-to-achieve-universal-fault-tolerant-quantum-computing-by-2030},
  note         = {Accessed: 17 June 2026}
}

@article{onah2026optimal,
  title   = {Optimal, Qubit-Efficient Quantum Vehicle Routing via Colored-Permutations},
  author  = {Onah, Chinonso and Michielsen, Kristel},
  journal = {arXiv preprint arXiv:2604.04570},
  year    = {2026},
  doi     = {10.48550/arXiv.2604.04570},
  url     = {https://arxiv.org/abs/2604.04570}
}

\end{document}